\documentclass[preprint,12pt]{elsarticle}

\usepackage{amssymb}
\usepackage{amsmath}
\usepackage{geometry}                
\usepackage{graphicx}
\usepackage{amssymb}
\usepackage{epstopdf}
\usepackage{color}
\usepackage{listings}
\usepackage{xcolor}
\usepackage{pdfpages}
\usepackage{lineno}
\usepackage{hyperref}
\usepackage{selinput}

\usepackage[title,titletoc]{appendix}

\usepackage{caption}
\DeclareCaptionFont{white}{\color{white}}
\DeclareCaptionFormat{listing}{\colorbox{gray}{\parbox{\textwidth}{#1#2#3}}}
\newcommand{\pp}{pp}

\newcounter{bla}

\journal{Computer Physics Communications}

\begin{document}

\begin{frontmatter}



\title{The JETSCAPE framework}


\author[a]{J.~Putschke}

\author[a,b]{K.~Kauder}

\author[n]{E.~Khalaj\corref{author}}

\author[o]{A.~Angerami}

\author[c]{S.~A.~Bass}

\author[a]{S.~Cao}

\author[d]{J.~Coleman}

\author[e]{L.~Cunqueiro}

\author[c]{T.~Dai}

\author[f]{L.~Du}

\author[p,q,r]{H.~Elfner}

\author[f]{D.~Everett}

\author[c]{W.~Fan}

\author[g,h]{R.~J.~Fries}

\author[i]{C.~Gale}

\author[j]{Y.~He}

\author[f]{U.~Heinz}

\author[k,l]{B.~V.~Jacak}

\author[k,l]{P.~M.~Jacobs}

\author[i]{S.~Jeon}

\author[c]{W.~Ke}

\author[g,h]{M.~Kordell~II}

\author[a]{A.~Kumar}

\author[j]{T.~Luo}

\author[a]{A.~Majumder}

\author[f]{M.~McNelis}

\author[k,l]{J.~Mulligan}

\author[e]{C.~Nattrass}

\author[l]{D.~Ollinychenko}

\author[a,i]{D.~Pablos}

\author[k,l]{L.-G.~Pang}

\author[i]{C. Park}

\author[c]{J.-F. Paquet}

\author[m]{G.~Roland}

\author[b]{B.~Schenke}

\author[n]{L.~Schwiebert}

\author[a]{C.~Shen}

\author[a]{C.~Sirimanna}

\author[a,o]{R.~A.~Soltz}

\author[a]{Y.~Tachibana}

\author[a,f]{G.~Vujanovic}

\author[j,k,l]{X.-N.~Wang}

\author[d]{R.~L.~Wolpert}

\author[c]{Y.~Xu}

\author[g,h] {Z.~Yang}

\cortext[author] {Corresponding author.\\\textit{E-mail address:} mekhalaj@wayne.edu}
\address[a]{Department of Physics and Astronomy, Wayne State University, Detroit MI 48201.}
\address[b]{Department of Physics, Brookhaven National Laboratory, Upton NY 11973.}
\address[c]{Department of Physics, Duke University, Durham, NC 27708.}
\address[d]{Department of Statistics, Duke University, Durham, NC 27708.}
\address[e]{Department of Physics and Astronomy, University of Tennessee, Knoxville TN 37996.}
\address[f]{Department of Physics, The Ohio State University, Columbus OH 43210.}
\address[g]{Department of Physics and Astronomy, Texas A\&M University, College Station TX 77843.}
\address[h]{Cyclotron Institute, Texas A\&M University, College Station TX 77843.}
\address[i]{Department of Physics, McGill University, Montreal QC, Canada H3A 2T8.}
\address[j]{Key Laboratory of Quark and Lepton Physics (MOE) and Institute of Particle Physics, Central China Normal University, Wuhan 430079, China.}
\address[k]{Department of Physics, University of California, Berkeley CA 94720.}
\address[l]{Nuclear Science Division, Lawrence Berkeley National Laboratory, Berkeley CA 94720.}
\address[m]{Department of Physics, MIT, Cambridge MA 02139.}
\address[n]{Department of Computer Science, Wayne State University, Detroit MI 48202.}
\address[o]{Lawrence Livermore National Laboratory, Livermore, CA 94550.}
\address[p]{GSI Helmholtzzentrum f\"{u}r Schwerionenforschung, 64291 Darmstadt, Germany.}
\address[q]{Institute for Theoretical Physics, Goethe University, 60438 Frankfurt am Main, Germany.}
\address[r]{Frankfurt Institute for Advanced Studies, 60438 Frankfurt am Main, Germany.}


\begin{abstract} 
The JETSCAPE simulation framework is an overarching computational envelope for developing complete event generators for heavy-ion collisions.   
It allows for modular incorporation of a wide variety of existing and future software that simulates different aspects of a heavy-ion collision. 
The default JETSCAPE package contains both the framework, and an entire set of indigenous and third party routines that can be used to directly compare with experimental data. 
In this article, we outline the algorithmic design of the JETSCAPE framework, define the interfaces and describe the default modules required to carry out full simulations of heavy-ion collisions within this package. 
We begin with a description of the various physics elements required to simulate an entire event in a heavy-ion collision, and distribute these within a flowchart representing the event generator and statistical routines for comparison with data. 
This is followed by a description 
of the abstract class structure, with associated members and functions required for this flowchart to work. We then define the interface that will be 
required for external users of JETSCAPE to incorporate their code within this framework and to modify existing elements within the default distribution. 
We conclude with a discussion of some of the physics output for both $p$-$p$ and $A$-$A$ collisions from the default distribution, and an outlook towards future releases. 
In the appendix, we discuss various architectures on which this code can be run and outline our benchmarks on similar hardware.
\end{abstract}


\end{frontmatter}



{\bf PROGRAM SUMMARY/NEW VERSION PROGRAM SUMMARY}

\begin{small}
\noindent
{\em Program Title:} JETSCAPE v1.3.                                      \\
{\em Licensing provisions(please choose one):} GPLv3.                                   \\
{\em Programming language:} C++.                                \\
{\em Computer:} Commodity PCs ({\it linux}), Macs, intermediate and large clusters. \\
{\em Supplementary material:} Short Manual at K.~Kauder [JETSCAPE Collaboration].
  Nucl.\ Phys.\ A {\bf 982}, 615 (2019)
  [arXiv:1807.09615 [hep-ph]].                                 \\
{\em Journal reference of previous version:} N/A.                  \\
{\em Does the new version supersede the previous version?:} N/A.   \\
{\em Reasons for the new version:} N/A. \\
{\em Summary of revisions:} N/A.*\\
{\em Nature of problem(approx. 50-250 words):}\\
Simulations of high energy heavy-ion collisions require a multitude of interacting elements, from simulations of the incoming nuclei, to the thermalization of the deposited energy-momentum, viscous fluid dynamical expansion, hadronization and freeze-out, as well as the production of hard partons, their propagation and interaction with the dense medium, escape and fragmentation into jets. To compare with high-statistics, event-by-event experimental data, requires a modular and extendable event generator, with state-of-the-art components modeling each aspect of the collision.\\
{\em Solution method(approx. 50-250 words):  }\\
A modular event generator is designed and released, and described in this article. Each factorizable physics component is set up as a separate module. The modules are connected and executed by an elaborate task based framework which initializes and executes the modules and conveys requisite information between them. The modules are set up using the property of inheritance within  C++ class structure. This allows users to design new modules by overloading base classes. A series of mature simulators which focus on specific aspects of the collision are used to overload the base classes in the default distribution.
\\
{\em Additional comments including Restrictions and Unusual features (approx. 50-250 words):} N/A. \\
   \\
* Items marked with an asterisk are only required for new versions
of programs previously published in the CPC Program Library.\\
\end{small}




\newpage
\tableofcontents
\newpage


\section{Introduction}

With the advent of heavy-ion collisions at the Large Hadron Collider (LHC), high energy nuclear physics, in particular the study of the Quark Gluon Plasma (QGP), transitioned from a discovery phase~\cite{Adams:2005dq,Adcox:2004mh,Arsene:2004fa,Back:2004je} to one of systematic exploration. 
Relativistic fluid dynamical simulations which described the evolution of the QGP, developed over a decade prior to the start of the LHC heavy-ion program, required only moderate enhancement in the effort to encompass the bulk behavior from $\sqrt{s} \sim 100$ GeV, at the Relativistic Heavy-Ion Collider (RHIC), to several TeV, at the LHC~\cite{Luzum:2009sb,Song:2010mg,Gale:2012rq}. However, with the incorporation of several new physics elements, such as fluctuating initial states, pre-equilibrium phases, hadronic afterburners etc., along with the need for temperature dependent shear and bulk viscosities, it became clear that there needed to be an event-by-event approach to compare theory predictions to experimental data. This needed to be far more extensive than prior simulators, and it needed an advanced statistical framework to determine (or reliably estimate) the multitude of unknown parameters (approximately 15 for a 3+1D simulation with bulk and shear viscosity) ~\cite{Novak:2013bqa}. 

Progress on the hard probe sector had been comparatively slower. Prior to the start of the LHC, there existed three clearly different formalisms for energy loss of high energy partons: A few-scattering-per-emission formalism and a multiple-scattering-induced-emission formalism based on perturbation theory, and others based on strong coupling. 
Each of these formalisms came in several flavors of their own: The few scattering formalism was implemented within the Higher Twist (HT) approach~\cite{Wang:2001ifa,Guo:2000nz,Majumder:2009ge} which included the parton's interaction with the bulk medium in terms of gluon matrix elements~\cite{Majumder:2012sh}, and the Gyulassy-Levai-Vitev (GLV) approach~\cite{Gyulassy:1999zd,Gyulassy:2000fs,Gyulassy:2000er}
which used the heavy static center model of Gyulassy and Wang (GW)~\cite{Wang:1991xy}. The multiple scattering induced emission approach consisted of two separate implementations, which also differed on how the dense medium was modeled: The Baier-Dokshitzer-Mueller-Peigne-Schiff (BDMPS) approach~\cite{Baier:1996sk,Baier:1998yf,Baier:1996kr} used the GW model, while the Arnold-Moore-Yaffe (AMY)~\cite{Arnold:2002ja,Arnold:2001ms,Arnold:2001ba} approach was based on a description of the medium using Hard-Thermal-Loop effective theory~\cite{Braaten:1989kk,Braaten:1989mz,Frenkel:1989br}. 

Beyond the systematics of how the medium is included in jet quenching calculations, another point of departure became the means by which multiple emissions are handled in each of these approaches: The higher-twist (and GLV) were by definition cast in the approximation of a thin medium, and were thus meant to be applied to the case where the virtuality of the shower was higher, where scattering in the medium was a correction to the vacuum shower. This lended itself rather naturally to a medium modified DGLAP~\cite{Wang:2001ifa,Majumder:2009zu,Kang:2014xsa} evolution for the fragmentation function of the leading hadron. This eventually led to the formulation of MATTER, a vacuum-like shower generator based on a medium modified Sudakov form factor~ \cite{Majumder:2013re,Cao:2017qpx}. In contrast to this is the rare emission approach of the BDMPS and AMY formalisms. These energy loss approaches were cast for jets with a virtuality of the order of that generated by multiple scattering in the medium. As such these used either a Poisson emission probability~\cite{Salgado:2003gb} or a rate equation~\cite{Turbide:2005fk,Jeon:2003gi,Qin:2009bk} to simulate the probability of rare gluon emissions. The rate equation approach became central to the formulation of the MARTINI event generator~\cite{Schenke:2009gb}.

In contrast to the perturbative approaches outlined above, in the period prior to the start up of the LHC, there arose several approaches to parton energy loss based on the AdS/CFT correspondence~\cite{Maldacena:1997re}. These assumed that the leading parton was strongly coupled to the QGP medium. These approaches modeled energy loss as the drag experienced by one end of a string fixed on a $d$-brane in 4 dimensions, whose remainder was continuously being drawn into a black hole situated at some depth in a 5th dimension~\cite{Herzog:2006gh,Gubser:2006bz,Gubser:2007xz,Chesler:2007an,Horowitz:2011wm}. Other approaches based on the broadening experienced by a light quark were also formulated~\cite{Liu:2006ug,Liu:2006he,CasalderreySolana:2007qw}. The dragging string approaches were included in the construction of an event generator called the HyBRID model~\cite{Casalderrey-Solana:2014bpa,Casalderrey-Solana:2015vaa}, which included string drag on each parton in the shower history from the vacuum event generator PYTHIA~\cite{Sjostrand:2007gs,Sjostrand:2014zea}. To date this remains the only event generator based on the strong coupling approach. 

With the ever increasing amount of data and complexity of observables from the LHC and RHIC, it became increasingly clear that a pure application of a single approach would not be able to explain the entirety of jet based observables, measured over a range of energies and centralities. There was a need to develop an overarching framework where different formalisms that are applicable to different epochs of the jet evolution can be realistically combined. The first set of attempts at such a framework were  mostly analytical approaches~\cite{Baier:2000mf,CaronHuot:2010bp}; these tended to be difficult to turn into a representative phenomenology. At the same time, the calculations of the soft sector, confronted with ever greater volume and variety of data had evolved into event generators~\cite{Shen:2014vra,Schenke:2010rr,Weller:2017tsr}. The evolution of event averaged energy loss calculations to Monte-Carlo generators included straightforward generalizations such as MATTER, MARTINI and the HyBRID model, which, like the energy loss calculations on which they are based, were initially cast as stand-alone codes. At the level of event generators, the situation is further complicated by the inclusion of ``mixed'' approach Monte-Carlos, such as the LBT based simulators~\cite{He:2015pra,Cao:2016gvr} which use the HT kernel in a rate equation, or Q-PYTHIA~\cite{Armesto:2009fj}, which applies the BDMPS kernel in the construction of a medium modified Sudakov, as well as bottom-up approaches such as JEWEL~\cite{Zapp:2008gi,Zapp:2012ak} which construct an energy loss methodology based on physical arguments and reproduce established formalisms within certain limits. The comparison between these efforts is further complicated by the fact that they use varying degrees of approximations for interaction with the bulk medium, from a parametrized medium in JEWEL, to a full 3+1D viscous fluid dynamical simulation with energy deposition in the case of LBT.

As a result, any global framework for calculations relevant to heavy-ion collisions must fulfill the following requirements: 

\begin{enumerate} 
\item Be in the form of an event generator, which will produce complete events that may be simultaneously analyzed and compared to an ever growing list of observables. \item Be as modular as possible, allowing for modification of physics elements at the most granular level. 
\item Be comprehensive in its content and yet remain easily extendable as new physics phenomena come into focus, as well as allow for new connections between different modules, hitherto assumed to be factorized.
\end{enumerate} 

In the remainder of this Article, we describe such a framework. The Jet Energy-loss Tomography with a Statistically and Computationally Advanced Program Envelope (JETSCAPE) suite contains both a modular event generator framework, state-of-the-art modules that simulate each sector of a heavy-ion collision, and Bayesian statistical routines for calibration and rigorous comparison with experimental data. In the rest of this Article, we focus on the framework (Sec.~\ref{framework}), along with a brief description of physics elements in the default distribution (Sec.~\ref{physics}).  A handful of results from jet and hadron energy loss simulations in a 2+1D hydro medium, as well as comparisons to \pp\ reference measurements are presented in Sec.~\ref{validation}. We conclude the introduction by pointing out that while jet energy loss is part of the acronym, JETSCAPE is by no means solely a jet event generator; it is also a state-of-the-art generator for the simulation of bulk dynamics in a heavy-ion collision, and can be used as such without invoking the production of jets. 
We also point out JETSCAPE is an ongoing project and new modules are being continually added. This manual describes the content of the package JETSCAPE v1.3. A discussion of upcoming versions is contained in Sec.~\ref{summary}.

\section{Physics of JETSCAPE}
\label{physics}

It is now widely accepted that jet quenching is a multi-scale phenomenon~\cite{Majumder:2010qh,Majumder:2015qqa,Caucal:2018dla}. Jets start out as single partons with a virtuality that far exceeds any scale in the medium, and progressively lose virtuality via sequential emissions. As the virtuality becomes comparable to the medium scale, the mode of energy loss must change. Some portion of the jet will remain weakly coupled while some portion may indeed become strongly coupled. 
The models for jet modification highlighted above form only a subset of all available approaches. Models differ in how QGP is modeled, in their approximations of the number of scatterings, in the amount of radiation they predict and in the nature of the coupling assumed between partons and the medium.

\begin{figure}[h]
 \begin{center}
   $\mbox{}$ \hspace{-1.05cm} \includegraphics[width=1.05\textwidth]{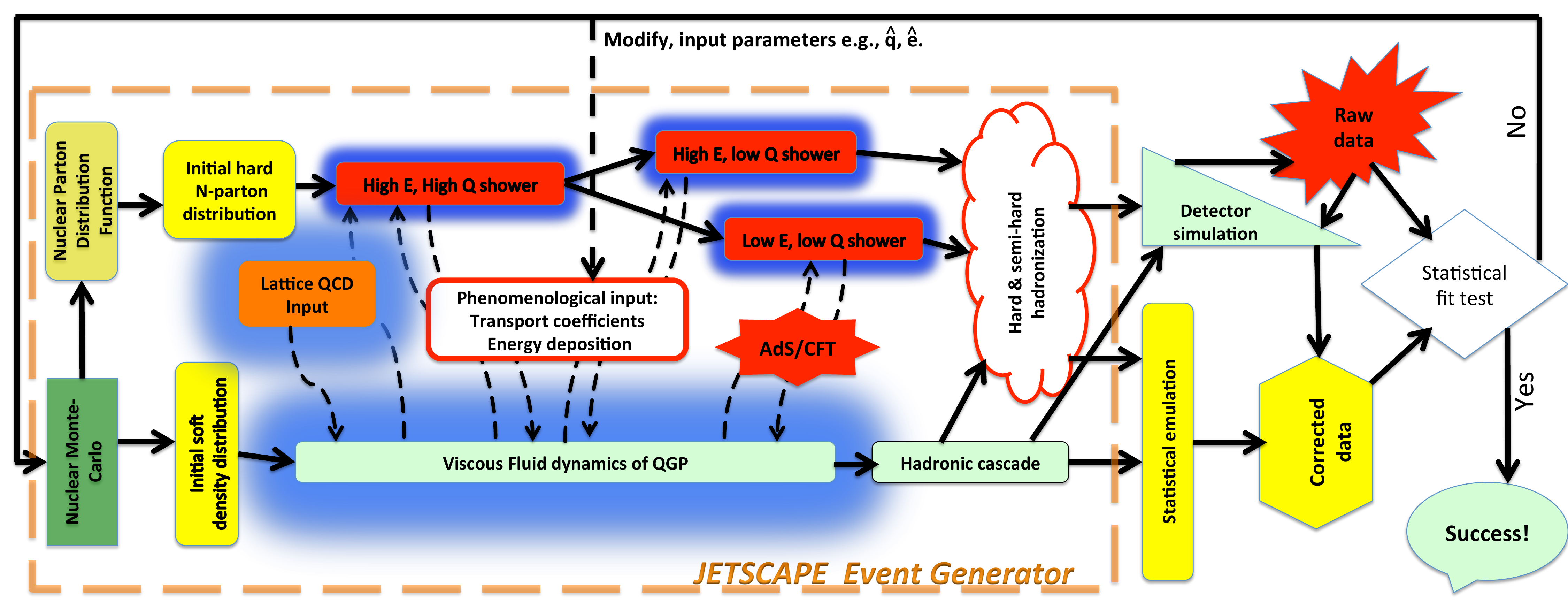}
\caption{A flow chart of the JETSCAPE event generator and associated statistical routines.}
\label{fig:flowchart}
\end{center}
\end{figure}

In order to sort through the available models and to determine the correct combination of models based on their physics capabilities requires an \emph{event generator framework}: A flexible and modular software framework which allows every user access to a variety of state-of-the-art modules which simulate separate portions of a heavy-ion collision, while the user can focus on only one or two modules related to his(her) expertise. Such a framework must include all modules that compose a soft sector event generator.
As a result, it must contain a pre-collision module that simulates the locations of the nucleons in a nucleus, to an initial state module which describes the evolution from the overlap between the two nuclei to the generation of an energy momentum tensor in local equilibrium, followed by a viscous fluid dynamical routine, followed by hadronization and a hadronic afterburner. There should be separate modules that generate hard scatterings, i.e, the progenitors of jets. 
This is followed by the jet quenching modules which have a controlled interaction with the soft sector, followed by modules that feed energy momentum back into the fluid medium and hadronize the escaping jet and exited medium. An illustration of the JETSCAPE framework is presented in the dashed box in Fig.~\ref{fig:flowchart}.

In Fig.~\ref{fig:flowchart}, each colored box within the dashed rectangle represents a software module that simulates a factorized portion of a heavy-ion collision. The various lines represent the controlled flow of information that constitutes the framework. The modules beyond the event-generator represent the various statistical routines that are required to compare the results of the event generator with experimental data. Each module within the event generator has parameters that control when the module starts and stops operation as well as how each operation depends on the input provided to each module. A large portion of these parameters are correlated. In a typical run, a user may encounter anywhere between 15 to 20 parameters that have to be set by comparison with a subset of experimental data. Advanced Bayesian techniques are used to carry out this calibration. In this article, we will focus on the design and operation of the JETSCAPE event generator. In the remainder of this section, we describe both the physics carried out by each of the boxes in the flow chart and briefly outline the typical simulation code that are default instances of these portions of codes. 

In the rest of the chapter, we are going to introduce the physics of various aspects of heavy-ion collisions which can be described by JETSCAPE. We are introducing the codes which are included in JETSCAPE as the default solution. Users are free to replace any of these modules with their own code.

\subsection{Initial State}

The initial state module of JETSCAPE (referred to as  Nuclear Monte-Carlo in the fig) has two important tasks. The first task is to 
calculate the initial entropy deposition into the soft medium and the components of its stress-energy tensor $T^{\mu \nu} (x,y,\eta_s, \tau)$.
This part serves as the long wave length initial condition for pre-equilibrium evolution and fluid dynamic expansion.
The second task is to compute the positions of the vertices for the initial hard processes $R_{bin}(x,y,\eta_s, \tau)$, which would then be passed to the jet production modules.

In relativistic heavy ion collisions at top RHIC and LHC energies, the nuclei are not only Lorentz contracted along the beam direction, but also time dilated to form
saturated gluon distributions that are described by the MC-KLN model \cite{McLerran:1993ni,McLerran:1993ka,Gelis:2010nm} and IP-Glasma \cite{Schenke:2012wb} models.
However, regardless of initial condition, the nucleon density distributions inside heavy nuclei are usually given by the generalized Woods-Saxon distribution 
with (or without) deformation,

\begin{equation}
\rho\left(r,\theta\right)=\frac{\rho_0}{1+\exp{\left(\frac{r-R(1+\beta_2Y_2^0(\theta)+\beta_4Y_4^0(\theta))}{a}\right)}}
\end{equation}
where $\theta$ is the polar angle with respect to the symmetry axis of the nucleus, $\rho_0$ is the nuclear density, $R_0$ is the effective radius, $Y_{2}^{0}(\theta)=\frac{1}{4}\sqrt{\frac{5}{\pi}} (3 \cos^2 \theta - 1)$ and $Y_{4}^{0}(\theta)=\frac{3}{16\sqrt{\pi}} (35 \cos^4 \theta - 30 \cos^2 \theta + 3)$ are spherical harmonic functions, $\beta_2$ and $\beta_4$ controls the deformation from the spherical shape.
For spherical nuclei such as ${}^{208} Pb$, $\beta_2 = \beta_4 = 0$.
JETSCAPE
In Monte Carlo Glauber Model \cite{Miller:2007ri}, the positions of the nucleons within a nucleus are sampled via a Monte-Carlo sampling routine with excluded volume to sample the locations of the centers
of the nucleons. It also assumes the nucleons to be hard spheres (modifiable assumption).
Using the radius parameters, the code will generate a random impact parameter $b$, it will then overlap the two nuclei and calculate the distribution of hard and soft interaction vertices (refer to \ref{sec:hard_n_parton_dist} for more details).

The default Nuclear Monte-Carlo model used in JETSCAPE is the TRENTO \cite{Moreland:2014oya} parametric initial condition model. This model is capable of mimicking the behavior of different classes of physical initial condition models, such as the Wounded Nucleon \cite{Bialas:1976ed} model, the MC-KLN model \cite{McLerran:1993ni,McLerran:1993ka,Gelis:2010nm}, the IP-Glasma \cite{Schenke:2012wb} model and the EKRT \cite{Niemi:2015qia} saturation models.
Notice that TRENTO is good at mimicking the ratio between geometric eccentricities $\epsilon_3$ and $\epsilon_2$ as a function of centrality, other than the detailed fluctuation structures (such as the color fluctuations raised in some initial state models). On the other hand, TRENTO is very efficient in generating millions of events whose initial total entropy and geometric eccentricities are consistent with experimental charged multiplicity and flow harmonics for various different collisions systems.
A priori, multiple different physics models exist for the description of the initial state and it will fall to the statistical analysis to verify or falsify these models in the context of the full JETSCAPE calculation.

In the JETSCAPE framework, we have provided several predefined collision systems (Au+Au $\sqrt{s_{NN}}=200$ GeV, Pb+Pb $\sqrt{s_{NN}}=2.76$ TeV and $\sqrt{s_{NN}}=5.02$ TeV collisions).
These systems have used the default entropy deposition parameter $p=0$ (IP-Glasma-like) and other default setups to produce several initial total entropy versus collision centrality tables.
As a bonus, it is quite easy to generate fluctuating initial conditions of one specific centrality range for these systems, using the {jetscape\_init.xml} configuration file.
In this mode, the inputs are \emph{collision\_system}, \emph{centrality\_min}, \emph{centrality\_max}.
There are also options in the configuration file to read initial conditions from external file, whose path is set in the xml file through \emph{initial\_profile\_path}.\\

{\bf Output:}
\begin{itemize}
\item spatial distribution of the vertices for the initial hard processes $R_{bin}(x,y,\eta_s,\tau)$

\item the distribution of energy density $e (x,y,\eta_s, \tau_0)$ or entropy density $s (x, y, \eta_s, \tau_0)$ at initial time $\tau_0$.
\end{itemize}

\subsection{Pre-hydrodynamic Stage}

Both phenomenological model-to-data comparisons ~\cite{Bernhard:2018hnz} and our understanding of QCD at high temperatures~\cite{Arnold:2002zm} suggest that the fireball created in a heavy ion collision can expand and deexcite for some time before its evolution is best described by viscous hydrodynamics. This stage of the dynamics can be called a `pre-hydrodynamic' phase. Usual initial state models for heavy ion collisions give some initial conditions for the stress-energy tensor $T^{\mu\nu}$ of the fireball directly from the particle production process. Any pre-hydrodynamic model should propagate the stress tensor from the initial energy deposition time $\tau_0 \approx 0$ to a later time, which we will call a `switching time' $\tau_s$, at which hydrodynamics can successfully describe the evolution of the stress tensor:
\begin{equation}
T^{\mu\nu}(\tau_0) \rightarrow T^{\mu\nu} (\tau_s).
\end{equation}  

It is not obvious what is the best approach for modeling the pre-hydrodynamic stage. Perturbative QCD arguments~\cite{Arnold:2002zm} rooted in the expectation that at high gluon density the effective coupling constant becomes small~\cite{McLerran:2008es} motivate a kinetic theory approach, where in the pre-hydrodynamic stage the parton distribution evolves according to the Boltzmann equation:
\begin{equation} \label{eq:boltz}
p \cdot \partial f(x;p) = C[f(x;p)].
\end{equation}
Here $p$ is a momentum four vector, $f(x;p)$ is the particle phase space distribution, and $C[f(x;p)]$ is the collision term which describes the change in particle phase space density from interactions. 
In kinetic theory, the stress tensor is defined as the following moment of the particle distribution function:
\begin{equation}
T^{\mu\nu} = \frac{g}{(2\pi)^3} \int \frac{d^3p}{E} p^{\mu} p^{\nu} f(x;p).
\end{equation}
Here $g$ is a degeneracy factor, and $E$ the energy. 

Taking the weak-coupling philosophy to the extreme, one can ignore the collision term in equation (\ref{eq:boltz}), which leads to the free-streaming limit~\cite{Liu:2015nwa}.  In this case $ p \cdot \partial f(x;p) = 0$. This approximation has the disadvantage that it can never thermalize the medium, and the matching to hydrodynamics gets more difficult for larger values of the switching time ~\cite{Liu:2015nwa}. A more physical model is Effective Kinetic Theory (EKT) ~\cite{Arnold:2002zm, Kurkela:2018vqr, Kurkela:2018oqw} which includes both scattering and fusion/splitting processes in the collision integrals for the quark and gluon distributions. This approach leads to thermalization and can be smoothly matched to dissipative hydrodynamics ~\cite{Kurkela:2018vqr}. Importantly, EKT dynamics is very close to free-streaming dynamics at early times \cite{Kurkela:2018vqr}, lending some justification to the free-streaming approximation at times $\tau < \tau_s$. 
In CLVisc hydro simulations with initial conditions from AMPT model, partonic cascade is used to take into account of the scattering among medium partons before matching the energy and momentum tensor into the hydro simulation.
Lacking a (3+1)-d implementation of the EKT approach, the default module resorts to the free-streaming approximation for the pre-hydrodynamic evolution. 

At the switching time $\tau_s$, the stress-tensor can be matched to viscous hydrodynamics by Landau-matching. The energy density $\epsilon$ and flow velocity $u^{\mu}$ are the eigenvalue and timelike eigenvector of the stress-tensor:
\begin{equation}
T^{\mu}_{\nu} u^{\nu} = \epsilon u^{\mu}.
\end{equation}  
A viscous hydrodynamic decomposition is given by 
\begin{equation}
T^{\mu\nu} = \epsilon  u^{\mu} u^{\nu} - (p+\Pi)\Delta^{\mu\nu} + \pi^{\mu\nu} 
\end{equation}
where $\Delta^{\mu\nu} \equiv g^{\mu\nu} - u^{\mu}u^{\nu}$ projects onto the space orthogonal to the flow, $p$ is the thermal pressure, $\pi^{\mu\nu}$ is the shear-stress tensor and $\Pi$ the bulk viscous pressure. This provides the initial conditions for viscous hydrodynamics at the switching time. 

{\bf Input:}

$\bullet$ Energy density $\epsilon(x, y, \eta_s; \tau_0)$ on grid points in $(x,y,\eta_s)$

{\bf Output:}

$\bullet$ Energy density $\epsilon(x, y, \eta_s; \tau_s)$ , flow-velocity $u^{\mu}(x, y, \eta_s; \tau_s)$, pressure $p(x, y, \eta_s; \tau_s)$, shear-stress $\pi^{\mu\nu}(x, y, \eta_s; \tau_s)$, and bulk pressure $\Pi(x, y, \eta_s; \tau_s)$ on grid points in $(x,y,\eta_s)$

\subsection{Fluid Dynamics}

Relativistic fluid dynamics is the most successful model that describes the momentum distribution of final state soft hadrons in high energy heavy ion collisions. The success of the model comes from the strong collective flow generated in fluid dynamic expansion of hot quark gluon plasma and dense hadronic matter. Starting from fluctuating initial conditions where the energy momentum tensor and net charges are provided by Nuclear Monte Carlo models, relativistic dissipative hydrodynamics solve a group of coupled partial differential equations to compute the time evolution of energy momentum tensor, conserved charges, shear viscous tensor and bulk viscosity on 3-dimensional lattice.

The expansion rate of the system is determined by the pressure gradient, with pressure as a function of energy density and net baryon chemical potential from the equation of state (EoS). For LHC and top RHIC energy collisions where the net baryon chemical potential is negligible, the EoS of strongly coupled QCD matter is determined by first principle lattice QCD computations, which makes relativistic hydrodynamics the best model on the market to describe the non-perturbative many body interactions between soft partons and hadrons.

There are many open sourced implementations such as vHLLE\cite{Karpenko:2013wva}, iEBE-VISHNU \cite{Shen:2014vra}, MUSIC \cite{Schenke:2010nt}, CLVisc\cite{Pang:2018zzo}, GPU-VH\cite{Bazow:2016yra} ... that can be plugged in the JETSCAPE framework for the fluid dynamic evolution module. These programs are developed to solve the same relativistic fluid dynamic equation, with different levels of approximations, for high energy collisions involving jet physics,
\begin{equation}
    \nabla_{\mu} T^{\mu\nu} = J^{\nu},\ {\rm with\ } T^{\mu\nu} = (\varepsilon + P + \Pi)u^{\mu} u^{\nu} - (P+\Pi)g^{\mu\nu} + \pi^{\mu\nu}
    \label{eq:hydro_eq}
\end{equation}
where $\nabla_{\mu}$ is the covariant derivative, $T^{\mu\nu}$ is the energy-momentum tensor, $J^{\nu}$ is the energy-momentum deposition to the medium by jet shower propagation,
$\varepsilon$ is the energy density, $P$ is the pressure given by EoS, $u^{\mu}$ is the fluid four-velocity obeying $u_{\mu}u^{\mu}=1$,
$\Pi$ is the bulk viscosity and $\pi^{\mu\nu}$ is the shear-stress tensor.

The evolution equations of the dissipative currents $\Pi$ and $\pi^{\mu\nu}$ can be written as follows,
\begin{eqnarray}
D \Pi = - \frac{1}{\tau_\Pi} (\Pi + \zeta \theta) - \frac{\delta_{\Pi\Pi}}{\tau_\Pi} \Pi \theta + \frac{\lambda_{\Pi \pi}}{\tau_\Pi} \pi^{\mu\nu} \sigma_{\mu\nu}
\label{eq1.5.1}
\end{eqnarray}
and
\begin{eqnarray}
\Delta^{\mu\nu}_{\alpha \beta} D \pi^{\alpha \beta} &=& - \frac{1}{\tau_\pi} (\pi^{\mu\nu} - 2 \eta \sigma^{\mu\nu}) - \frac{\delta_{\pi\pi}}{\tau_{\pi}} \pi^{\mu\nu} \theta - \frac{\tau_{\pi\pi}}{\tau_\pi} \pi^{\lambda \langle} \sigma^{\nu \rangle}\,_\lambda  \nonumber \\
&& + \frac{\phi_7}{\tau_\pi} \pi^{\langle \mu}\,_\alpha \pi^{\nu \rangle \alpha} + \frac{\lambda_{\pi \Pi}}{\tau_\pi} \Pi \sigma^{\mu\nu}
\label{eq1.5.2}.
\end{eqnarray}
where $D = u^\mu \nabla_\mu$, $\Delta_{\alpha\beta}^{\mu\nu} = 1/2 (\Delta^\mu_\alpha \Delta^\nu_\beta + (\Delta^\mu_\beta     \Delta^\nu_\alpha) - 1/3 \Delta^{\mu\nu} \Delta_{\alpha \beta}$ and $\Delta^{\mu\nu} = g^{\mu\nu} - u^\mu u^\nu$.
The second order transport coefficients are summarized in Table~\ref{table1.5.1}.
\begin{table}[htp]
\centering
\begin{tabular}{c|c|c|c|c|c}
\hline \hline
$\delta_{\Pi\Pi}$ & $\lambda_{\Pi \pi}$ & $\delta_{\pi\pi}$ & $\tau_{\pi\pi}$ & $\phi_7$ & $\lambda_{\pi \Pi}$ \\ \hline
 $\frac{2}{3}\tau_\Pi$ & $\frac{8}{5} (\frac{1}{3} - c_s^2) \tau_\Pi $ & $\frac{4}{3}\tau_\pi$ & $\frac{10}{7} \tau_\pi$ & $\frac{9}{70} \frac{4}{e + \mathcal{P}}$ & $\frac{6}{5}\tau_\pi$ \\ \hline
\end{tabular}
\caption{The second order transport coefficients derived kinetic theory at small mass limit \cite{Denicol:2014vaa}.}
\label{table1.5.1}
\end{table}

Zero net baryon density is assumed in the current framework which is appropriate for the scope of JETSCAPE.
Without conserved current in the system, the EoS is usually provided as a function of the local energy density $e$ in the form of a text table for the hydrodynamics code to read in. The numerical table can be in the format as suggested in Table~\ref{table1.5.2}.
\begin{table}[ht!]
\centering
\begin{tabular}{c|c|c|c}
\hline \hline
$e$ (GeV/fm$^4$) & $\mathcal{P}(e)$ (GeV/fm$^4$) & $s(e)$ (1/fm$^3$) & $T(e)$ (GeV) \\ \hline
\dots & \dots & \dots & \dots \\
\end{tabular}
\caption{An example of the EoS table format for the hydrodynamics code.}
\label{table1.5.2}
\end{table}

{\bf Input:}
\begin{itemize}
    \item Energy momentum tensor $T^{\mu\nu}(x, y, \eta_s; \tau_0)$ on grid points at a given longitudinal proper time $\tau_0$
    \item Choices of the transport coefficients, namely the specific shear viscosity $\eta/s(T)$, the specific bulk viscosity $\zeta/s(T)$, their relaxation time $\tau_\pi$, $\tau_\Pi$, and other second order transport coefficients.
    \item The equation of state (EoS) of the QCD matter $\mathcal{P}(e)$.
\end{itemize}

{\bf Output:}

\begin{itemize}
    \item The positions $x^\mu$, surface element vectors $d^3\sigma_\mu$, and hydrodynamic variables ($u^\mu$, $\pi^{\mu\nu}$, $\Pi$) on the isothermal hyper-surface in (3+1)D at the switching temperature $T_\mathrm{sw}$ to the hadronic transport
    \item Evolution of the hydrodynamic variable, such as $T$, $u^\mu$,  $\pi^{\mu\nu}$, $\Pi$ on the grid. (Optional)
\end{itemize}

We point out, that for the first year deliverable, the code will be run several times with a wide range of impact parameters, without the inclusion of hard partons or jets. The produced $E_{T}$ or number of charged particles will be used to bin the events in various centrality bins. Hard partons and jet energy loss will be run on these binned events, with no modification in centrality due to the presence of the jet.

\subsection{Hard N Parton Distribution}
\label{sec:hard_n_parton_dist}

In a $p$-$p$ collision, processes resulting in the production of a hadron $h$ with transverse momentum $p_T > 1$GeV (at any rapidity $y$, well within the range of beam rapidity) can be described using perturbative QCD, with the hard scattering cross section $\hat{\sigma}$ (of partons $a+b$ to $c + X$) factorized from the initial state Parton Distribution Functions (PDFs) for the two incoming partons ($G_{p\rightarrow a}, G_{p \rightarrow b} $), and the fragmentation function for the out-going parton ($D_{c \rightarrow h}$)~\cite{Collins:1989gx,Collins:1988ig,Collins:1985ue,Collins:1983ju,Collins:1981uw}:
\begin{eqnarray}
\frac{d^3 \sigma_{p + p \rightarrow h + X }}{dy d^2p_T} &=& \int dx_a \int dx_b 
G_{p\rightarrow a}(x_a, \mu^2_i) G_{p \rightarrow b}(x_b, \mu^2_i) \nonumber \\
&\times& \frac{d \hat{\sigma}_{a+b \rightarrow c + X} (Q^2)}{d\hat{t}} 
\frac{D_{c \rightarrow h}(z, \mu^2_f)}{\pi z}.
\end{eqnarray}

In the equation above $\hat{t} = (p_c - p_a)^2$ and the away-side parton $X$ is integrated out. 
The incoming momentum fractions $x_a, x_b$ denote the fractions of the forward (or light-cone) momentum of the protons carried by the partons $a$ and $b$, while $z$ denotes the fraction of the transverse momentum of the outgoing parton $c$ carried by the hadron $h$  which is produced in the fragmentation of the parton. Each of these non-perturbative functions $G(x,\mu^2_i), D(x,\mu^2_f)$ is factorized from the hard scattering cross section at the scale $\mu_i^2, \mu_f^2$, while the hard scattering cross section is renormalized at the scale $Q^2$. Typically, one sets $\mu_i^2 = \mu_f^2 = Q^2 = p_T^2$, which represents that both initial and final state radiation up to this scale is resummed within the soft functions $G$ and $D$. These perturbative resummations, as well as the calculation of the hard cross section can now be carried out order by order in perturbation theory. Processes that can be described using such a factorized pQCD approach are referred to as hard processes. 

In the PYTHIA event generator~\cite{Sjostrand:2014zea,Corke:2009pm,Sjostrand:2007gs,Bengtsson:1987kr}, hard scattering is ensured by requiring that the $p_T$ exchanged between partons $a,c$ be within a range $\hat{p}_T^{min} \leq\hat{p}_T \leq\hat{p}_T^{max}$. The factorized initial state is generated by starting from the PDF at the scale $\mu_i^2 = \hat{p}_T^2$ and then generating the initial state radiation by evolving back down to the soft scale of 1~GeV. This is followed by several hard interactions between the partons leading to the formation of $N$ hard partons with $p_T > 1$GeV. Typically these are then subjected to final state radiation starting at the scale $p_T$ and evolving down to the soft scale of $1$GeV. In neither the initial state or final state portions of the PYTHIA shower is there any notion of location, especially in the direction of propagation of the shower. 

In the case of jet production in a heavy-ion collision, one requires the transverse location of the hard scatterings and the location of the partons in the outgoing shower within the evolving medium. In the current implementation of hard scattering within JETSCAPE, the PYTHIA portion of the process is terminated at the generation of the $N$ hard partons. Locations of the hard scatterings in the transverse plane (at $t,z\!=\!0$) are determined using the initial state simulation as described below. The spatial location of the partons in the outgoing shower are determined by the JETSCAPE framework. Each shower initiating parton, unless modified, propagates from the starting point at $t=z=0$ along light like trajectories in the direction  $\vec{v} = \vec{p}/E$. The framework retains the starting point of each parton. As each of the energy loss modules introduce splits, the origins and momenta of each of the daughter partons is retained by the framework. At every time step, the framework checks with the energy loss modules to determine whether or not there is a split. In which case the current parton is placed in the history of the shower and replaced with the new partons in the current shower.

As the two nuclei collide, there is an event-by-event fluctuating 2-D probability distribution for binary nucleon nucleon scattering. This probability distribution, calculated by the initial state model (usually TRENTO in JETSCAPE), is typically sampled to calculate the locations of hard partonic scattering in a nucleus nucleus collision.  There are several considerations related to this process: A single nucleon in a large nucleus, is different from a nucleon in isolation, as a result, its parton distribution is expected to be different. These differences, typically measured in Deep-Inelastic Scattering (DIS) on a large nucleus or in $p$-$A$ collisions are described by a Nuclear Parton Distribution Function (NPDF). Hard partonic scatterings are probabilistically disfavored and are thus rare. As a nucleon proceeds through a large nucleus, it undergoes several soft interactions prior to the hard interaction that leads to the formation of a jet pair. As a result, one can expect both a change in the NPDF as well as a correlation between jet production and the underlying event activity~\cite{Kordell:2016njg}.
Within the JETSCAPE framework, the modification to the PDF, caused by the multiple soft collisions, is so far ignored. As a result, the default distribution cannot be used to describe the centrality dependence of jet cross sections in $p-A$ collisions where such effects are expected to be considerable. The modification of the initial state PDF due to static nuclear interactions, can be simulated with the use of NPDFs.

\subsection{Energy Loss}
\label{ELoss}

This portion of the code will propagate and progressively modify the collection of hard partons, taken from the N-hard parton distribution module, through the dense medium, and onward to the hadronization modules. The user has considerable flexibility in this portion of the code to interlace various different energy loss schemes (or design their own jet modification routines). The current interface will hand a single parton to the energy loss routine in question and accept $n \geq 0$ partons in return. The basic role of the energy loss routine is to determine the momenta of the $n$ partons. The JETSCAPE-1.0 code does not contain source terms ($J^{\mu} = 0$) which codify the transfer of energy-momentum to the fluid dynamical medium. Energy deposition and equilibration is carried out in an approximate fashion using a modifiable recoil mechanism (see below). 

The current version of the code provides multiple different energy loss routines for the user: MATTER, MARTINI, LBT, and AdS/CFT. These describe the propagation and medium induced shower from the final state partons that emanate from the truncated PYTHIA generator. While there is a prescribed method of how to interlace these codes together to obtain meaningful results, there is no mechanism that forbids un-physical setups that a user may invent. In the following we describe the default setup that is advocated by the collaboration. 

The default distribution will ascribe an upper limit of the initial virtuality of the partons. These will then be introduced into the MATTER event generator. In MATTER, a virtuality-ordered shower is initiated by a single hard parton produced at a point $r$ with a forward light-cone momentum $p^{+} = (p^0 + \hat{n}\cdot \vec{p} )/\sqrt{2}$ where $\hat{n} = \vec{p}/| \vec{p} |$ represents the direction of the jet. One may sample a Sudakov form factor to determine the actual virtuality ($t = Q^{2}$) of the given parton~\cite{Majumder:2013re,Majumder:2014gda}, 
\begin{eqnarray}
\label{eq:matter}
\Delta(t,t_{0}) &=& \exp \left[- \int\limits_{t_{0}}^{t}  \frac{dQ^{2}}{Q^{2}} \frac{\alpha_{s} (Q^{2})}{2\pi} \int\limits_{t_{0}/t}^{(1- t_{0}/t)}  dz P(z) \right. \\
& \times &
\left. \left\{ 1 +  \int\limits_{0}^{\zeta^{+}_{\mathrm{MAX}}}  d\zeta^{+} \frac{\hat{q} ( r + \zeta)  }{Q^{2} z(1-z)}  \Phi ( Q^{2}, p^{+}, \zeta^{+}  ) \right\} \right], \nonumber
\end{eqnarray}
where $\Phi$ represents a sum over phase factors that depend on $\zeta^+$, $p^+$, and $Q$. The transport coefficient $\hat{q}$ is evaluated at the location of scattering $\vec{r} + \hat{n} \zeta^{+}$, $P(z)$ is the vacuum splitting function, and $\zeta^+_{\mathrm{MAX}}$ is the maximum length ($\sim 1.3 \tau_f^+$) we use to sample the actual splitting time of the given parton with $\tau_f^+$ as the mean light-cone formation time $\tau^+_f = 2 p^+/Q^2$~\cite{Cao:2017qpx}. With $Q^2$ determined, $z$ can be sampled using the splitting function $P(z)$, and the transverse momentum of the produced daughter pair can be calculated with the difference in invariant mass between the parent and daughters. This process is iterated until $Q^2$ of a given parton reaches a predetermined value of $Q_0^2$. 

Below $Q_0^2$, one may load other energy loss modules such as LBT, MARTINI and AdS/CFT. In LBT~\cite{Cao:2017hhk,Chen:2017zte,Luo:2018pto,He:2018xjv}, the phase space distribution of a given parton is evolve according to the linear Boltzmann equation. For the elastic scattering process, the scattering rate is evaluated with the leading-order matrix elements for all possible ``$12\rightarrow34$" processes between the given jet parton ``1" and a thermal parton ``2" present in the medium background. For the inelastic scattering, or the medium-induced gluon radiation process, the rate is calculated by integrating over the medium-induced gluon spectrum
\begin{equation}
 \label{eq:gluonnumber}
 \Gamma^\mathrm{inel.} = \int dxdk_\perp^2 \frac{dN_g}{dx dk_\perp^2 dt},
\end{equation}
where the differential spectrum of radiated gluon is taken from the higher-twist energy loss formalism \cite{Guo:2000nz,Majumder:2009ge,Zhang:2003wk}:
\begin{eqnarray}
\label{eq:gluondistribution}
\frac{dN_g}{dx dk_\perp^2 dt}=\frac{2\alpha_s C_A \hat{q} P(x)k_\perp^4}{\pi \left({k_\perp^2+x^2 m^2}\right)^4} \, {\sin}^2\left(\frac{t-t_i}{2\tau_f}\right),
\end{eqnarray}
where $x$ and $k_\perp$ are the fractional energy and transverse momentum of the emitted gluon with respect to its parent parton, $\alpha_s$ is the strong coupling constant, $C_A=N_c$ is the gluon color factor, $P(x)$ is the splitting function, $\hat{q}$ is the transport coefficient, $t_i$ denotes the production time of the given parton, and $\tau_f={2Ex(1-x)}/{(k_\perp^2+x^2m^2)}$ is the formation time of the radiated gluon with $E$ and $m$ as the parton energy and mass respectively. With these scattering rates, the Monte Carlo method is applied to determine whether scattering happens within a given time step. The elastic and inelastic scattering rates are implemented together that guarantee unitary of scattering probabilities. After a particular scattering channel is selected, the 4-momenta of the incoming and outgoing partons are then sampled using the differential rate (or cross section) of the selected channel. Note that within LBT, thermal partons that constitute the medium background can be scattered out of the medium and become part of the jet. These partons are known as ``recoiled partons", and are fully tracked in LBT and allowed to scatter again with the medium. Meanwhile, when a thermal parton is scattered out of the medium background, a hole (or negative parton) is left inside the medium. This is denoted as ``back reaction" and is also fully tracked in LBT so that the energy-momentum conservation of the system can be guaranteed. The subtraction between recoiled partons and back reaction in the final state mimics the effects of jet-induced medium excitation without implementing the sophisticated simulation of energy deposition into the subsequent hydrodynamic evolution. To implement separate analysis on regular particles and back reaction particles, one may utilize the status code associated with each particle: 0 for jet partons and recoiled partons, and -1 for those back reaction partons. The same physics of these recoil and back reaction processes have also been implemented in the MATTER module and can be switched on by setting ``recoil = 1" in the xml file for MATTER.

In MARTINI, the elastic scattering process~\cite{Schenke:2009ik} is implemented in a similar way to LBT. The radiative energy loss is implemented according to the AMY formalism~\cite{Arnold:2002ja,Arnold:2002zm}. The time evolution of the jet momentum distribution is then governed by a set of coupled rate equations as follows:
\begin{align} 
\label{eq:rateMAR}
	\frac{dP_q(p)}{dt} = & \int\limits_k P_q(p+k)\frac{d\Gamma^{q}_{qg}(p+k,k)}{dkdt}
	- P_q(p)\frac{d\Gamma^{q}_{qg}(p,k)}{dkdt} +2P_g(p+k)\frac{d\Gamma^g_{q\bar{q}}(p+k,k)}{dkdt},\nonumber\\
	\frac{dP_g(p)}{dt} = & \int\limits_k P_q(p+k)\frac{d\Gamma^{q}_{qg}(p+k,p)}{dkdt} +P_g(p+k)\frac{d\Gamma^{g}_{gg}(p+k,p)}{dkdt}\nonumber\\
	&-P_g(p)\left(\frac{d\Gamma^g_{q\bar{q}}(p,k)}{dkdt} + \frac{d\Gamma^g_{gg}(p,k)}{dkdt}\theta(2k-p)\right),\nonumber
\end{align}
in which $d\Gamma^a_{bc}(p,k)/dkdt$ is the transition rate for a parton $a$ of energy $p$ to emit a parton $c$ of energy $k$ and become a parton $b$. The factor of 2 in front of $d\Gamma^g_{q\bar{q}}$ is from the fact that $q$ and $\bar{q}$ are distinguishable. For the $g \rightarrow gg$ process, the $\theta$ function is for avoiding double counting of final states. Here $P_q(p)$ and $P_g(p)$ are the energy distributions of quarks and gluons respectively. In the current version of MARTINI, the radiative energy loss mechanism has been improved by implementing the effects of finite formation time as well as the running coupling. The formation time of the emitted gluon increases with $\sqrt{p}$, within which the hard parton and the emitted gluon stay in a coherent state. This interference suppresses the emission rate at early times after a hard parton is produced. For the renormalization scale of running coupling constant $\alpha_s(\mu)$, the root mean square of the momentum transfer $\sqrt{\langle p^2_\perp \rangle}$ between the two partons -- parametrized as $\sqrt{\langle p^2_\perp \rangle} = (\hat{q}p)^{1/4}$ -- is used, where $\hat{q}$ is the jet transport coefficient and $p$ the energy of the jet parton~\cite{Young:2012dv}.

A hybrid model~\cite{Casalderrey-Solana:2015vaa,Casalderrey-Solana:2014bpa} based on strongly coupled holographic techniques is also provided within the JETSCAPE framework. Such model implicitly assumes a factorization between the hard scale governing the perturbative parton splittings, the virtuality $Q$, and the scale at which those partons interact with the medium, assumed to be dominated by the medium temperature $T$, which is $T\sim \Lambda_{\rm QCD}$ and thus falls within a non-perturbative regime. Between two successive splittings of a hard parton, a drag is applied on the parton based on the following energy loss rate computed within the AdS/CFT framework~\cite{Chesler:2014jva,Chesler:2015nqz}:
\begin{equation}
\frac{dE}{dx} = -\frac{4}{\pi}E_\mathrm{in}\frac{x^2}{x^2_\mathrm{stop}}\frac{1}{\sqrt{x^2_\mathrm{stop}-x^2}},
\end{equation}
where 
\begin{equation}
x_\mathrm{stop}=\frac{1}{2\kappa_\mathrm{sc}}\frac{E^{1/3}_\mathrm{in}}{T^{4/3}}
\end{equation}
is known as the stopping distance with $\kappa_\mathrm{sc}$ being the strong coupling parameter of the model. This energy loss rate determines the amount of energy and momentum that flows from the parton hard modes into the medium hydrodynamic modes.

In a later version of the code, we will consider the transfer of energy from the shower to the soft medium. In this first version, the energy loss calculation is slated to take place after the hydro has completed. The energy loss codes depend on local intrinsic properties of the medium, such as temperature $T(x, y, z, t)$ and entropy density $s(x, y, z, t)$, etc, given by hydrodynamics in Milne coordinates where $\tau=\sqrt{t^2 - z^2}$ and $\eta_s = \frac{1}{2}\ln\frac{t+z}{t-z}$. They commence with one list of partons, at higher average virtuality located at the collision point, and end with another list of partons with virtuality $\mathrm{\Lambda_{QCD}}$, with locations at exit points in the expanding medium. Note that several partons may not exit the medium, in which case their entire energy will thermalize in the medium.

The framework can be used to test the sensitivity of results to different physical processes by composing these modules in different ways,
\begin{enumerate}
\item Continue with PYTHIA final state showers to the final hadronic state: This method is best suited for the reproduction of native PYTHIA simulations within the JETSCAPE framework. The primary goal is to compare and contrast predictions for a variety of $p$-$p$ observables between different tunes of PYTHIA.
\item Continue with PYTHIA final state showers to the final partonic state: In several instances, such as the sole use of MARTINI, LBT and the Hybrid approach, the input consists of the PYTHIA partonic shower, evolved down to the minimal scale. However, as PYTHIA does not retain or report space-time locations of the partons, there is no clear starting point in space-time where MARTINI or LBT energy loss should be considered to start. As such one starts at a default initial time $\tau_0$, where the entire developed shower has been expected to be formed. The hybrid model attaches to each parton in the history of the shower and modifies its energy according to the drag formula discussed in Sec.~\ref{ELoss}.
\item Switch to a MATTER shower at $\tau=0$: This is typically done by turning off final state radiation and hadronization in PYTHIA and passing the final state partons to MATTER. Within MATTER, which carries out a virtuality based DGLAP shower~\cite{Altarelli:1977zs,Dokshitzer:1977sg,Gribov:1972ri,Gribov:1972rt}, on these partons, assumes a starting spacetime point given by the binary nucleon-nucleon collision at $\tau = 0$. MATTER carries out vacuum like showers in the absence of the medium, and smoothly transitions to medium modified showers in the presence of a medium. This is described in Sec.~\ref{ELoss}.
\end{enumerate}

{\bf Input:}

$\bullet$ \,\,\,\,  N hard partons with 4-momentum $p_x^i,p_y^i,p_z^i,E^i$ formed at locations $x^i,y^i,z^i$.
The temperature profile [$T(x,y,z,t)$] of the expanding medium [or $s(x,y,z,t)$];

{\bf Output:}

$\bullet$ \,\,\,\, M final (close to on-shell) partons with 4-momentum $p_x^j,p_y^j,p_z^j,E^j$ formed at locations $x^j,y^j,z^j$.
\\
\\

\subsection{Soft Matter Particlization}

This module converts fluid cells into individual hadrons through the Cooper-Frye Formula \cite{Cooper:1974mv}. Monte-Carlo samples of hadrons are generated on the hydrodynamic hyper-surface and are ready to feed into hadronic transport module for further scatterings and resonance decays.

The particle momentum distribution is computed using Cooper-Frye formula,
\begin{equation}
E \frac{dN}{d^3 p} = \int_\Sigma p^\mu d^3 \sigma_\mu (f_0(p \cdot u, T) + \delta f(p \cdot u, T, \pi^{\mu\nu}, \Pi)).
\end{equation}
Here the out-of-equilibrium corrections from shear and bulk viscous corrections are
\begin{equation}
\delta f^\mathrm{shear} (p \cdot u, T, \pi^{\mu\nu})  = f_0(p \cdot u, T) (1 \pm f_0(p \cdot u, T)) \frac{p^\mu p^\nu \pi_{\mu\nu}}{2 T^2 (e + \mathcal{P})}
\end{equation}
and
\begin{equation}
\delta f^\mathrm{bulk} (p \cdot u, T, \Pi)  = f_0 (1 \pm f_0) \left( -\frac{\Pi}{T \hat{\zeta}(T)}\right) \left(-E \left(\frac{1}{3} - c_s^2 \right) + \frac{m^2}{3} \frac{1}{E}\right).
\end{equation}
Here the coefficient $\hat{\zeta}(T) = \zeta(T)/\tau_\Pi$.

{\bf Input:}

\begin{itemize}
    \item The (3+1)-d hyper-surface fluid elements from the hydrodynamic evolution. The required quantities are $x_\mathrm{fo}^\mu$ (fm), $d^3\sigma_\mu$ (fm$^3$), $u^\mu$, $e_\mathrm{sw}$ $\left(\frac{\mathrm{GeV}}{\mathrm{fm}^3}\right)$, $\mathcal{P}_\mathrm{sw}$ $\left(\frac{\mathrm{GeV}}{\mathrm{fm}^3}\right)$, $T_\mathrm{sw}$ (GeV), $\pi^{\mu\nu}$ $\left(\frac{\mathrm{GeV}}{\mathrm{fm}^3}\right)$, and $\Pi$ $\left(\frac{\mathrm{GeV}}{\mathrm{fm}^3}\right)$.
\end{itemize}

{\bf Output:}
With OSCAR2013 format, the following information are included in the output, 
\begin{itemize}
    \item t (fm), x (fm), y (fm), z (fm), mass (GeV), E (GeV), px (GeV), py (GeV), pz (GeV), PDG code, id, charge
\end{itemize}

\def\code#1{\texttt{#1}}
\newcommand{\cred}[1]{{\color{red} #1}}
\newcommand{\cblue}[1]{{\color{blue} #1}}

\subsection{Hard Jet Hadronization}

JETSCAPE v1.3 uses default string hadronization provided by PYTHIA, based on the Lund string model \cite{Andersson:1983ia,Sjostrand:1985ys}. If  PYTHIA is used as the final state
event generator the color flow information can be directly passed on from the final parton shower to PYTHIA string 
hadronization. If MATTER, or one of the in-medium parton shower modules is used, strings need to be defined through some procedure
and handed over to PYTHIA string fragmentation. JETSCAPE v1.3 provides two ways of accomplishing this:
the Colored Hadronization Module and the Colorless Hadronization Module 
described in this subsection.
The labels ``colored" and ``colorless" describe whether color flow is tracked in the final state parton system.
Most in-medium shower Monte Carlo codes, currently do not track color flow
in the shower. In fact, compared to showers in the vacuum, color is constantly exchanged with the 
colored medium and thus becomes randomized quickly \footnote{This argument is true in the multiple scattering limit. Actually, it has been argued that a parton dipole can maintain its color correlations until a decoherence time $\tau_{coh}$ \cite{CasalderreySolana:2011rz}, a consideration especially relevant to the radiated soft spectrum through the physics of angular ordering.}. In that situation, the Colorless Hadronization Module should be the default choice, while in $p+p$ collisions, where color flow information is in principle available, the Colored Hadronization Module
should be the preferred one.

The Colored Hadronization Module assumes that each parton carries an unique color label which has been assigned under the large $N_c$ limit approximation through the several $1\rightarrow 2$ splittings that happened throughout the vacuum shower evolution. Each set of partons originating from a specific hard parton globally carry the color of that initial hard parton, which is color neutralized into a color singlet by attaching as many ``fake" remnants flying down the beam pipe, at very high rapidities, as necessary. In this way, each of the ($\geq 2$) showers is hadronized separately, and no color reconnection among the products of the hard scattering and the underlying event is considered. We have checked that the introduction of these ``fake" remnants does not alter the physics at mid rapidity, and thus fulfils the purpose of achieving color neutral objects while correctly providing the set of rapidity filling soft hadrons observed in experiments. 

The Colorless Hadronization Module takes a parton system and constructs strings based on a minimization criteria.
More specifically, the module establishes links among partons by minimising distances in $(\eta,\phi)$ space, called $\Delta R$.
Strings can be established throughout the full recorded parton event, also between different showers in the same event. Junctions
are not supported in the current version. Strings are established through the following algorithm. First, find the number of strings, by counting the number of quarks and antiquarks (the endpoints of the strings). If an odd number of quarks is found, a quark-like remnant flying along the beam direction is added.  If there are no quarks at all, then two quark-like remnants flying along the beam direction, in opposite directions, are added.
Next, we find the pair of quark and anti-quark whose $\Delta R$ is minimal. Continue until all string endpoints are established.
Now we decide which gluons will be attached to a specific string. To this end, for a given gluon we find the string which minimizes the $\Delta R$ from that gluon to the geometric centre of the string in the laboratory frame.
The gluons are linked in the string sequentially, starting from one of the endpoints and choosing those with the least $\Delta R$ distance with respect to the last link. It continues until there are no more gluons for that string, and then we repeat for each string.
Once the color flow for all strings has been constructed, partons are passed to PYTHIA in such a way that all partons in the event hadronize simultaneously.

It is important to note that once the number of partons increases considerably, which is typically the situation for a parton shower evolving within the medium, the probability of having very collinear, consecutive links in a given string increases, leading to spurious hadron production due to the arbitrarily small mass of the configuration. To prevent this from happening, we manually ensure that the tri-momentum difference between consecutive links in a string is greater than some soft scale $\mathcal{O}(\Lambda_{\rm QCD})$.

{\bf Input:}

\begin{itemize}

\item List of on-shell partons (``positive" partons are hadronized separately from the ``negative" ones).

\end{itemize}

{\bf Output:}

\begin{itemize}

\item  List of hadrons (labelled as either ``positive" or ``negative").

\end{itemize}


\subsection{Hadronic Cascade}

The evolution of the thermalized bulk is simulated by the relativistic fluid
dynamics.  However, it cannot account for a later stage of the fireball
evolution, where the medium is more dilute and is out of full chemical
equilibrium. At this stage there is no quark-gluon plasma, only hadronic
rescatterings and decays. These are simulated by a hadronic cascade, which is
also often called hadronic afterburner. Combining hydrodynamics with a hadronic
afterburner is a well-established approach, known as a hybrid
approach~\cite{Hirano:2012kj,Petersen:2008dd,Werner:2010aa,Ryu:2012at}.

The physical inputs to hadronic cascade are hadron properties, such as masses,
decay widths, branching ratios; and experimentally measured scattering
cross-sections. Particular observables, for which hadronic afterburner plays a
substatial role (up to 50\% effects) are, for example, proton $p_T$-spectra and
flows.  Kaons and pions are usually less affected.

Some of the hadronic cascade realizations are SMASH \cite{Weil:2016zrk}, UrQMD
\cite{Bass:1998ca}, JAM \cite{Nara:1999dz}, PHSD \cite{Bratkovskaya:2011wp},
GiBUU \cite{Buss:2011mx}, there is also a number of others. Currently only
SMASH is due for inclusion in the JETSCAPE framework (as part of JETSCAPE v1.4\footnote{While SMASH is currently not part of JETSCAPE v1.3, the cascade base class is included, and SMASH will be part of the upcoming release JETSCAPE v1.4. Hence we discuss this module here briefly, and refer the reader to the SMASH publications for further details.}), 
but there is no conceptual reason, why
other transport codes cannot be included too.

\vspace{3mm}
\noindent
{\bf Input:}

\begin{itemize}
  \item List of hadrons, specifically their positions $x^{\mu}_i = (t_i,
        \vec{r}_i)$, momenta $p^{\mu}_i = (E_i, \vec{p}_i)$, and PDG
        codes~\cite{Agashe:2014kda} encoding the hadron species.  The coordinate system
        is Cartesian. The times $t_i$ do not need to be identical. The reference frame
        of the hadronic cascade simulation is the reference frame of the input.
  \item End time of the evolution $t_{end}$, specified in the config.
\end{itemize}

\vspace{3mm}
\noindent
{\bf Output:}

\begin{itemize}
  \item Final list of hadrons at time $t_{end}$ after all rescatterings and
        decays that they have undergone.  This list includes, as input, positions
        $x^{\mu}_i$, momenta $p^{\mu}_i$ and PDG codes.
\end{itemize}


\section{The JETSCAPE Framework Design}
\label{framework}

The key challenge in JETSCAPE's framework design was to provide a
simple and robust, yet transparent and extensible computational infrastructure that encompasses the needs of all physics
considerations during the complex simulation of partonic energy loss in heavy-ion collisions. 
The design goal of JETSCAPE's framework was to provide an, as close as possible, one-to-one computational representation of the current \emph{state of the art} understanding of the different physics aspects of heavy-ion collisions, as outlined schematically in Fig.\ \ref{fig:flowchart}. The design choices, that have been made, facilitate such a one-to-one mapping of the physics, as well as provide modularity and extensibility. They represent the different physics aspects (filled colored boxes in Fig.\ \ref{fig:flowchart}) as \code{tasks} and the communication, exchange/access to relevant physics quantities (arrows in Fig.\ \ref{fig:flowchart}), via the \code{signal/slot} paradigm \cite{sigslot}.

JETSCAPE's task implementation is inspired by ROOT's \code{TTask} \cite{root}. Every simulated step is attached to a tree of tasks that can in turn
recursively control a collection of subordinate tasks. The top level
object is, in principle, no different from the lowest level sub-task.
Generically speaking, each task has to respect the common interface
and implement a process when intimated to initialize, execute, and
finalize itself. Specific groups, such as energy loss,
specialize this concept further by adding a few additional methods to
a derived interface base class. The task design provides the required infrastructure, which is both robustly modular, and yet simple to understand and trivial to extend.

All inter-task communications are facilitated using methods provided by
the interface classes. Internally, communication needs are then
implemented via the signal/slot paradigm\footnote{The reader may
browse through the code and observe that this paradigm is not yet
obeyed in all places, but even where not noted, any different
communication (such as by passing pointers) should be considered
deprecated.}. The signal/slot paradigm in the context of JETSCAPE 
provides several desired features: improved safety by not providing pointers to different tasks, and more importantly, it allows to clearly define a transparent communication interface between different stages in the simulation of heavy-ion collision in which the allowed exchanged information is defined by our understanding of the relevant physics as discussed in detail in Chapter \ref{physics}.
This implementation is however transparent to users
and authors of physics modules, and can be expanded in the framework at a more fundamental
level if needed.

To start this chapter, we will first discuss the above mentioned two basic building blocks of the JETSCAPE framework.
We then will be able to discuss and understand the workflow of an example JETSCAPE program, 
followed by a review of the extensibility features.
Finally, we detail the basic data types and
the safety mechanisms as well as other functionalities that the framework utilizes.


\subsection{Framework Modularity: The basic building blocks}
\label{sec:modular}

The JETSCAPE framework is modular in the sense that one can attach, remove, or
modify one part of the code without changing any other part of the
code. Throughout our design, modularity is realized by
independent sections of functionality that we call {\bf modules}.
The independence of each module is realized by the task design (\code{JetScapeTask}),
in combination with the management of communications and data exchanges by the framework via the signal/slot paradigm (\code{JetScapeSignalManager}).
Such design is also well suited to provide a simple extensible interface which will be discussed in more detail in Sec.\ \ref{sec:extens}.
In the following section, we will discuss in more detail the implementation of these two basic building blocks of the JETSCAPE framework.

\subsubsection{JETSCAPE Modules: The \code{JetScapeTask}}
\begin{figure}[h]
  \centering
  \includegraphics[width=1\textwidth]{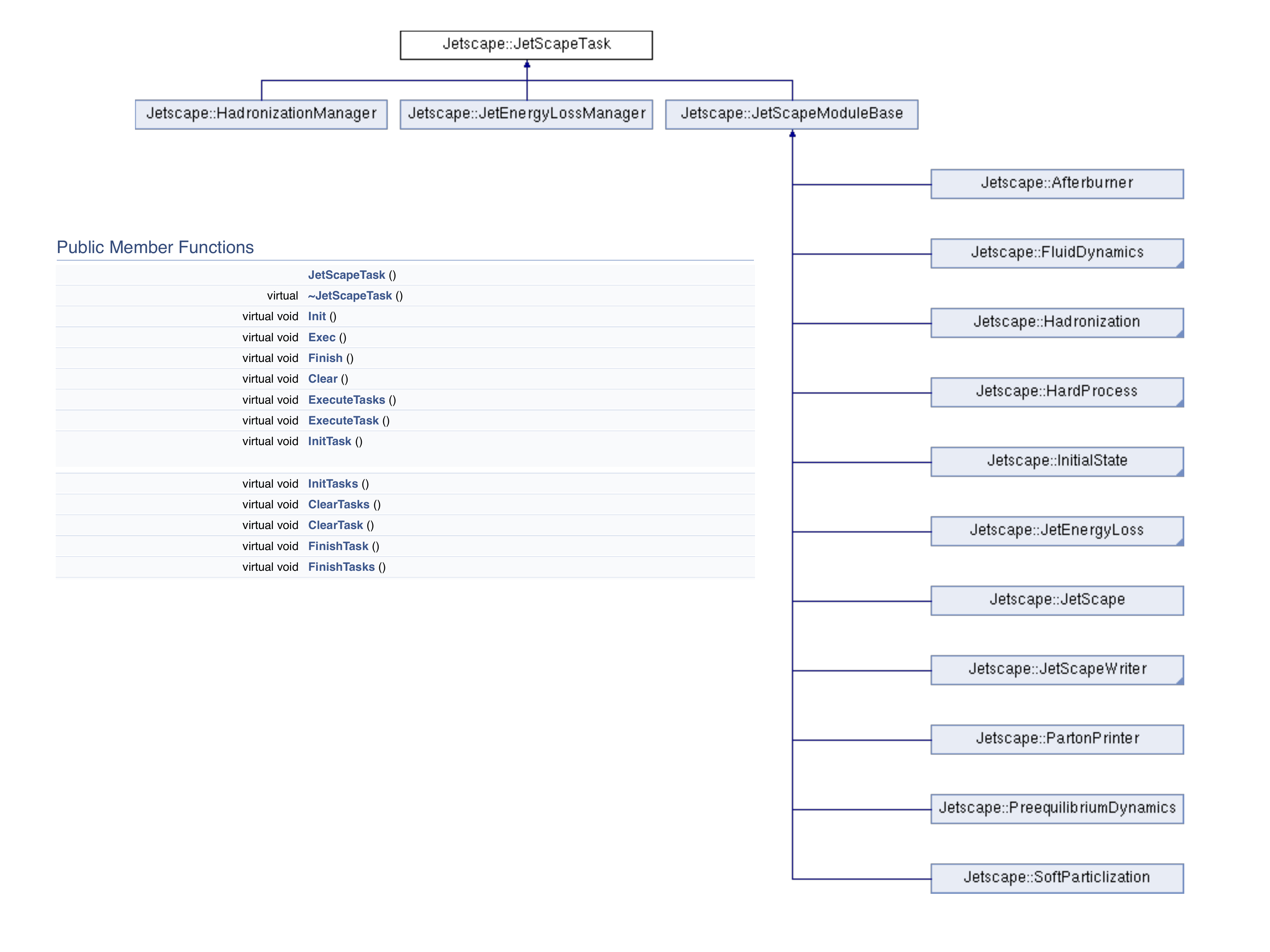}
  \caption{\code{JetScapeTask} inheritance diagram.}
  \label{fig:task_inher}
\end{figure}

The JETSCAPE framework is designed to cover all physics aspects of
heavy ion collisions (see Fig.\ \ref{fig:flowchart}). In order to make this possible, we have defined a
unified interface for all aspects, which we call {\bf tasks}. The
unified interface is realized using a base class called \code{JetScapeTask}, and all the modules corresponding to different stages
of the simulation are derived from this base class (see Fig.\ \ref{fig:task_inher}).

The \code{JetScapeTask}
base class provides virtual methods to initialize, execute and finish
JETSCAPE tasks. All physics and non-physics modules that are run by
the JETSCAPE framework are a subclass of \code{JetScapeTask} and override
those virtual methods in their own specific way.

The implementation of the \code{JetScapeTask} base class defines a vector of type
\code{JetScapeTask} that enables recursive execution of tasks: Each task
may have its own subtasks, subtasks may have further ``sub-subtasks", and so on,
the level of hierarchy may get as deep as necessary. At each level of this recursive
execution, these tasks override the virtual methods provided by
the framework.

The modularity of the JETSCAPE framework is achieved by each subtask being allowed to overwrite the methods specified within \code{JetScapeTask}, but still being constrained to remain within the ambit of methods
outlined in \code{JetScapeTask}.

In terms of methods, \code{JetScapeTask} defines virtual methods
\code{Init()}, \code{Exec()}, \code{Finish()} and
\code{Clear()} that are overridden by user defined modules.
The
\code{Init()} method is used by tasks to do initializations.
It is called once during the lifetime of the task, initializing all relevant variables and flags (from the provided \code{xml} file) needed to ensure proper execution of the attached physics modules (for more details see Sec.\ \ref{sec:xml}).
Note however that modules created via \code{Clone()} may circumvent
this automatic initialization; this is the case for the copies of jet
energy loss modules used for each showering parton.
The \code{Exec()} handles the real execution of a task.
It is called exactly once per generated event.
This method may not always contain the core functionality of a
module. For example, energy loss calculations happen multiple times
per event (they are a per-parton or potentially per time-step process),
and corresponding modules should have their core functionality in
\code{DoShower()} or \code{DoEnergyLoss()}.
The \code{Clear()} methods are overridden for the
purpose of releasing memory/clearing pointers or performing necessary operations needed for the proper execution of
modules after each event. Whereas the
\code{Finish()} methods are overridden for the
purpose of releasing memory/clearing pointers or performing other operations at the end of a
module's life cycle.
All of the above methods
are called by the framework recursively for each task using the
accumulative methods \code{InitTasks()}, \code{ExecTasks()},
\code{FinishTasks()} and \code{ClearTasks()}. Accumulative methods
are handled by the framework and do not concern users of the
framework.

In order to separate the concerns of physics and non-physics modules,
we add another base class in between physics modules and
\code{JetScapeTask} that is called \code{JetScapeModuleBase}. This
base class is derived from \code{JetScapeTask}, so it ensures the
unified interface. Moreover, this base class is derived form
Signal/Slot base class (explained in Section~\ref{sec:sigmgr}), as a result, it can also connect different modules via the Signal/Slot
mechanism~\cite{sigslot} in order to provide the necessary communication/access to data. Another functionality of this base class is
to ensure task/thread safety for a random number seed, or more
specifically, access to a properly seeded Mersenne Twister generator
using the \code{GetMt19937Generator()} method.

\subsubsection{Module Communication: The \code{JetScapeSignalManager}}
\label{sec:sigmgr}
In a typical program flow of the JETSCAPE framework (see Sec.\ \ref{programflow}), modules need to exchange data beyond the
sequential execution of the attached tasks in order to perform the relevant physics (see Fig.\ \ref{fig:flowchart}).
\begin{figure}[tb]
  \centering
  \includegraphics[width=0.5\textwidth]{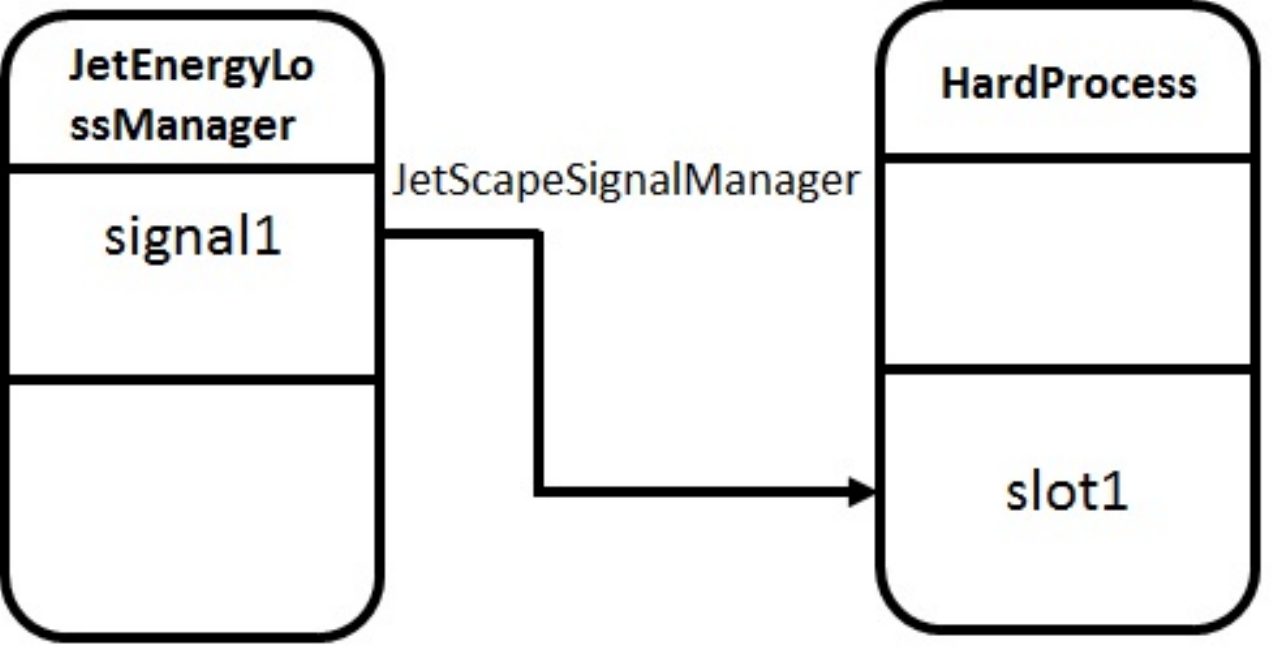}
  \caption{An example of managing connection using signals and slot between modules by \texttt{JetScapeSignalManager}.}
  \label{fig:sigslot}
\end{figure}
For example, the energy loss manager needs to receive the
shower-initiating partons from the hard-scattering module. We
visualize this communication, managed by \code{JetScapeSignalManager},
in Fig.~\ref{fig:sigslot}.

In order to implement those kinds of communication between modules,
the \code{Jet\-ScapeSignalManager} class uses the signals and slots
paradigm library developed in the Qt project and implemented in the
sigslot package by S. Thompson~\cite{sigslot}. A signal is sent
when the execution of a particular module reaches a specific point. A
slot function is a response to a particular signal. A signal
can connect to one or more slot functions. The
\code{JetScapeSignalManager} class is a singleton that manages
connections between signals and slot functions.

For the case of the signals and slots connection between \code{JetEnergyLossManager} and hard scattering modules \code{HardProcess}, we define a signal for the \code{JetEnergyLoss} module to establish
the connection to the \code{GetHardPartonList()} method of the hard
scattering module. Then we define a connection method in the
\code{JetScapeSignal\-Manager} class that is called
\code{Connect\-GetHard\-PartonListSignal()}. This method accepts a
pointer to the energy loss manager module as input, and invokes the
slot function using the \code{connect()} method of the signal from
\code{JetEnergyLossManager}. When the connection is established between the signal
and the slot method, data can be exchanged, which is the list of
shower-initiating partons in this case.

\code{JetScapeSignalManager} manages all relevant instances of signals and
slots connections (created during the \code{Init()} phase of the JETSCAPE framework see Sec.\ \ref{programflow}) that are necessary for data exchange between
JETSCAPE modules to ensure that for each physics process all the needed data are available. The list of all connection methods are summarized in Table \ref{table:sigslottable}.

\begin{table}[htp]

\footnotesize
\begin{center}
\begin{tabular}{c|c|c|c}
Module 1 & Module 2 & Signal & Slot \\
\hline
\code{JetEnergyLossManager} &  \code{HardProcess} & \code{GetHardPartonList()} &  \code{GetHardPartonList()} \\ \hline
\code{JetEnergyLoss} &  \code{FluidDynamics} & \code{jetSignal()} &  \code{UpdateEnergyDeposit()}  \\ \hline
\code{JetEnergyLoss} &  \code{FluidDynamics} & \code{edensitySignal()} &  \code{GetEnergyDensity()}  \\ \hline
\code{JetEnergyLoss} &  \code{FluidDynamics} & \code{GetHydroCellSignal()} &  \code{GetHydroCell()}  \\ \hline
\code{JetEnergyLoss} &  \code{JetEnergyLoss} & \code{SentInPartons()} &  \code{DoEnergyLoss()}  \\ \hline
\code{Hadronization} &  \code{Hadronization} & \code{TransformPartons()} &  \code{DoHadronization()}  \\ \hline
\code{HadronizationManager} &  \code{HardProcess} & \code{GetHadronList()} &  \code{GetHadronList()}  \\ \hline
\code{HadronizationManager} &  \code{JetEnergyLoss} & \code{GetFinalPartonList()} &  \code{SendFinalStatePartons()}  \\ \hline
\end{tabular}
\end{center}
\caption{The list of all connection methods between JETSCAPE modules provide by the \code{JetScapeSignalManager}.}
\label{table:sigslottable}
\end{table}%



\subsection{A typical JETSCAPE Program Flow}
\label{programflow}

\begin{figure}[t]
  \centering
  \includegraphics[width=0.99\textwidth]{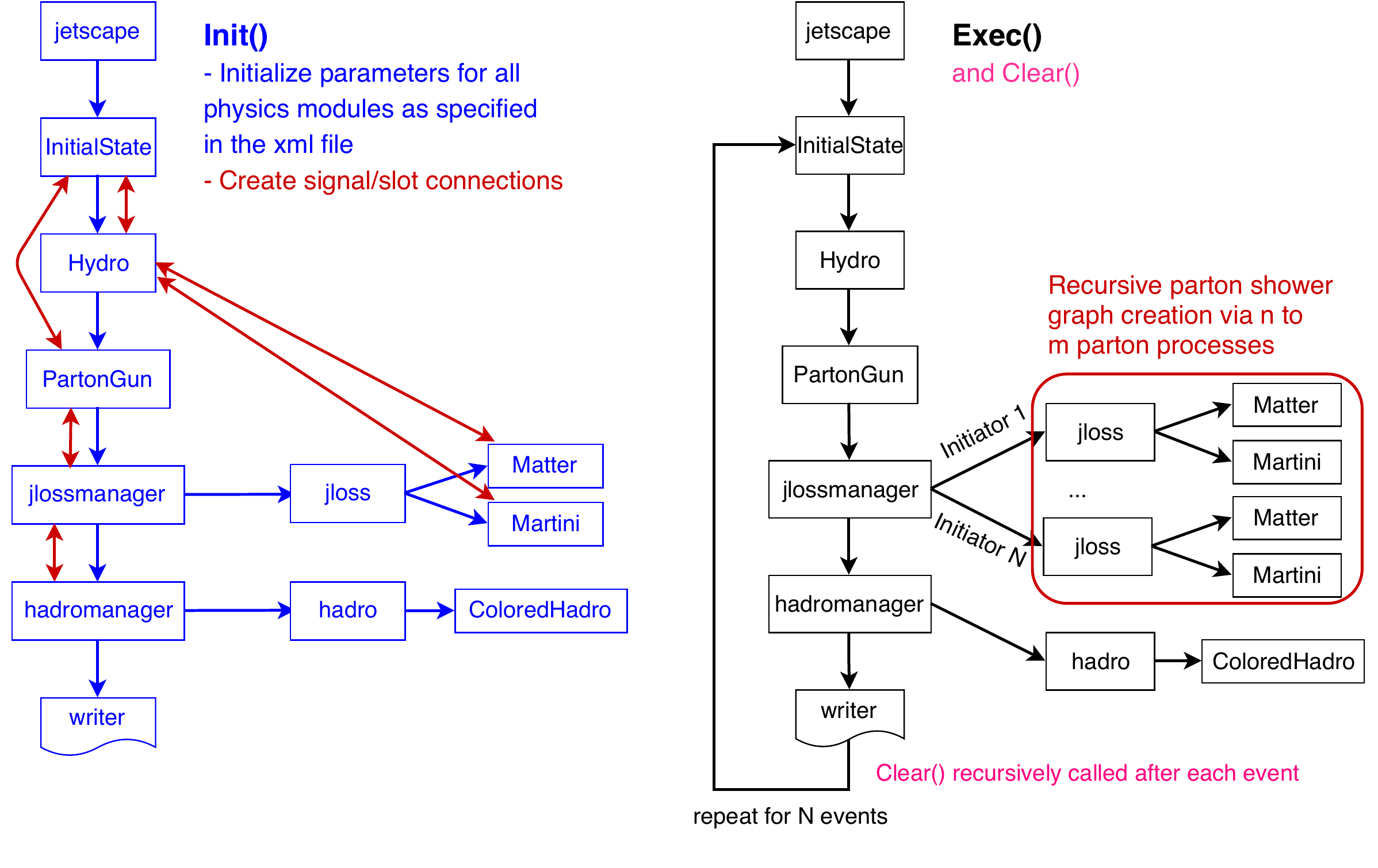}
  \caption{Example workflow of the \code{Init()} (left side) and \code{Exec()} (and \code{Clear()}) (right side) phase of the task based JETSCAPE framework (the not extensively used Finish() phase is omitted). One should be aware that the created signal/slot connections in the \code{Init()} phase are of course present and utilized in the \code{Exec()} phase, but are not show in the figure for simplicity. }
  \label{fig:workflow}
\end{figure}

After the introduction of the \code{JetScapeTask} and the communication between different modules facilitated by the \code{JetScapeSignalManager} in Sec.\ \ref{sec:modular}, we can now discuss a typical workflow emphasizing the \code{Init()} and \code{Exec()} phase of the tasks as illustrated in Fig.~\ref{fig:workflow}, representing a variation of the workflow as defined in \code{examples/brickTest.cc} (see Sec.\ \ref{sec:prog} for the source code). 
With this we can examine how the framework is able to simulate all aspects of the
collision of ions, from the initial overlap to the explosive expansion including partonic energy loss followed by 
the evaporation into conventional matter (assuming the proper implementation of the relevant physics aspects are implemented in the corresponding physic modules\footnote{The mentioned \code{examples/brickTest.cc} only provides modules with a very simplistic implementation of the different physics aspects.}).

The program starts with the framework's main task, called \code{jetscape}. All relevant physics modules have to be attached to the main task or to their relevant manager tasks (for details see Sec.\ \ref{sec:manager}). In the \code{Init()} phase all attached modules are initialized recursively and all relevant variables and flags are read in and set from the provided \code{xml} file (for more details see Sec.\ \ref{sec:jsxml} and Sec.\ \ref{sec:xml}). In addition in this initialization phase the \code{JetScapeSignalManager} provides all the relevant signal/slot connections for data exchange between the modules (see Sec.\ \ref{sec:sigmgr}). An example is shown in Fig.\ \ref{fig:workflow} (left panel), such a connection ensures proper execution of the attached physics modules (following our understanding of the different physics aspects in heavy-ion collisions as discussed in Sec.\ \ref{physics}).

At runtime, in the \code{Exec()} phase (see Fig.\ \ref{fig:workflow} (right side)) the main task \code{jetscape} calls
 an \emph{initial state} module (\code{InitialState}) to
simulate the initial overlap of the ions. Then, it runs a
\emph{hydrodynamics} module (\code{Hydro}) for medium evolution. Next,
a \emph{hard process} module (\code{PartonGun}) creates
shower-initiating partons at vertices determined by the initial
state. These are then sent to an \emph{energy loss manager} module
(\code{jlossmanager}) which performs energy loss calculations in
multiple time steps. 
Note that in the current design of JETSCAPE, hydrodynamical
calculations are already completed for all times at this stage, and
the medium space-time information is queried, not calculated on demand.

Multiple models (\code{Matter, Martini}) combine in the \code{jloss}
container, which the manager clones for each shower initiator.
At each time step,
the framework sends a parton list to all the attached energy loss modules,
which do the energy loss calculations and send back a list of partons
to the framework. Internally the JETSCAPE framework stores the parton shower as a directed graph (see Sec.\ \ref{sec:showergraph} for more details). The energy loss modules decide if they can do
calculations with the received parton from the framework, according to their implemented kinematic range of applicability.
When the
parton showering is done, the framework sends a list of final state
partons to the hadronization manager module. Then, the hadronization
module (\code{hadroModule} or \code{colorless}) produces stable hadrons from the final state partons. All the
modules write their output into the output file using the attached \code{writer}
module(s). 

The above discussion should make it apparent that the JETSCAPE framework provides a program workflow (shown in Fig.\ \ref{fig:workflow}) which indeed represents a one-to-one computational representation of our current understanding concerning the different physics aspects of heavy-ion collisions, as outlined schematically in Fig.\ \ref{fig:flowchart} and discussed in more detail in Sec.\ \ref{physics}. This one-to-one mapping of the physics of heavy-ion collisions combined with the strict modular design (see Sec.\ \ref{sec:modular}) allows the user to implement their physics with as little as possible overhead into JETSCAPE. This will be discussed in the next section.

\subsubsection{A Note on Ordering}
\label{sec:note-ordering}

Modules will be executed (specifically, finish their
\code{Exec()} method) in the order in which they were attached to the
top (\code{jetscape}) task. In practice, that means modules need to be
attached in a physically meaningful order\footnote{In future releases it is foreseen to enforce a physically meaningful order at the framework level.} . For example, hard process and
hydro modules have to be attached after the initial state. In JETSCAPE 1.0 however, the
order between the two does not matter since they don't rely on each other.  

Following with this logic, the energy loss modules have to follow the hydro modules and the hard process. Hadronization modules have to be added after both the hydro and energy loss have completed. There is a possibility to add a hadronic afterburner (SMASH~\cite{Petersen:2018jag}, default in JETSCAPE 2.0) after the completion of the hydro module. If this is added, it too must be added before the energy loss and hard hadronization modules.



\subsection{Framework Extensibility}
\label{sec:extens}
The JETSCAPE framework introduces an extensible design that enables
users to implement modules (see Sec.\ \ref{sec:modular}) that can be executed by the framework. In
order to achieve this, the framework provides a set of base classes
for each physics or non-physics aspect of the simulation (see Fig.\ \ref{fig:task_inher}). Users can
extend the framework by implementing modules to cover initial state,
hard process, hydrodynamics, energy loss, or hadronization. Due to the design of the JETSCAPE framework (see Sec.\ \ref{sec:modular}) the user only has to focus on implementing the physics in his/her realm of expertise while adhering to the interface provided by the module base classes. All communication/data exchange and proper integration, initialization as well as execution is handled by the JETSCAPE framework.  Moreover,
users can extend the framework by implementing non-physics modules
such as JETSCAPE writer or reader modules. In the rest of this
section, we discuss in detail how to extend the framework by implementing different
types of modules.

\subsubsection*{Implementing energy loss modules}
\label{sec:impl-energy-loss}
\begin{figure}[h]
  \centering
  \includegraphics[width=5.52in,height=2.47in]{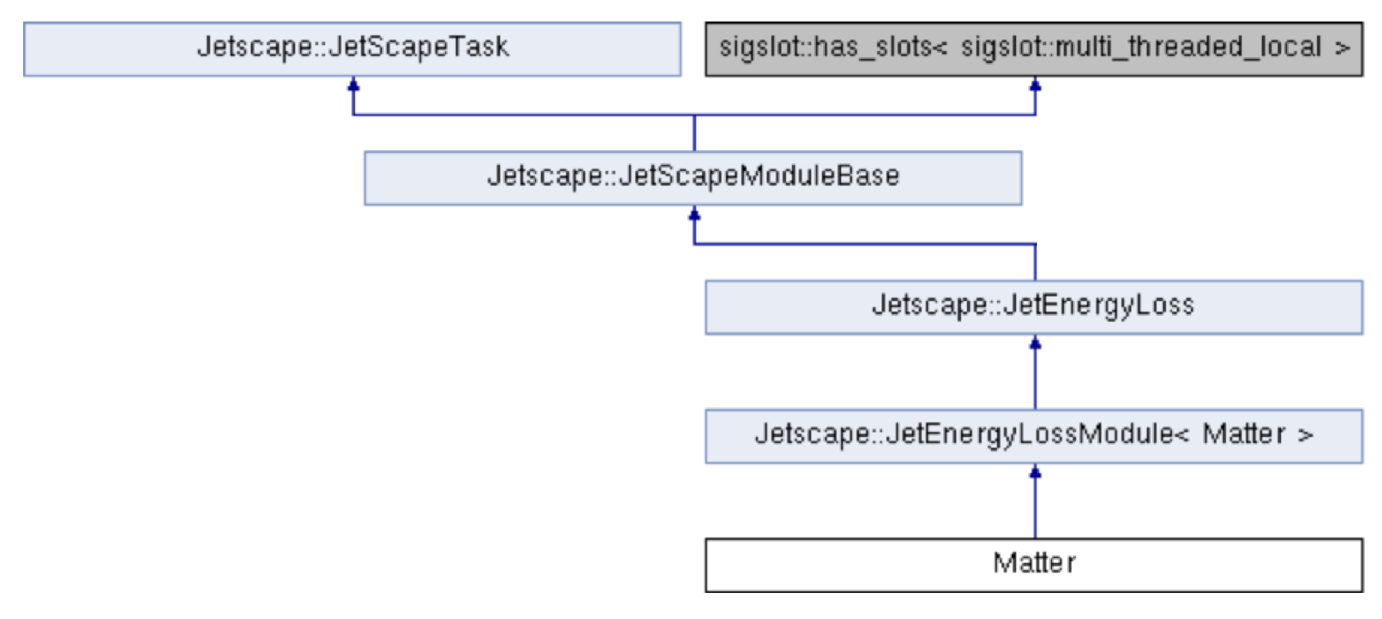}
  \caption{JETSCAPE class structure for implementing an energy loss module.}
  \label{fig:jstask}
\end{figure}

In Fig.~\ref{fig:jstask}, we
show the class structure for an energy loss module called \code{Matter}. On
the top left, we have the base class \code{JetScapeTask}. The
class \code{JetScapeModuleBase} is a subclass of both
\code{JetScapeTask} and signal/slots. As we mentioned before, the
\code{JetScape\-Module\-Base} adds more physics-related properties to
\code{JetScapeTask} like a properly seeded PRNG for reproducible event
generation.
The next level of the hierarchy is the \code{JetEnergyLoss} class that
mainly performs parton showering and aggregation of the shower into a
graph.
It also provides the infrastructure to make
\code{GetHydroCellSignal()} available as the means to obtain medium information.
At each time step, \code{JetEnergyLoss} sends a parton to all attached
energy loss modules and receives a list of partons that are added to
the parton shower graph. In order to implement an energy loss
module, the added module must be a subclass of a template class
called \code{JetEnergyLossModule} that provides a method to be
overridden for energy loss calculations. In Fig.~\ref{fig:jstask},
this class is called \code{JetEnergyLossModule<Matter>} that is
specific to the Matter energy loss module. Inside the energy loss
module class, the user needs to override several methods. The most
important method to override is called \code{DoEnergyLoss()}.This
method contains the main jet energy loss calculations for the added
module. The \code{DoEnergyLoss()} method uses a vector of input
partons named \code{pIn} and inserts the produced partons in a
output vector called \code{pOut} therefore representing a generic interface for a $n\rightarrow m$ parton process.
The module
\code{JetEnergyLoss} fills the input vector of partons and reads
the output vector or add its partons to the parton shower graph\footnote{Utilizing the generic directed graph structure provided by the JETSCAPE framework, partons which are not produced by a $1\rightarrow n$ process, for example medium recoils, are attached as incoming edges/partons to the node/vertex via appending them to the \code{pIn} vector and represent new root nodes in the graph.} (see Sec.\ \ref{sec:showergraph}). The user
can override the \code{Init()} method for initialization and the
\code{WriteTask()} for producing further module specific output. The \code{JetEnergyLoss} base class ensures that the parton shower itself is stored and saved to file in the currently supported output formats (see. Sec.\ \ref{sec:impl-writ-modul}).

\subsubsection*{Implementing hydrodynamics modules}
\label{sec:impl-hydr-modul}

\begin{figure}[h]
  \centering
  \includegraphics[width=5.37in,height=1.83in]{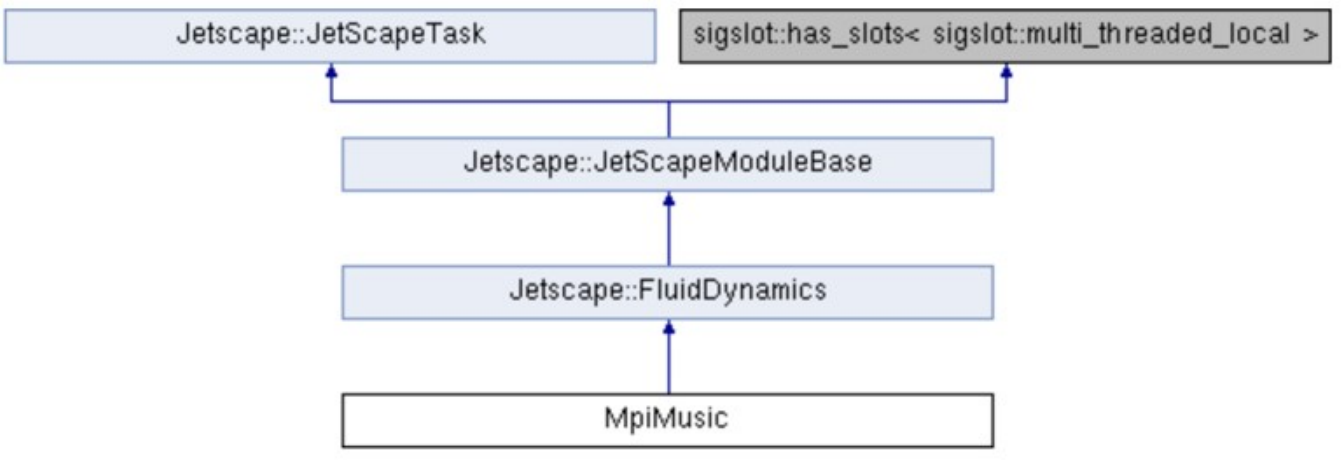}
  \caption{JETSCAPE class structure for implementing a hydrodynamics module.}
  \label{fig:hydrostr}
\end{figure}

In Fig.~\ref{fig:hydrostr},
we show the class structure for a hydrodynamics module called
MpMusic. Each hydrodynamics module must be derived from a class
called \code{FluidDynamics}, which is a sub class of
\code{JetScapeModuleBase}. The \code{FluidDynamics} class
provides functionalities related to hydro cell and temperature that
can be reused by a hydrodynamics module. The added hydrodynamics
module must override \code{Initialize\-Hydro()},
\code{EvolveHydro()}, and \code{Get\-Hydro\-Info()} methods. The
main method to override is \code{Evolve\-Hydro()}, because this
method performs the hydrodynamics calculations. Users override
\code{Initialize\-Hydro()} to initialize the hydrodynamics module
and \code{Get\-Hydro\-Info()} for get cell information. 
Module developers are encouraged to take guidance from the existing
model implementations in \code{src/jet/}.

\subsubsection*{Implementing hard process modules}
\label{sec:HardProcess-Impl}

\begin{figure}[h]
  \centering
  \includegraphics[width=5.37in,height=1.82in]{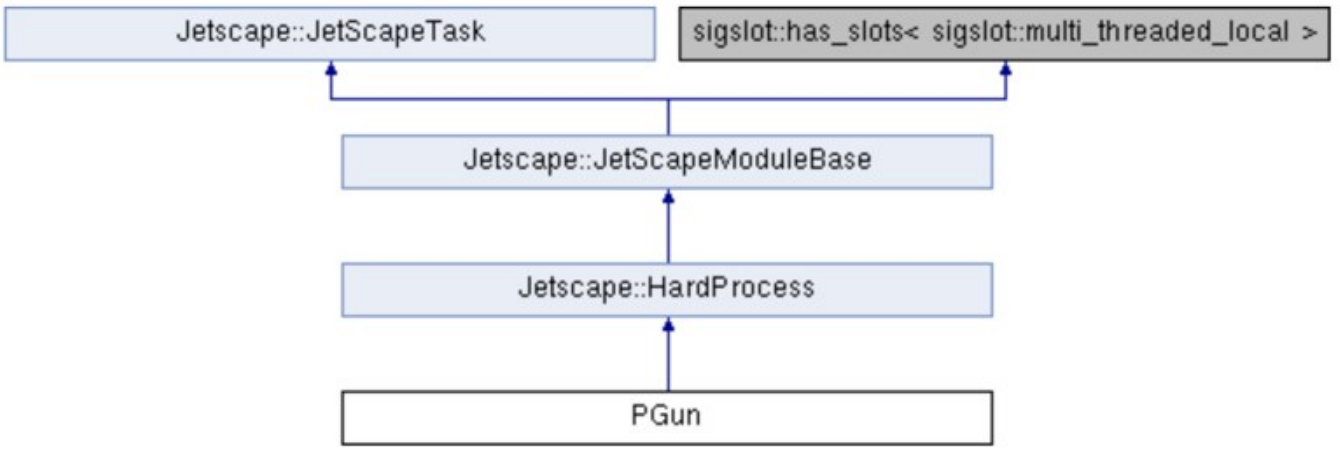}
  \caption{JETSCAPE class structure for implementing a hard process module.}
  \label{fig:hardprocess}
\end{figure}

In Fig.~\ref{fig:hardprocess}, we show the class structure for a hard
process module named \code{PGun} (``parton gun'').
Every hard process module in the JETSCAPE
framework must be a subclass of the class \code{HardProcess}. This
class provides common functionalities among hard process modules,
such as adding partons to a list and obtaining said
list. The hard process module then must override methods
\code{InitTask()} and \code{Exec()}. In the \code{InitTask()}
method, one handles initializations and reading parameters from an
XML file. In the \code{Exec()} method, a user writes code to prepare
the list of hard partons to be sent to energy loss modules.
Module developers are encouraged to take guidance from the existing
``Gun'' modules in \code{src/initialstate/}.

\subsubsection*{Implementing initial state modules}
\label{sec:impl-init-state}

\begin{figure}[h]
  \centering
  \includegraphics[width=5.37in,height=1.81in]{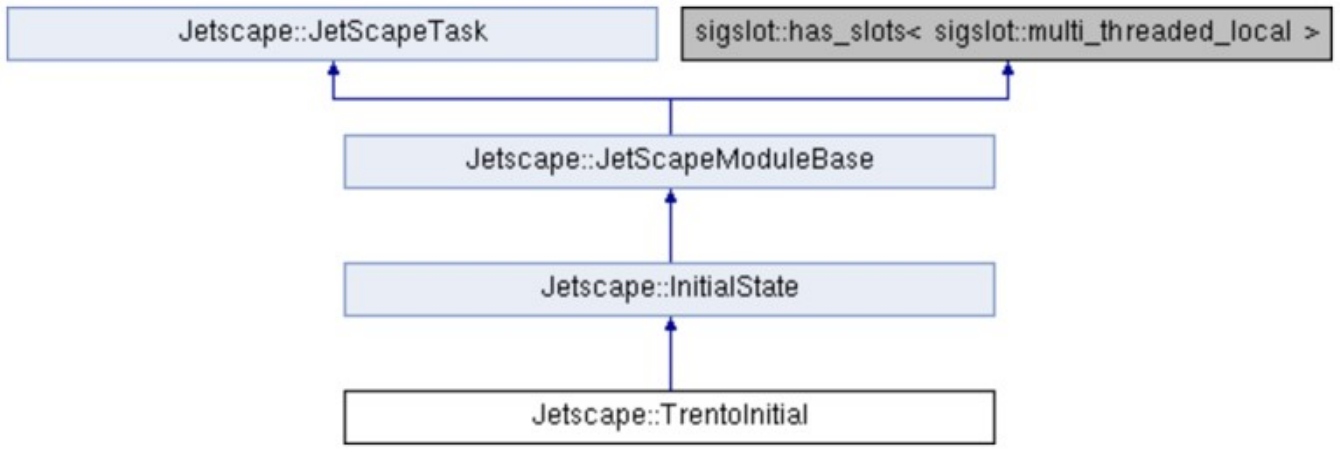}
  \caption{JETSCAPE class structure for implementing an initial state module.}
  \label{fig:initialstate}
\end{figure}

In Fig.~\ref{fig:initialstate}, we show the class structure for an
initial state module named \code{TrentoInitial}. Each initial state module
must be a sub class of \code{InitialState} class, which utilizes
common reusable functionalities for all the initial state
modules.
Inside the initial state module class, a developer should override
\code{Exec()} to populate the data fields
\code{\detokenize{entropy_density_distribution_}} and
\code{\detokenize{num_of_binary_collisions_}}, as well as ideally
override the 
methods \code{GetNpart(), GetNcoll(), GetTotalEntropy()}.

\subsubsection*{Implementing hadronization modules}
\label{sec:impl-hadr-modul}

\begin{figure}[h]
  \centering
  \includegraphics[width=5.15in,height=1.82in]{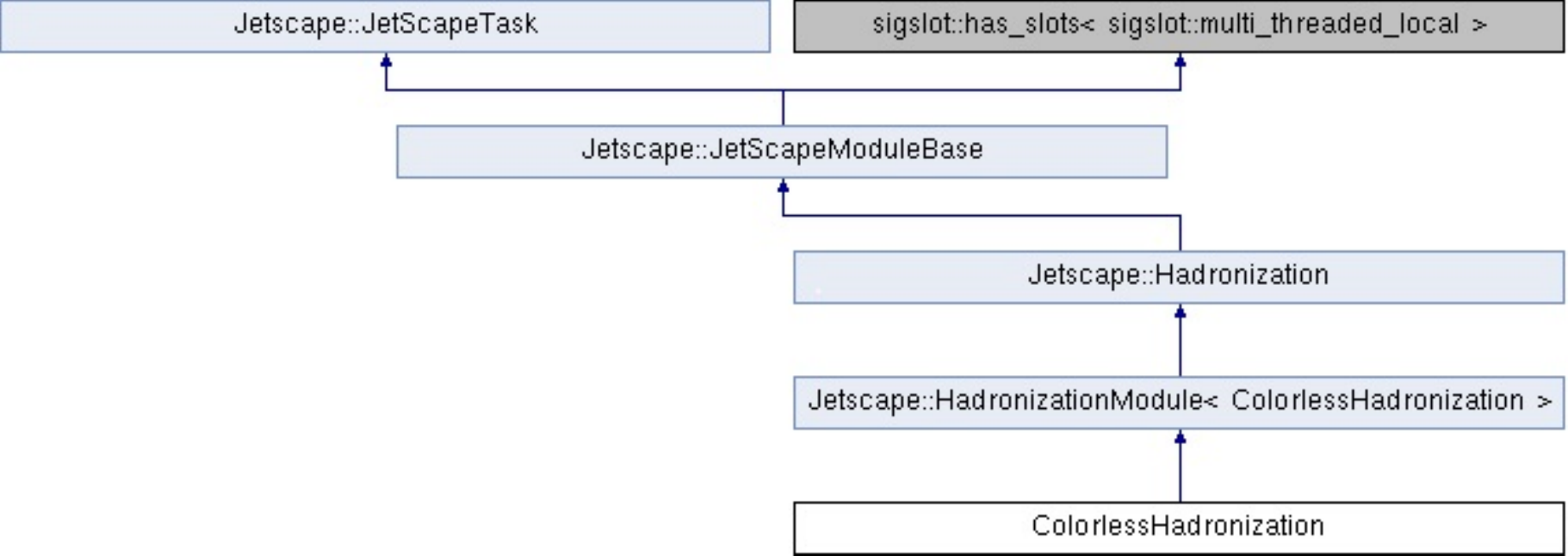}
  \caption{JETSCAPE class structure for implementing a hadronization module.}
  \label{fig:hadro}
\end{figure}

In Fig.~\ref{fig:hadro}, we show the class structure for a
hadronization module named \code{Color\-less\-Ha\-dro\-ni\-za\-tion}. A hadronization
module is a subclass of a class called
\code{Ha\-dro\-ni\-za\-tion\-Module}. This class is a template class in the
JETSCAPE framework that enables communication between the framework
and the hadronization module. The class  \code{Hadronization\-Module} is
a subclass of another class called \code{Hadronization}, which is a
subclass of \code{JetScape\-Module\-Base}.
The core functionality to be override in \code{Hadronization}
is the method \code{DoHadronization()} that
accepts a list of final state partons parton (or rather, a vector of such lists aggregating
all showers in the event)
and generates hadrons (and remaining partons) to the output list of
hadrons. By default, the JETSCAPE framework contains two hadronization
modules: colored and colorless string fragmentation based on PYTHIA.

\subsubsection*{Implementing writer modules}
\label{sec:impl-writ-modul}

\begin{figure}[h]
  \centering
  \includegraphics[width=5.37in,height=1.81in]{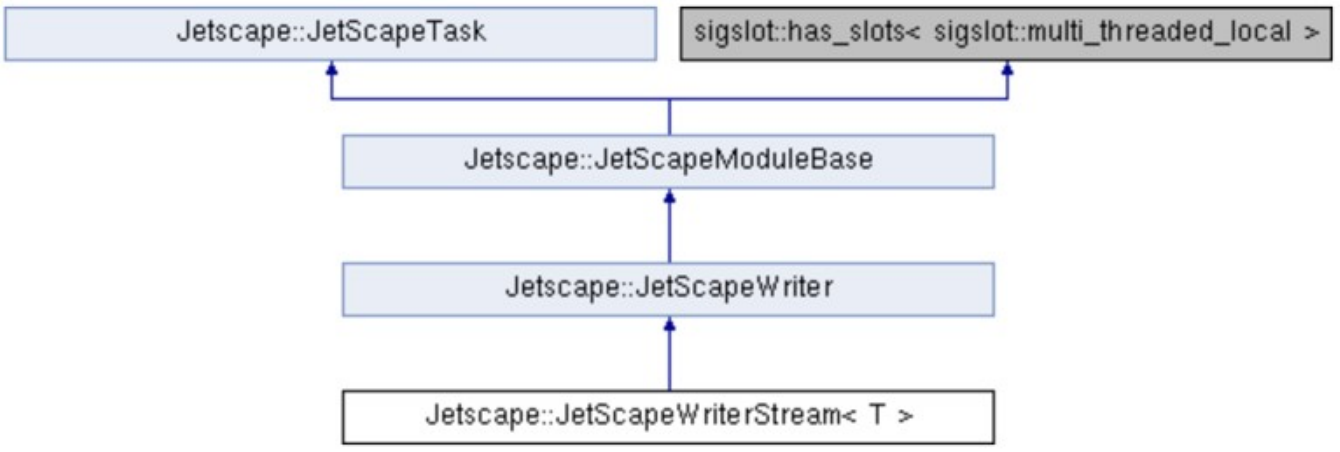}
  \caption{JETSCAPE class structure for implementing a writer module.}
  \label{fig:writer}
\end{figure}

In Fig.~\ref{fig:writer}, we show the class structure for a writer
class in the JETSCAPE framework. A writer class must be a subclass of
the \code{JetScape\-Writer} class. \code{JetScape\-Writer} provides
different write methods for various data types like parton and
hadron. For stream output writers, users define a template class by
replacing \code{T} in Fig.~\ref{fig:writer} with an output stream
class.

Writer classes are the main interface between running JETSCAPE and
processing the output. The JETSCAPE framework includes a HepMC3.0
writer. While the specifications for version 3.0 are not finalized as
of this writing, its added focus on heavy-ion collisions, and its
simple interface to ROOT made it our general format of choice\footnote{Nonetheless, due to some issues caused by HepMC3.0 we will provide a HepMC2.0 writer in one of the next maintenance releases.}. 

JETSCAPE also includes stream
classes for \code{ofstream} and \code{ogzstream} for ASCII and gzipped
ASCII which serve as the
default output format with maximal added information.
However, they are also a starting point for
customized writers to be adapted to a user's needs, and a few simple
changes can produce any output format, compact or verbose, the user
desires.

As discussed in \ref{sec:showergraph}, parton showers are internally
represented as directed graphs, thus preserving the entire shower
history. This graph structure is reflected in the default ASCII output
format, and the provided \code{JetScapeReader} classes will properly
reconstruct the graph structure. HepMC also utilizes a graph
structure and the writer class takes advantage of that. Note, however,
that HepMC always requires incoming and outgoing edges for every
vertex, whereas JETSCAPE graphs normally start and end with a vertex;
the HepMC output therefore duplicates some edges.

Hadronization does not easily lend itself to a graph structure. The
default output therefore only provides a list of hadrons which also
contain soft hadrons from Cooper-Frye freezeout. In the HepMC output,
hadrons are instead all connected to a disjoint hadronization vertex.

\subsubsection{A Note on Analysis Tasks}
\label{sec:note-analysis-tasks}

It is envisioned that JETSCAPE be mostly used in a two step process,
where a large number of events is created first and then analysis
follows in a second step from the created output files. However, the
task structure lends itself naturally to adding any task related to
performing analysis on the fly during event creation. A dedicated base
class for this may be added in a future release, but in the meantime,
users interested in such a workflow are advised to derive from the
writer base class which has access to all information, and look to
\code{JetScape\-Writer\-HepMC} for guidance on how to access it.



\subsection{Data Types}
The JETSCAPE framework introduces data structures to be utilized in
various aspects of the heavy ion collision simulations. These data
structures are implemented as base classes or classes, and they are
shared between different modules that are executed by the
framework. In this section, we discuss several important data
structures that are defined and used in the JETSCAPE framework.

\subsubsection{Base Class for Particles}
\begin{figure}[h]
  \centering
  \includegraphics[width=4.12in,height=1.42in]{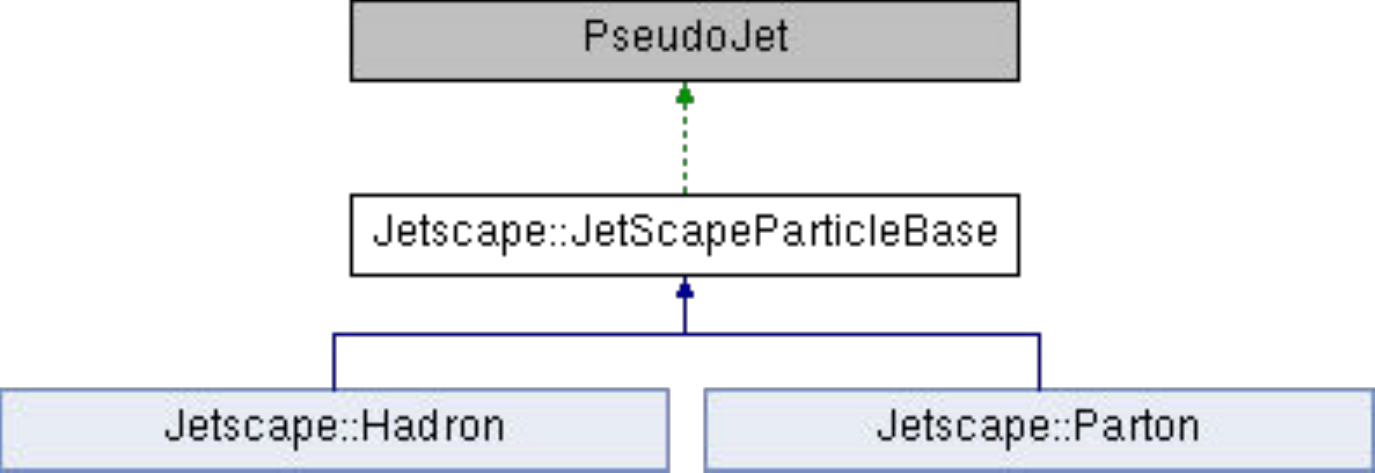}
  \caption{Class structure for \texttt{JetScapeParticleBase}.}
  \label{fig:particles}
\end{figure}

The JETSCAPE framework defines the class \code{JetScapeParticleBase}
as a base class for all the particles that will be generated during
the simulation. This base class privately inherits from
\code{PseudoJet} from FastJet's \code{fjcore}~\cite{Cacciari:2011ma} package.
By inheriting privately, we disable some of the
properties of \code{PseudoJet} that are not needed or could negatively
interfere with the often explicitly virtual nature of particles inside
the framework.
We add
several other properties such as particle identifier
\code{\detokenize{pid_}},
particle label \code{\detokenize{plabel_}},
particle status \code{\detokenize{pstat_}},
particle mass \code{\detokenize{mass_}},
particle position as a four vector
\code{\detokenize{x_in_}}, and a jet four vector without gamma factor
\code{\detokenize{jet_v_}}.
To facilitate detection of pathological cases where multiple energy
loss modules try to act on the same parton, we also add a boolean
\code{\detokenize{controlled_}} and a
string \code{\detokenize{controller_}}.
All added properties are
protected, which means the subclasses of \code{JetScapeParticleBase}
have access to them. We have derived two classes from
\code{JetScapeParticleBase}, \code{Parton} and \code{Hadron}. In
Fig.~\ref{fig:particles} we show the class structure for
\code{JetScapeParticleBase} and its derived classes \code{Parton}
and \code{Hadron}.
Both classes offer constructors using a custom \code{FourVector} class
or using $p_T, \eta, \phi$, and energy.

\begin{itemize}

\item \code{Parton} class: one derived class from
  \code{JetScapeParticleBase} is the \code{Parton} class.
  In addition to inherited properties from the base class, the
  \code{Parton} class has a mean formation time
  \code{\detokenize{mean_form_time_}}, 
  an event-by-event formation time \code{\detokenize{form_time_}},
  various fields related to color,
  a pointer to the shower that the parton belongs to
  \code{\detokenize{pShower_}} and its position within the shower
  graph (\code{\detokenize{edgeid_}}).
  
\item \code{Hadron} class: In the context of JETSCAPE, hadrons are
  much simpler objects than partons; at this point, the \code{Hadron}
  class merely adds the decay width (\code{\detokenize{width_}}) to
  the base class.
\end{itemize}

\subsubsection{Data Structure to Model Parton Shower}
\label{sec:showergraph}
A directed graph is a natural representation of a partonic shower
evolution, and the \code{PartonShower} class is derived from a graph
base class from the Graph Template Library (GTL) to conform to this
paradigm and make available existing implementations of
graph-theoretic tools for efficient and potentially new approaches to
studying the shower. Similar to any graph,
\code{PartonShower} has a list of nodes and a list of edges. In
our model, a parton is an edge in the shower graph
and splitting of a parton happens at a vertex (node) of the graph. To
keep the showering information as a graph, the \code{PartonShower}
class defines a map with partons as keys \code{pMap} and another map
with vertices as keys \code{vMap}.
The \code{PartonShower} class
provides functionalities for different modules to query the showering
graph, including getting the number of partons, getting the final
state partons, and getting parents of a parton. 



\subsection{Framework's Internals}
In this section we discuss several important internal mechanisms/utilities of the
JETSCAPE framework in more detail. 
First we describe the \code{JetScapeModuleMutex} class that ensures mutual exclusion between
various modules.

\subsubsection{The \rm \code{JetScapeModuleMutex}}
Due to the modular design of the JETSCAPE framework, users may attach
more than one module to the main task, for each aspect of
simulation. It is possible that two or more of the attached modules
are \emph{not} mutually exclusive. This means those modules cannot be attached to
the main task at the same time. For example, two energy loss modules,
MARTINI and LBT, work in the same region of energy, virtuality and
temperature, so they cannot both be attached to the main task. In
general, it is the user's responsibility to ensure only mutually
exclusive modules are attached at the same time, but to make the
JETSCAPE framework safer to use, we added the base class
\code{JetScapeModuleMutex} that provides the functionality to indicate
mutually exclusive modules.

As a base class, \code{JetScapeModuleMutex} provides a virtual
method \code{CheckMutex} that needs to be implemented for each
JETSCAPE module that needs a mutex. To implement this, the author of a
module needs to introduce a mutex class that is derived from
\code{JetScapeModuleMutex} and overrides the \code{CheckMutex}
method. In the overridden method, the author of the module indicates
his or her module is mutually exclusive with which other modules. 

Defining a mutex class for a module is optional, and it is the
author's decision. Moreover, utilizing the mutex can be turned off in
the XML file. For different types of modules we introduce the XML tag
\code{mutex} and the value can be ON or OFF. We included mutex
classes for MARTINI, LBT and AdSCFT energy loss modules, and we named
them MartiniMutex, LBTMutex, and AdSCFTMutex, respectively. 

\subsubsection{JETSCAPE Manager Tasks}
\label{sec:manager}

The \code{JetScape\-Signal\-Manager} is an internal singleton that
manages all signal/slot connections. This section is instead concerned with a
specific task type, exemplified in \code{Jet\-Energy\-Loss\-Manager}
and \code{Hadronization\-Manager}. We use this pattern when partons
should be sent to a list of alternative tasks for handling 
dependent on their properties. This is explicitly the case for jet energy loss,
as described in Sec.~\ref{programflow}, and is foreseen for
hadronization where alternatives such as recombination will be part of
future releases.

Their \code{Exec{}} routine consists of creating signals and slots and
then sending and receiving via these slots.
This creation is centrally registered in
\code{JetScape\-Signal\-Manager},
so Developers who wish to extend the framework with similar base
physics functionalities need to also change
\code{src/framework/JetScape\-Signal\-Manager.h} and \code{.cc}. We do not,
however, foresee the necessity to do so for most users and recommend
contacting the developers for assistance.

\subsubsection{The \code{JetScapeXML} Singleton} 
\label{sec:jsxml}

This singleton is currently a thin interface that grants access to the
global XML configuration file as a \code{tinyxml2::XMLElement}
object~\cite{tinyxml}. XML documents can be traversed using
\code{FirstChildElement()} and, and elements can be evaluated using
\code{GetText()} and \code{QueryDoubleText()}.

In the current release, it is the responsibility of the physics
modules to extract information using the above mechanisms in
\code{Init()} or \code{InitTask()}.
Module authors are advised to consider existing modules, such as
\code{src/initialstate/pgun.cxx} for usage examples.
Note that
it is rather easy to create segmentation faults by carelessly
accessing non-existing tags in a mal-formed configuration XML file.
We plan to expand this class with higher-level methods in the future. 

\subsubsection{The \code{JetScapeLogger} Singleton}

This singleton provides unified and thread-safe logging capabilities
via macros. It should be treated like a regular \code{std::ostream},
except without using \code{std::endl}. Four recommended default
streams are available, all of which output to \code{stdout}:
\begin{description}
\item[\code{WARN}] This stream is used for errors and warnings of serious
  conditions. It cannot be disabled, so it should be used very sparingly.
\item[\code{JSDEBUG}] This stream should be used during development only. 
  It can be disabled with  \code{JetScapeLogger::Instance()->SetDebug(false);}
\item[\code{INFO}] This stream is recommended for general
  process and setting information that is only shown once or important
  information that is shown once per event.
  It can be disabled with \code{JetScapeLogger::Instance()->SetInfo(false);}
\item[\code{VERBOSE(N)}] This stream is used for increasing levels of
  additional information where \code{N} is an \code{unsigned short}. The
  verbosity level can be set with a static member function:
  After \code{JetScapeLogger::Instance()->SetVerboseLevel(vlevel)},
  only messages with $\code{N}<\code{vlevel}$ will be displayed.
\end{description}

The following usage of verbosity levels is recommended:
\begin{description}
\item[1-5]: Technical information on an event level, for example
  the random seed. 
\item[6-9]: Less relevant event-level information, or particle level information.
  Typically, this may just belong in JSDEBUG, but was found relevant
  to stay available after development was completed.
\end{description}

The implementation as macros prevents protection in a namespace,
which combined with the relatively generic macro names may lead to
clashes with external packages. The macros will be renamed in the
next release, and future releases may reimplement the logger class
to be namespace-safe.




\section{Physics validation of JETSCAPE} 
\label{validation}

Besides the proper technical implementation of the JETSCAPE framework as outlined in sec.\ \ref{framework}, we want to briefly discuss in this chapter that the JETSCAPE framework and the provided modules are capable of reproducing the physics of interest. To validate the physics of JETSCAPE we focus on two crucial aspects as discussed below: Adequate description of the \pp\ reference baseline and first results of the charged hadron and jet nuclear modification factor at LHC energies in a multi-stage energy loss approach including a 2+1D hydro medium.

 \subsection{\pp\ Baseline Reference}
 \label{ppRef}

\begin{figure}[t]
  \centering
  \includegraphics[width=.7\textwidth]{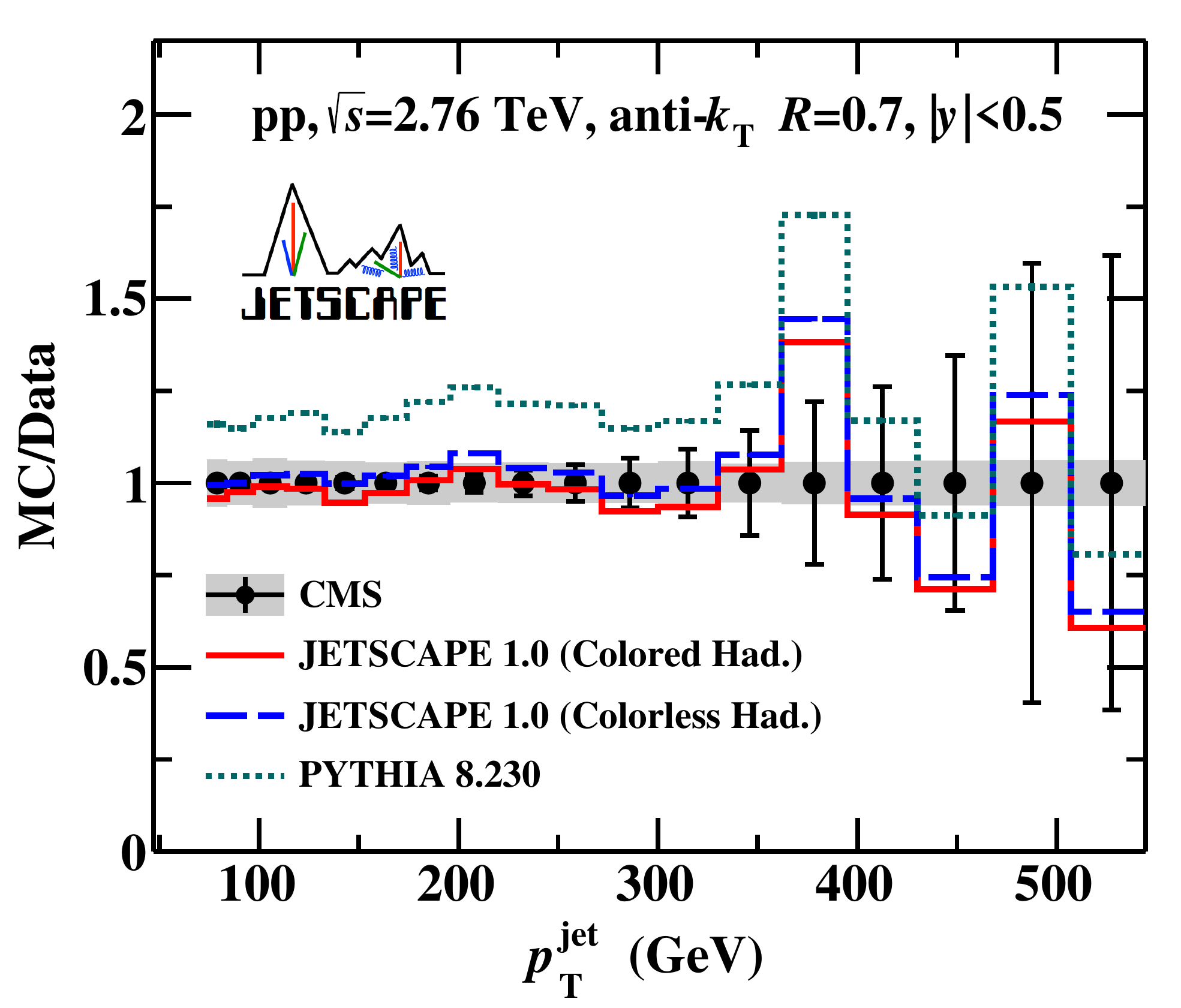}
  \caption{Ratios of inclusive cross sections $d\sigma/dp_T dy$ for full jets in $\sqrt{s} = 2.76$ TeV p+p collisions
  with jet radius $R=0.7$ and jet rapidity $|y|<0.5$.
  Three different Monte Carlo calculations, JETSCAPE Colored Hadronization (solid red line),
  JETSCAPE Colorless Hadronization (dashed blue line), and default PYTHIA 8 (dotted green line), are divided by
  the cross section measured by the CMS experiment \cite{Khachatryan:2015luy}.
  Statistical errors (black error bars) and systematic errors (grey band) errors are plotted with the data. The statistical errors of the
  Monte Carlo calculations are negligible.}
  \label{fig:ppRef}
\end{figure}

It is an important requirement that the JETSCAPE framework can provide a reliable baseline by reproducing
key observables in \pp\ collisions. This should be achieved to a degree of accuracy that is comparable to
existing Monte Carlo generators.
The tune JETSCAPE PP19 is based on JETSCAPE 1.0 and has been developed as a first step in this direction \cite{pppaper}.
It does well with a wide variety of \pp\ data (more details will be provided in an upcoming publication) but further fine tuning is possible and will have to be addressed in the future.
We will briefly summarize our study, by focusing on one important benchmark observable, the jet cross section, which gives a proof-of-principle as well as first benchmarks in the \pp\ sector.
The modules used for the JETSCAPE PP19 tune are (see sec.\ \ref{physics} for more details): Hard processes
are generated with PYTHIA 8 while final state showers are handled by MATTER with $\hat q =0$ and both Colored
Hadronization and Coloreless Hadronization can be used.
Only one parameter has been optimized to fit data, the initial parton virtuality assumed by MATTER. It is set to half of the transverse momentum of the parton handed over from PYTHIA. Obviously, a significantly more sophisticated fine tuning of parameters is possible and might be addressed in the future.

As a crucial benchmark observable we show in Fig.\  \ref{fig:ppRef} the ratio of the inclusive jet cross section
$d^2\sigma/dp_T dy$ as a function of jet
transverse momentum $p_T$, with data from the CMS collaboration \cite{Khachatryan:2015luy}. The collision energy in this
case is $\sqrt{s} = 2.76$TeV, and jets with $R=0.7$ and jet rapidity $|y|<0.5$ have been used. The three lines correspond to JETSCAPE
with Colored Hadronization, JETSCAPE with Colorless Hadronization and default PYTHIA 8. Very good agreement with experimental data over a wide range of jet energies is observed at LHC energies. The upcoming dedicated publication discusses in more detail inclusive jet cross sections for several RHIC and LHC collision energies as well as for various jet radii and
rapidity ranges. It also presents calculations for hadron cross sections, jet shape and fragmentation functions at those energies.
In summary, JETSCAPE shows similar agreement for many observables when compared to experimental data with respect to existing \pp\ Monte Carlo generators.

%

\subsection{Multi-stage Energy Loss in a 2+1D hydro at the LHC}
\label{pbpb}

 \begin{figure}[tbh]
  \begin{center}
   \includegraphics[width=0.48\textwidth]{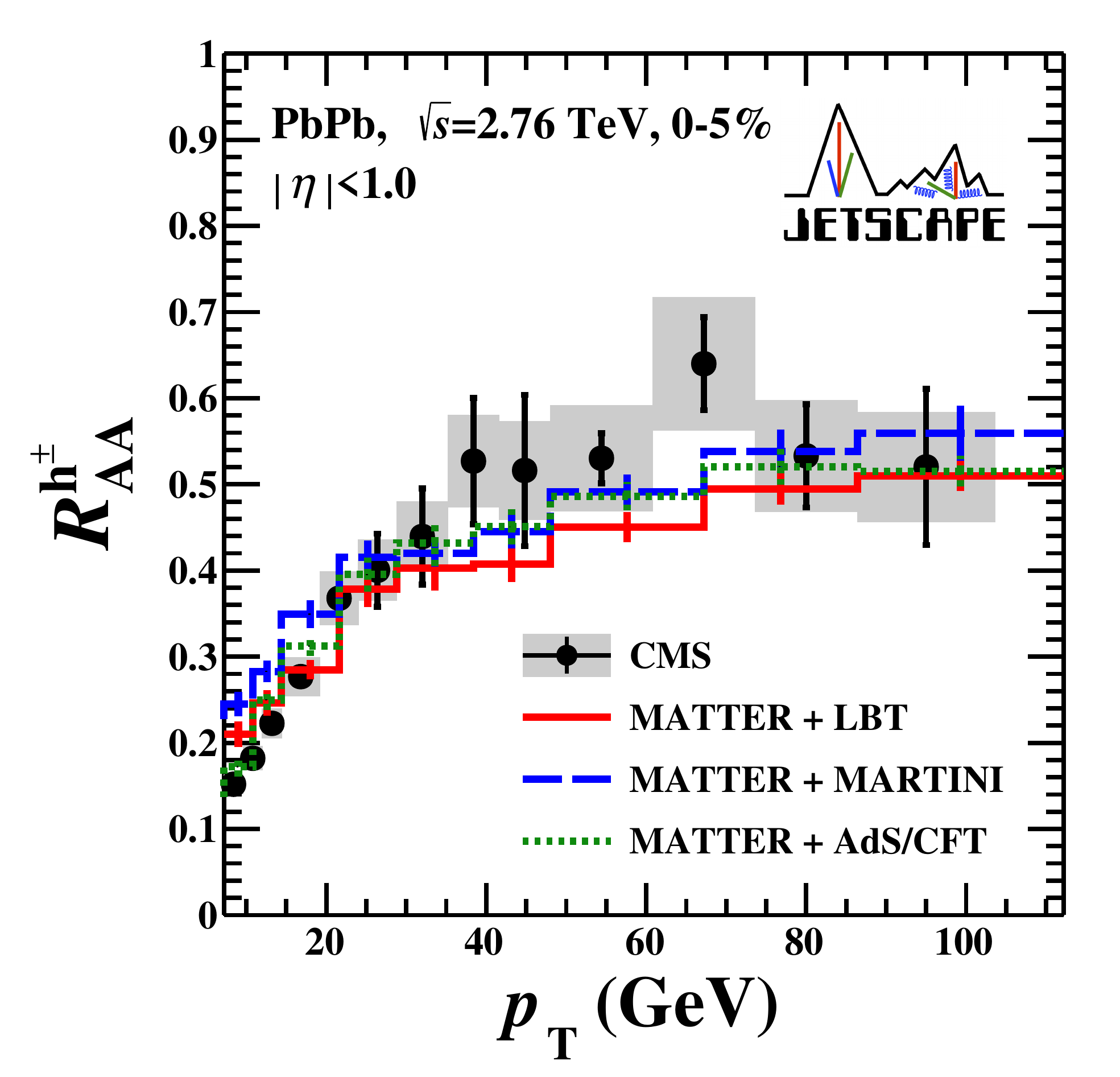}
    \includegraphics[width=0.48\textwidth]{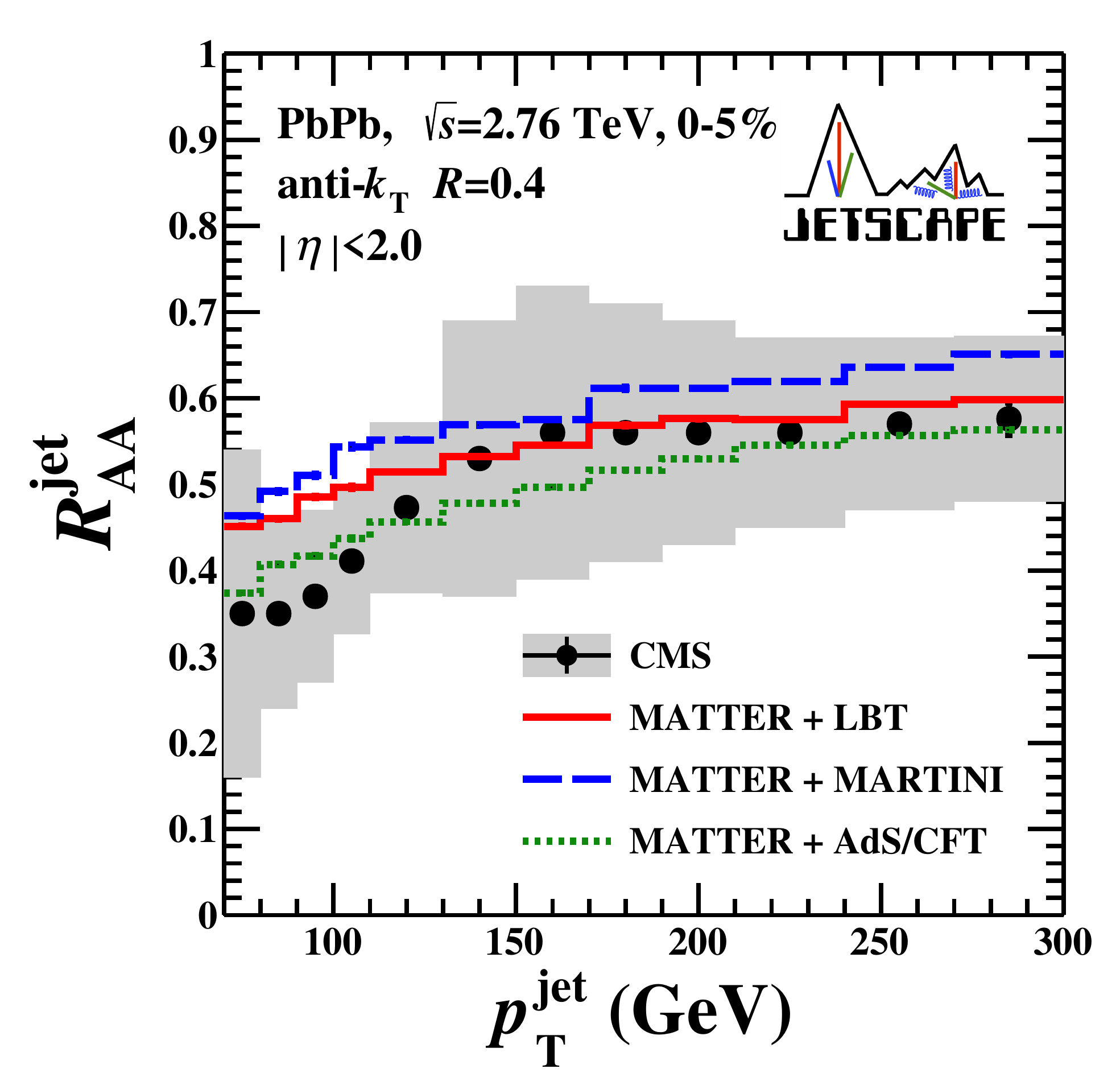}
  \caption{(Color online) Charged hadron $R_{AA}$ (left) and inclusive jet $R_{AA}$ (right) for $0$-$5\%$ PbPb collisions at $2.76$ TeV. In each figure, the JETSCAPE calculations that couple MATTER with MARTINI, LBT, and AdS/CFT are compared to measured data~\cite{CMS:2012aa,Khachatryan:2016jfl}.
   }
    \label{fig:raa}
    \end{center}
 \end{figure}

First results of charged hadron yield suppression, $R_{AA}$, and inclusive jet $R_{AA}$ at LHC energies in a multi-stage energy loss approach will be shown (see Fig.\  \ref{fig:raa}).
Simulations of Pb-Pb collisions are performed by the MATTER vacuum and medium shower coupled to the three different time-ordered parton shower models; LBT, MARTINI, and AdS/CFT.
Switching between different energy loss modules is done independently for each parton.
The switching parameter $Q_0$ is set to $2$ GeV.
The event-averaged hydrodynamic background is provided by 2+1D VISHNew~\cite{Shen:2014vra} with T$\raisebox{-0.4ex}{R}$ENTo~\cite{Moreland:2014oya} initial conditions.

In Fig.\ \ref{fig:raa} we present the JETSCAPE calculation using combinations of MATTER with LBT, MARTINI and AdS/CFT, which gives reasonable descriptions of the measured data from the CMS experiments~\cite{CMS:2012aa,Khachatryan:2016jfl}.
We also observed that the three different combinations of energy loss modules implemented in the JETSCAPE framework yield compatible results for single hadron and jet observables.

\newpage

\section{Summary and Outlook}
\label{summary}

In this article we introduced JETSCAPE: An open source, thread-safe, modular, extendable and user-friendly modern-event-generator framework. The default event-generator includes state-of-the-art modules that simulate each physics component, with emphasis on the complex description of partonic energy loss in heavy-ion collisions. We validated the physics of JETSCAPE to ensure an adequate description of the \pp\ reference baseline, as well as describing nuclear modification measurements at the LHC incorporating a 2+1D hydrodynamic module.

The modular design of the JETSCAPE framework provides a one-to-one computational representation of the current understanding of the different physics aspects of heavy-ion collisions (as represented schematically in Fig.\ \ref{fig:flowchart}).
The modular, intuitive, user-friendly work flow and state-of-the-art physics modules give JETSCAPE the potential to become the tool of choice in the heavy-ion physics community to implement and explore new physics.
Furthermore, we think that JETSCAPE will also be helpful for experimentalists to study the sensitivity of new experimental observables to different physics aspects, as well as comparing to existing measurements.

It should be pointed out that the evolution of JETSCAPE is ongoing. JETSCAPE v1.3 is still an early release and several physics modules discussed here are not yet included. The SMASH module is currently being tested and is due for inclusion in v1.4.
Currently, photons produced in the high virtuality phase (MATTER) are not calculated. Photons produced in the low virtuality phase (LBT/MARTINI) are not retained in the final list of particles. This addition of photons in the energy loss phase and associated processing by the framework is currently in development and projected for inclusion in v1.6.
The combined hadronization of hard and soft sectors via recombination, added to string fragmentation in a hybrid hadronization model is in preliminary testing and slated for inclusion in v2.0, which constitutes the next major release.

JETSCAPE v2.0 will also include several new additions: Heavy flavor energy loss routines in the high virtuality sector will be added to MATTER. Bayesian statistical routines for calibration and rigorous comparison with experimental data will be provided. The framework will also be extended to allow a consistent and concurrent treatment of energy-momentum deposition from the jet to the QGP fluid, and its associated excitation of hydrodynamic modes. To achieve the second goal, JETSCAPE will parallelize computing intensive portions of the code to GPU processors (see Appendix \ref{benchmark} for current benchmarking results of JETSCAPE).

Even with these enhancements, it will be challenging for JETSCAPE v2.0 to describe LHC events with several 
($n\geq4$) jets with $p_T \gtrsim 20$GeV, where the wakes of jets might intersect. Incorporation of such non-linear effects in jet energy deposition and excitation of the QGP will require major updates in the framework, fluid dynamics sector, and some of the energy loss modules. So far, the focus of JETSCAPE simulations has been on the hadronic sector for both soft and hard observables, with jet related hard photons. There is no current module to compute electro-magnetic signatures of the QGP. More advanced topics such as Onia suppression or regeneration, chiral magnetic or vortical effects etc., are also not planned in the immediate upcoming releases.

The JETSCAPE event generator, as provided, contains several new modules, which have never been combined in prior analyses, e.g., TRENTO + FreeStreaming + MUSIC + SMASH, and/or MATTER + (LBT/MARTINI) + AdS/CFT + PYTHIA hard scattering and hadronization etc. As such, the package is provided to the user un-calibrated (untuned), either in the soft or hard sector. To use the package effectively, the user will first need to calibrate the soft sector. This involves using a subset of experimental data to constrain the values of 10-15 parameters. Once calibrated, the soft sector can be used to predict other experimental results, and provide the background for jet quenching. The use of a particular combination of jet modification modules will also require prior calibration using a subset of data from the hard sector.

A systematic calibration of the 3+1D version of the soft sector as well as the hard sector will be performed in the future. Bayesian techniques will be used to establish methodical constraints on the physical parameters of the models while varying the parameters over a large range of possible values. This analysis is part of the Collaboration's mandate, and the results will be made available for use by the community. This numerically demanding calibration will be performed using resources made available from the XSEDE project~\cite{ecss,xsede}. The statistical package developed for this calibration will be released as part of the v2.0 release. The impatient user may avoid the calibration of the soft sector entirely by using a pre-calibrated medium using the \code{hydroFileTest} or \code{hydroJetTest} setups. This will invoke older hydrodynamic profiles, obtained from a separate Bayesian calibration~\cite{Bernhard:2018hnz}, which are publicly available. As for the hard sector, preliminary calibrations may be obtained from Refs.~\cite{Tachibana:2018yae,Park:2019sdn}.

\section*{Acknowledgments} 
This  work  was  supported  in  part  by  the National Science Foundation (NSF) within the framework of the JETSCAPE collaboration, under grant numbers ACI-1550172 (G.R.), ACI-1550221(R.J.F., M.K. and Z.Y.), ACI-1550223 (D.E., M. M. and U.H.), ACI-1550225 (S.A.B., J.C., T.D., W.F., W.K., R.W. and Y.X. ), ACI-1550228 (L.-G.P., X.-N.W., P.J., J.M.), and ACI-1550300 (S.C., L.C., K. K., A.K., M.E.K., A.M., C.N., D.P., J.P., L.S., C.Si. and R.A.S.). It was also supported in part by the NSF under grant numbers PHY-1207918 (M.K.), 1812431 (R.J.F., M.K. and Z.Y.) and by the US Department of Energy, Office of Science, Office of Nuclear Physics under grant numbers \rm{DE-AC02-05CH11231} (D.O., Y.H. and X.-N.W.), \rm{DE-AC52-07NA27344} (R.A.S.), \rm{DE-SC0012704} (C.S.), \rm{DE-SC0013460} (S.C., A.K., A.M., C.S. and C.Si.), \rm{DE-SC0004286} (L.D., U.H.), \rm{DE-SC0012704} (K.K.), \rm{DE-FG02-92ER40713} (J.P.) and \rm{DE-FG02-05ER41367} (T.D., W.K., J.-F.P., S.A.B. and Y.X.). The work was also supported in part by the National Science Foundation of China (NSFC) under grant number 11521064, Ministry of Science and Technology (MOST) of China under Projects number 2014CB845404 (Y.H., T.L. and X.-N.W), and in part by the Natural Sciences and Engineering Research Council of Canada (C.G., S.J., C.P. and G.V.), the Fonds de recherche du Qu\'ebec Nature et technologies (FRQ-NT) (G.V.) and by the Office of the Vice President for Research (OVPR) at Wayne State University (Y.T.).  This work used the Extreme Science and Engineering Discovery Environment (XSEDE), which is supported by National Science Foundation grant number ACI-1548562. Computations were made in part on the supercomputer \emph{Guillimin} from McGill University, managed by Calcul Qu\'ebec and Compute Canada. The operation of this supercomputer is funded by the Canada Foundation for Innovation (CFI), NanoQu\'ebec, R\'eseau de M\'edicine G\'en\'etique Appliqu\'ee~(RMGA) and FRQ-NT. Computations were also carried out on the Wayne State Grid funded by the Wayne State OVPR. Data storage was provided in part by the OSIRIS project supported by the National Science Foundation under grant number OAC-1541335. C.G. gratefully acknowledges support from the Canada Council for the Arts through its Killam Research Fellowship program. C.S. gratefully acknowledges a Goldhaber Distinguished Fellowship from Brookhaven Science Associates.

\clearpage

\begin{appendices}
\section{Framework's Input and Output}

\newcommand{\trento}{{\rm T\raisebox{-0.5ex}{R}ENTo}}

\subsection{Input Parameters From XML File}
\label{sec:xml}
\subsubsection{Initial State Module {\rm \texttt{<IS>}}}

\vspace{10pt} \paragraph{general parameters}
\vspace{-5pt}
\begin{description}
\setlength{\parskip}{0pt}
\item[\texttt{<grid\_max\_x>}:]
the range of $x$ ([-grid\_max\_x, grid\_max\_x]) in [fm].
\item[\texttt{<grid\_max\_y>}:]
the range of $y$ ([-grid\_max\_y, grid\_max\_y]) in [fm].
\item[\texttt{<grid\_max\_z>}:] If the hydro is in $(\tau, x, y, \eta_s)$ coordinates, this term denotes the range of $\eta_s$ ([-grid\_max\_z, grid\_max\_z]).
If the hydro is in $(t, x, y, z)$ coordinates, this term denotes the range of $z$ ([-grid\_max\_z, grid\_max\_z]) in [fm].
\item[\texttt{<grid\_step\_x>}:]
the size of the grid in $x$ in [fm].
\item[\texttt{<grid\_step\_y>}:]
the size of the grid in $y$ in [fm].
\item[\texttt{<grid\_step\_z>}:] If the hydro is in $(\tau, x, y, \eta_s)$ coordinates, this term denotes the size of the grid in $\eta_s$.
If the hydro is in $(t, x, y, z)$ coordinates, this term denotes the size of the grid in $z$ in fm.
\end{description}

\vspace{10pt} \paragraph{options}
\vspace{-5pt}
\begin{description}
\setlength{\parskip}{0pt}
\item[\texttt{<initial\_profile\_path>}:]
the path of the initial profile.~(option to read initial conditions from saved file.)
\end{description}

\vspace{10pt} \paragraph{\trento {\rm \texttt{<Torento>}}}

\vspace{-5pt}
\begin{description}
\setlength{\parskip}{0pt}
\item[\texttt{use\_module}:]
the attribute to specify the setting for the collision system (\texttt{"pre\_defined"} or \texttt{"user\_defined"}).
\item[\texttt{<pre\_defined>}:] 
the flag to use the pre-defined default collision system (AuAu~at~200~GeV, PbPb~at~2.76~TeV and at~5.02~TeV).
\begin{itemize}
\setlength{\itemsep}{0pt}
   \item[] \texttt{collision\_system}: the colliding system (e.g.~\texttt{"auau200"},~\texttt{"pbpb2760"},~\texttt{"pbpb5020"}).
    \item[] \texttt{centrality\_min}: the minimum value of the centrality in [\%].
        \item[] \texttt{centrality\_max}: the maximum value of the centrality in [\%].
\end{itemize}
\item[\texttt{<user\_defined>}:] 
the flag to use the user-defined collision system. Events are generated in 0-100 \% centrality range. 
\begin{itemize}
\setlength{\itemsep}{0cm}
   \item[] \texttt{projectile}: the name of projectile nucleus (e.g.~\texttt{"Au"},~\texttt{"Pb"}).
    \item[] \texttt{target}: the name of target nucleus (e.g.~\texttt{"Au"},~\texttt{"Pb"}).
     \item[] \texttt{sqrts}: the centre-of-mass energy per nucleon pair in [GeV].
     \item[] \texttt{cross\_section}: the nucleon-nucleon cross section in [fm$^2$] (1~fm$^2$=10~mb).
\end{itemize}
\end{description}

Notice that the current framework does not provide interface to change many physical parameters used in TRENTO,
which will be done in the future. Such parameters are,
\begin{itemize}
\item Nuclear density distribution with parameters (e.g radius and skin thickness in Woods-Saxon distribution).
\item width of nucleon 
\item entropy deposition parameter $p$ (TRENTO only)
\item multiplicity fluctuation parameter $w$ (TRENTO only)
\end{itemize}

\subsubsection{Parameters for Hard Process {\rm \texttt{<Hard>}}}

\vspace{10pt} \paragraph{Parton Gun {\rm \texttt{<PGun>}}}
\vspace{-5pt}
\begin{description}
\setlength{\parskip}{0pt}
\item[\texttt{<pT>}:] the initial $p_{\rm T}$ of the parton in [GeV/$c$].
\end{description}

\vspace{10pt} \paragraph{Pythia Gun {\rm \texttt{<PythiaGun>}}}
\vspace{-5pt}
\begin{description}
\item[\texttt{<pTHatMin>}:] the minimum $p_{\rm T}$ in the rest frame of the initial hard scattering process.
\item[\texttt{<pTHatMax>}:] the maximum $p_{\rm T}$ in the rest frame of the initial hard scattering process.
\item[\texttt{<eCM>}:] the centre-of-mass energy per nucleon pair in [GeV].
\item[\texttt{<LinesToRead>}:] the tag to add lines for initialization of pythia.\\
e.g. \vspace{-5pt}
\begin{verbatim}
<LinesToRead>
    HardQCD:all = on
</LinesToRead>
\end{verbatim}
\vspace{-8pt}
Note: if the tag exists it cannot be empty.
\end{description}

\subsubsection{ Energy Loss Modules {\rm \texttt{<Eloss>}}}

\vspace{10pt} \paragraph{general parameters}
\vspace{-5pt}
\begin{description}
\setlength{\parskip}{0pt}
\item[\texttt{<deltaT>}:]
the time step for the jet evolution in the lab frame in [fm/$c$]. 
\item[\texttt{<maxT>}:]
the maximum time for the jet evolution in the lab frame in [fm/$c$]. 
\end{description}

\vspace{10pt} \paragraph{{{MATTER}} {\rm \texttt{<Matter>}}}
\vspace{-5pt}
\begin{description}
\setlength{\parskip}{0pt}
\item[\texttt{<Q0>}:] the virtuality of a parton to switch from {MATTER} to the transport energy loss module in [GeV].
\item[\texttt{<T0>}:] the temperature to switch from the transport energy loss module to {MATTER} in [GeV]. The value must be the same as that in \texttt{<hydro\_Tc>} in the transport energy loss module.
\item[\texttt{<vir\_factor>}:] the factor to be multiplied by the $p_{\rm T}$ of the initial parton in {MATTER} to obtain the maximum virtuality of the parton. 
\item[\texttt{<in\_vac>}:] the flag to turn off and on the medium effect in {MATTER} ( \texttt{1}:~in vacuum, \texttt{0}:~in medium)
\item[\texttt{<recoil\_on>}:] the flag to turn on and off the recoils in {MATTER} (\texttt{1}:~on, \texttt{0}:~off)
\item[\texttt{<broadening\_on>}:] the flag to turn on and off the broadening effect in {MATTER} (\texttt{1}:~on, \texttt{0}:~off). 
If \texttt{<recoil\_on>} is \texttt{1} (recoil is on), the broadening effect is automatically turned off regardless of this flag's setting.
\item[\texttt{<brick\_med>}:] the flag to use the static uniform medium (brick) in {MATTER} (\texttt{1}:~yes, \texttt{0}:~no). 
\item[\texttt{<brick\_length>}:] the length of the brick in [fm].
\item[\texttt{<hydro\_Tc>}:] the temperature below which the medium effect is turned off in {MATTER} in [GeV].
\item[\texttt{<qhat0>}:] the value of $\hat{q}_0$ in {MATTER} in [GeV$^2$/fm]. 
If a negative value is set here, $\alpha_{\rm s}$ is used to calculate $\hat{q}$. 
\item[\texttt{<alphas>}:] the value of $\alpha_{\rm s}$ in {MATTER}. 
To use the value of $\alpha_{\rm s}$ being set here, 
set the value in \texttt{<qhat0>} to a negative value. 
\end{description}

\vspace{10pt} \paragraph{{{LBT}} {\rm \texttt{<Lbt>}}}
\vspace{-5pt}
\begin{description}
\setlength{\parskip}{0pt}
\item[\texttt{<Q0>}:] the virtuality of a parton to switch from {MATTER} to {LBT} in [GeV].
\item[\texttt{<in\_vac>}:] the flag to turn off and on the medium effect in {LBT} ( \texttt{1}:~in vacuum, \texttt{0}:~in medium). 
\item[\texttt{<only\_leading>}:] the flag to turn off the tracking of any radiated partons and recoils in {LBT} (\texttt{1}:~track only the partons received from {MATTER}, \texttt{0}:~track all the partons in jet). 
\item[\texttt{<hydro\_Tc>}:] the temperature below which the medium effect is turned off in {LBT} in [GeV]. 
The value must be the same as that in \texttt{<T0>} in {MATTER}.
\item[\texttt{<alphas>}:] the value of $\alpha_{\rm s}$ in {LBT}. 
\end{description}

\vspace{10pt} \paragraph{{{MARTINI}} {\rm \texttt{<Martini>}}}
\vspace{-5pt}
\begin{description}
\setlength{\parskip}{0pt}
\item[\texttt{<Q0>}:] the virtuality of a parton to switch from {MATTER} to {MARTINI} in [GeV].
\item[\texttt{<alphas>}:] the value of $\alpha_{\rm s}$ in {MARTINI}. 
\item[\texttt{<pcut>}:] the momentum cut in [GeV]. The tracking of the particles with momentum in the local rest frame of the medium fluid below this value is turned off in {MARTINI}. 
\item[\texttt{<hydro\_Tc>}:] the temperature below which the medium effect is turned off in {MARTINI} in [GeV]. 
The value must be the same as that in \texttt{<T0>} in {MATTER}.
\item[\texttt{<path>}:] 
the path of the directory containing the tables used in {MARTINI}. 
\end{description}

\vspace{10pt} \paragraph{{{AdS/CFT}} {\rm \texttt{<AdSCFT>}}}
\vspace{-5pt}
\begin{description}
\setlength{\parskip}{0pt}
\item[\texttt{<Q0>}:] the virtuality of a parton to switch from {MATTER} to {AdS/CFT} in [GeV]. 
\item[\texttt{<in\_vac>}:] the flag to turn off and on the medium effect in {AdS/CFT} ( \texttt{1}:~in vacuum, \texttt{0}:~in medium)
\item[\texttt{<kappa>}:] the value of $\kappa$ in {AdS/CFT}. 
\item[\texttt{<T0>}:] the temperature below which the medium effect is turned off in {AdS/CFT} in [GeV]. 
The value must be the same as that in \texttt{<T0>} in {MATTER}. 
\end{description}

\subsubsection{Preequilibrium Dynamics Module {\rm \texttt{<Preequilibrium>}}}

\vspace{10pt} \paragraph{general parameters}
\vspace{-5pt}
\begin{description}
\setlength{\parskip}{0pt}
\item[\texttt{<tau0>}:]
the starting proper time $\tau$ of the preequilibrium dynamics in the relativistic $\tau$-$\eta_{\rm s}$ coordinates in [fm/$c$].
\item[\texttt{<taus>}:]
the switching proper time $\tau$ from preequilibrium dynamics to hydrodynamics (Landau Matching) in the relativistic $\tau$-$\eta_{\rm s}$ coordinates in [fm/$c$].
\end{description}

\vspace{10pt} \paragraph{{{Freestreaming}} {\rm \texttt{<FreestreamMilne>}}}
\vspace{-5pt}
\begin{description}
\setlength{\parskip}{0pt}
\item[\texttt{<freestream\_input\_file>}:]
the path of the input file for the {Freestreaming} Module. 
\end{description}

\subsubsection{Hydrodynamics Module {\rm \texttt{<Hydro>}}}

\vspace{10pt} 
\paragraph{Brick (static uniform/Bjorken expanding medium)\,\rm\texttt{<Brick>}}
\vspace{-5pt}
\begin{description}
\setlength{\parskip}{0pt}
\item[\texttt{bjorken\_expansion\_on}:]
the attribute for the flag to turn on and off the Bjorken expansion of the brick medium (\texttt{"true"} or \texttt{"false"}).
\item[\texttt{start\_time}:]
the attribute for the initial time for the Bjorken expansion $t_{\rm Bj}$ in [fm/$c$]. \item[\texttt{<T>}:] 
the temperature of the brick medium $T$ in [GeV]. 
If the Bjorken expansion is on, the temperature of the brick medium evolves according to $T(t) = T (t_{\rm Bj}/t)^{1/3}$. 
\end{description}

\vspace{10pt} 
\paragraph{Hydro from File\,\rm\texttt{<hydro\_from\_file>}}
\vspace{-5pt}
\begin{description}
\setlength{\parskip}{0pt}
\item[\texttt{<read\_in\_multiple\_hydro>}:] 
the number of the hydro profile. 
\item[\texttt{<hydro\_files\_folder>}:] 
the path of the hydro profile. 
\item[\texttt{<hydro\_type>}:] 
the type of the hydro profile file ( \texttt{1}:~VISHNew, \texttt{2}:~MUSIC).
\item[\texttt{<VISH\_file>}:] 
the path of the hydro profile from VISHNew in hdf5 format. 
\item[\texttt{<load\_viscous\_info>}:] 
the flag to use the information of medium viscosity from VISHNew for jet energy loss calculation (\texttt{1}:~yes, \texttt{0}:~no). 
\item[\texttt{<MUSIC\_input\_file>}:] 
the path of input file for MUSIC (used to specify the grid information).
\item[\texttt{<MUSIC\_file>}:] 
the path of the hydro profile from MUSIC in plain binary format. 
\item[\texttt{<T\_c>}:] 
the transition temperature between QGP and Hadron Resonance Gas in~[GeV]. 
\item[\texttt{<read\_hydro\_every\_ntau>}:] 
the flag to read MUSIC hydro profile every step of $\tau$ (\texttt{1}:~yes, \texttt{0}:~no). 
\end{description}

\vspace{10pt} 
\paragraph{MUSIC\,\rm\texttt{<MUSIC>}}
\vspace{-5pt}
\begin{description}
\setlength{\parskip}{0pt}
\item[\texttt{<MUSIC\_input\_file>}:] 
the path of input file for MUSIC.
\item[\texttt{<Perform\_CooperFrye\_Feezeout>}:] 
the flag to perform particlization via Cooper-Fry formula at freezeout (\texttt{1}:~yes, \texttt{0}:~no). 
\end{description}

\subsubsection{Jet Hadronization {\rm \texttt{<JetHadronization>}}}

\vspace{10pt} 

\begin{description}
\setlength{\parskip}{0pt}
\item[\texttt{<eCMforHadronization>}:] 
the centre-of-mass energy per nucleon pair in [GeV]. 
\end{description}

\subsubsection{Soft Particlization {\rm \texttt{<SoftParticlization>}}}

\vspace{10pt} 
\paragraph{iSpectraSampler\,\rm\texttt{<iSS>}}
\vspace{-5pt}
\begin{description}
\setlength{\parskip}{0pt}
\item[\texttt{<hydro\_mode>}:] 
the type of the hydro freezeout surface file ( \texttt{0}:~VISHNew, \texttt{1}:~$(2+1)$-D MUSIC, \texttt{2}:~$(3+1)$-D MUSIC).
\item[\texttt{<iSS\_input\_file>}:] 
the path of input file for iSpectraSampler.
\item[\texttt{<iSS\_working\_path>}:] 
the working path of iSpectraSampler.
\item[\texttt{<number\_of\_repeated\_sampling>}:] 
the number of sampling per one hydro freezeout surface. 
\item[\texttt{<Perform\_resonance\_decay>}:] 
the flag to perform resonance decay after the particlization (\texttt{1}:~yes, \texttt{0}:~no). 
\end{description}

\subsection{Input Format of External Hydro Evolution History}

Jet shower propagation requires the local temperature $T$, entropy density $s$ and fluid velocity $u^{\mu}$, $\ldots$ of any given spatial coordinate $(x^1, x^2, x^3)$ at any given time $x^0$. In other words, it requires the whole hydrodynamic evolution history. Usually it is computationally expensive to compute the evolution history for one hydrodynamic event starting from a given fluctuating initial condition. One feasible option is to associate multiple jet shower propagation events with the same hydrodynamic evolution history, in non-concurrent-running mode where the energy momentum deposition into the medium is treated as perturbation. In this mode, it is quite convenient to connect JetScape with third party hydrodynamic modules by reading the stored external evolution history from the hard disk. To reduce the effort in writing a different interface for each different hydrodynamic program with different output formats, we recommend users to use the unified data format to store the external hydrodynamic evolution history. 

We propose to use the hdf5 file format to store each evolution history and its associated parameters.
The first thing that one should tell Jetscape in the xml configuration file is the folder of the external hydrodynamic evolution history.
In order to reduce the complexity, the external evolution history file should be "Event-N/JetData.h5" for each event with index "N" under the designated folder.
 
\begin{verbatim}
<hydro_from_file>
    <name>Hydro from file </name>
    <hydro_files_folder>./test_hydro_files</hydro_files_folder>
</hydro_from_file>
\end{verbatim}

{\bf Input:}

$\bullet$ One hdf5 file that stores the whole evolution history and associated parameters.
The required descriptive parameters for the evolution history are,
\begin{description}
    \item[Event/XL:] the  lower boundary in the x direction [${\rm fm}$] 
    \item[Event/XH:] the upper boundary in the x direction [${\rm fm}$]
    \item[Event/YL:] the  lower boundary in the y direction [${\rm fm}$] 
    \item[Event/YH:] the upper boundary in the y direction [${\rm fm}$] 
    \item[Event/Tau0:] the initial thermalization time [${\rm fm}$] 
    \item[Event/dTau:] the time step [${\rm fm}$] 
    \item[Event/DX:] the x step spacing [${\rm fm}$] 
    \item[Event/DY:] the y step spacing [${\rm fm}$] 
\item[Event/OutputViscousFlag:] 0 for ideal hydro, 1 for viscous hydro
\end{description}
The required data arrays in the evolution history are,
\begin{description}
    \item[Event/e:] energy density [${\rm GeV/fm^3}$] 
    \item[Event/s:] entropy density [${\rm fm^{-3}}$]
\item[Event/Vx:] the x component of fluid velocity 
\item[Event/Vy:] the y component of fluid velocity
\item[Event/Temp:] the local temperature [${\rm GeV}$]
\item[Event/P:] the local pressure [${\rm GeV/fm^3}$]
\item[Event/Pi00:] the 00 component of shear viscous tensor [${\rm GeV/fm^3}$]
\item[Event/Pi01:] the 01 component of shear viscous tensor [${\rm GeV/fm^3}$]
\item[Event/Pi02:] the 02 component of shear viscous tensor [${\rm GeV/fm^3}$]
\item[Event/Pi03:] the 03 component of shear viscous tensor [${\rm GeV/fm^3}$]
\item[Event/Pi11:] the 11 component of shear viscous tensor [${\rm GeV/fm^3}$]
\item[Event/Pi12:] the 12 component of shear viscous tensor [${\rm GeV/fm^3}$]
\item[Event/Pi13:] the 13 component of shear viscous tensor [${\rm GeV/fm^3}$]
\item[Event/Pi22:] the 22 component of shear viscous tensor [${\rm GeV/fm^3}$]
\item[Event/Pi23:] the 23 component of shear viscous tensor [${\rm GeV/fm^3}$]
\item[Event/Pi33:] the 33 component of shear viscous tensor [${\rm GeV/fm^3}$]
\item[Event/BulkPi:] the bulk viscosity [${\rm GeV/fm^3}$]
\end{description}

For hydrodynamic expert: notice that the $\pi^{03}$, $\pi^{13}$, $\pi^{23}$ and $\pi^{33}$ terms should be saved with $\tau$ factors such
that they all have the same dimension ${\rm GeV/fm^3}$.

 \subsection{Output data format in JETSCAPE framework}
 The JETSCAPE framework provides three different classes namely {\bf JetScapeWriterAscii, JetScapeWriterAsciiGZ, and JetScapeWriterHepMC}  to facilitate storing the event information into the hard disk in ASCII, Compressed binary format, and the HepMC format, respectively.
 In the event record we save information such 
 \begin{itemize}
 
 \item Hard scattering cross section (sigmaGen), hard scattering cross section error (sigmaErr), and weight factor for the hard scattering 
 \item Parton List from the Hard Process
 \item Particle properties such as PID, particle status, four-momenta for particles in the Parton shower (GTL format) for each shower initiating parton 
 \item Final state particles at the hadronic level.
 \end{itemize}
For demonstration, we discuss here the event information saved in ASCII format. The entries in the event record with the header { \bf sigmaGen, sigmaErr}, and {\bf weight }  represent the hard scattering cross section (in mb), error in the hard scattering cross section (in mb), and weight of the hard scattering, respectively. The header {\bf HardProcess Parton List} represents the list of the particles produced from the hard scattering. It is followed by label {\bf PythiaGun} which represents the name of the module used for the hard scattering. The entries followed by this header represent quantities such as internal label (for future use), PID, particle status, $p_{\mathrm{T}} , \eta, \phi_{p},E, x,y,z$, and $t $ for the particle produced from the hard scattering, where the momenta are in GeV and the spatial coordinates in fm.
\begin{lstlisting}[frame=single, language=C++]
0 Event
# sigmaGen 6.18e-06
# sigmaErr 6.18e-06
# weight 1
# HardProcess Parton List: PythiaGun
0 21 0 135 -0.04 3.42 135 0.2 1 0 0
0 2  0 110 -1.66 0.31 303 0.2 1 0 0
0 21 0 12  -0.91 0.33 19  0.2 1 0 0
\end{lstlisting}

The subsequent block with the header { \bf  Energy loss Shower Initating Parton } contains the shower information for every parton produced in the hard process. First, we list the particle information for the shower initiating parton. We follow the same format as in the hard process parton list discussed above. A parton shower is described in momentum space by assigning a label to each particle vertex represented by $x,y,z,$ and $t$. The row with label {\bf $\bf [i]=>[j]$ P} represents that the particle at vertex {$\bf  [j]$}  is a daughter of particle at the vertex { $ \bf[i]$}. The label {\bf P} means the daughter is a parton. The entries followed by the label {$\bf [i]=>[j] $} represent the particle information in the same format as for the hard process parton list. 
\begin{lstlisting}[frame=single, language=C++]
# Energy loss Shower Initating Parton: JetEnergyLoss
0 21 0 135 -0.04 3.42 135 0.2 1 0 0
# Parton Shower in JetScape format to be 
used later by GTL graph:
[0] V 0 0 0 0
[1] V 0 0 0 0
[2] V 0 0 0 0.1
[3] V 0 0 0 0.1
[4] V 0 0 0 0.6
[5] V 0 0 0 0.6
[6] V 0 0 0 1.2
[7] V 0 0 0 1.2
[0]=>[1] P 0 21 0 119 -0.04 3 135 0.2 1 0 0
[1]=>[2] P 0 21 0 120 0.004 3 120 0.10 0.97 -0.004 0.1
[1]=>[3] P 0 21 0 14 -0.36 2 15 0.10 0.97 -0.004 0.1
[3]=>[4] P 0 21 0 11 -0.29 2 11 -0.03 1 -0.17 0.6
[3]=>[5] P 0 21 0 3 -0.60 2 3 -0.03 1 -0.17 0.6
[2]=>[6] P 0 21 0 7 0.002 3 7 -0.90 0.53 0.0004 1
[2]=>[7] P 0 21 0 112 0.004 3 112 -0.90 0.53 0.0004 1
\end{lstlisting}

We also save the final state particles at the hadronic level in a format shown below.
Each entry represents a final state hadron in a format (index, hadron label, internal index, PID, particle status, $p_{\mathrm{T}} , \eta, \phi_{p},E, x,y,z,t $)
\begin{lstlisting}[frame=single, language=C++]
# Hadronization module: 
# Final State Hadrons
[0] H 0 211 0 21.86 -1.65 0.29 59.38 0 0 0 0
[1] H 0 -211 0 3.06 -1.62 0.32 8.08 0 0 0 0
[2] H 0 211 0 0.76 -1.24 0.44 1.45 0 0 0 0
[3] H 0 -211 0 5.21 -1.66 0.26 14.21 0 0 0 0
[4] H 0 211 0 1.11 -1.82 4.72 3.56 0 0 0 0
[5] H 0 111 0 0.37 -1.90 5.96 1.30 0 0 0 0
[6] H 0 111 0 0.99 -1.0 4.52 1.54 0 0 0 0
[7] H 0 -211 0 0.62 -1.44 3.79 1.41 0 0 0 0
[8] H 0 211 0 0.31 0.64 4.49 0.40 0 0 0 0
[9] H 0 211 0 2.47 -0.05 3.38 2.48 0 0 0 0
[10] H 0 -211 0 0.82 -0.42 3.51 0.91 0 0 0 0
\end{lstlisting}

In the end, we report the summary about the full JETSCAPE run.
We save the {\bf nTried, nSelected, nAccepted, sigmaGen, sigmaErr, eCM, pTHatMin, pTHatMax} quantities that represent the same quantities defined in the PYTHIA8.
\begin{lstlisting}[frame=single, language=C++]
# EVENT GENERATION INFORMATION
# nTried    = 35
# nSelected = 5
# nAccepted = 5
# sigmaGen  = 5.22636e-06
# sigmaErr  = 1.46469e-06
# eCM  = 2760
# pTHatMin  = 110
# pTHatMax  = 120
# /EVENT GENERATION INFORMATION
\end{lstlisting}

Users familiar with the above output format can write their own scripts to  extract the relevant information from the file saved in the hard disk containing full event record. Nevertheless, we do provide two sample scripts namely { \bf JetscapeFinalStateParton.cc } and {\bf JetscapeFinalStateHadron.cc} in the {\bf JETSCAPE/examples} directory which can be used within the JETSCAPE framework to extract the final state particles at the hadronic level and the partonic level from the data saved in the ASCII format.
In this method, we load the data saved in ASCII format using {\bf JetscapeReaderASCii} function and traverse through different events using {\bf Next()}. For a given event, one uses {\bf GetHadrons()} to get the final state particles at the hadronic level.
To access the final state parton one can use {\bf GetPartonShowers()}. Then, for each shower, one can call a function {\bf GetFinalPartons.at(i) } to access $i^{\mathrm{th}}$ final state partons in a given shower. Running the {\bf ./JetscapeFinalStateParton} and {\bf ./JetscapeFinalStateHadron}, we get the list of final state partons and hadrons saved in data file in ASCII format.
\begin{lstlisting}[frame=single, language=C++]
0 211 0 59.38 20.89 6.44 -55.20
1 -211 0 8.08 2.89 0.98 -7.48
2 211 0 1.45 0.69 0.33 -1.22
3 -211 0 14.21 5.03 1.34 -13.22
4 211 0 3.56 0.008 -1.11 -3.37
5 111 0 1.30 0.35 -0.11 -1.24
6 111 0 1.54 -0.18 -0.97 -1.17
7 -211 0 1.41 -0.49 -0.38 -1.25
8 211 0 0.40 -0.06 -0.30 0.21
9 211 0 2.48 -2.40 -0.58 -0.14
10 -211 0 0.91 -0.76 -0.30 -0.35
\end{lstlisting}
In above box each entry represents a final state hadron which are with information  (index, PID, particle status, $E , Px, Py, Pz$).

\lstset{
  frame=single,
  breaklines=true,
  showstringspaces=false,
}

\section{Download and Install JETSCAPE Framework}
In this section we describe downloading JETSCAPE framework from GitHub, its installation process on Mac and Linux operating systems, and running of code examples.

\subsection{Download}

The JetScape Framework can be downloaded using from GitHub using the following command:

\begin{lstlisting}[language=bash]
git clone https://github.com/JETSCAPE/JETSCAPE.git
\end{lstlisting}

This command creates a local copy of the code on you machine. The root directory contains CMakeLists.txt, AUTHORS, COPYING, and README files. The structure of the sub directories in the GitHub repository is as follows:

\begin{itemize}

\item \emph{cmakemodules} directory: containing cmake files for finding several packages used by the framework;
\item examples directory: containing different test programs and executables;
\item external\_packages directory: containing third party packages used by the framework;
\item src directory: the main directory containing source code of the framework and modules. This directory contains several sub directories for corresponding parts of the framework. The sub directories are: framework that contains source code of the classes defined and used by the JETSCAPE framework, hadronization that contains source code for hadronization modules, hydro that contains source code for hydrodynamics modules, initialstate that contains source code for initial state modules, jet that includes source code for energy loss modules, preequilibrium that contains source code for free streaming modules and, reader that contains the JETSCAPE reader source code.

\end{itemize}

\subsection{Installation with Docker}

Docker is a software tool that allows one to deploy an application in a portable environment. A docker ``image" can be created for the application,
allowing any user to run a docker ``container" from this image. We have prepared a docker image for the JETSCAPE environment, 
which allows you to use JETSCAPE on Mac, Windows and Linux without installing a long list of pre-reqs or worrying about interference with software you already have installed. 
Step-by-step instructions are provided below.

For those unfamiliar with Docker: To illustrate what this will look like, consider the following standard workflow. In a terminal on your machine (call it Terminal 1), you will clone JETSCAPE -- 
this terminal is on your ``host" machine -- just a standard, typical terminal. In another terminal (call it Terminal 2), you will invoke a command that runs a pre-defined docker container. 
Terminal 2 then lives entirely inside this docker container, completely separated from your host machine. It can only access the files that are inside that pre-defined docker container -- and not any 
of the files on your host machine -- unless we explicitly share a folder between them. The standard workflow that we envision is the following: You will share the folder containing 
JETSCAPE between the host machine and the docker container. Then, anytime you want to build or run JETSCAPE, you must do it inside the docker container. 
But anytime you want to edit text files (e.g. to construct your own configuration file), or analyze your output files, you can do this from your host machine (which we recommend). 
Simple as that: Depending which action you want to do, perform it either on the host machine, or in the docker container, as appropriate -- otherwise it will not work. \\

\noindent \textbf{Step 1: Install Docker} \\

\noindent \textit{Mac}
\begin{enumerate}
	\item Install Docker Desktop for Mac: https://docs.docker.com/docker-for-mac/install/
	\item Open Docker, go to Preferences $\rightarrow$ Advanced and:
	\begin{enumerate}
		\item Set CPUs to max that your computer has (sysctl -n hw.ncpu),
		\item Set memory to what you are willing to give Docker. \\
	\end{enumerate}
\end{enumerate}

\noindent \textit{Linux}
\begin{enumerate}
	\item Install Docker: https://docs.docker.com/install/
	\item Allow your user to run docker (requires admin privileges):
	\begin{lstlisting}[language=bash]
	sudo groupadd docker
 	sudo usermod -aG docker $USER
	\end{lstlisting}
	Log out and log back in. \\
\end{enumerate}

\noindent \textbf{Step 2: Run JETSCAPE}

The docker container will contain only the pre-requisite environment to build JETSCAPE, but will not actually contain JETSCAPE itself. 
Rather, we will create a directory on our own machine with the JETSCAPE code, and share this directory with the docker container. 
This will allow us to build and run JETSCAPE inside the docker container, but to easily edit macros and access the output files on our own machine.

\begin{enumerate}
	\item Make a directory on your machine (which will be shared with the docker container), and clone JETSCAPE into it.
	\begin{lstlisting}[language=bash]
mkdir ~/jetscape-docker
cd ~/jetscape-docker		
git clone https://github.com/JETSCAPE/JETSCAPE.git
	\end{lstlisting}
	\item Create and start a docker container that contains the JETSCAPE pre-reqs: \\
	
	\indent \textbf{Mac}:
	\begin{lstlisting}[language=bash]
docker run -it -v ~/jetscape-docker:/home/jetscape-user 
--name myJetscape jetscape/base:v1
	\end{lstlisting}
	
	\bigskip
	\indent \textbf{Linux}:
	\begin{lstlisting}[language=bash]
docker run -it -v ~/jetscape-docker:/home/jetscape-user 
--name myJetscape --user $(id -u):$(id -g) 
jetscape/base:v1
	\end{lstlisting}
	
	\bigskip
	This is what the \texttt{docker run} command does:
	\begin{itemize}
		\item \texttt{docker run} creates and starts a new docker container from a pre-defined image \texttt{jetscape/base:v1}, which will be downloaded if necessary.
                	\item \texttt{-it} runs the container with an interactive shell.
		\item \texttt{-v} mounts a shared folder between your machine (at \texttt{\$HOME/jetscape-docker}) and the container (at \texttt{/home/jetscape-user}), 
		through which you can transfer files to and from the container. You can edit the location of the folder on your machine as you like.
		\item \texttt{--name} (optional) sets a name for your container, for convenience. Edit it as you like.
		\item \texttt{--user \$(id -u):\$(id -g)} (only needed on linux) runs the docker container with the same user permissions as the current user on your machine 
		(since docker uses the same kernel as your host machine, the UIDs are shared). Note that the prompt will display ``I have no name!", which is normal.
	\end{itemize}
	
	\item Build JETSCAPE as usual:
	\begin{lstlisting}[language=bash]
        cd JETSCAPE
        mkdir build
        cd build
        cmake ..
        make -j4    # Builds using 4 cores
	\end{lstlisting}

\end{enumerate}
That's it! You are now inside the docker container, with JETSCAPE and all of its prerequisites installed. You can run JETSCAPE executables or edit and re-compile code. 
Moreover, since we set up the jetscape-docker folder to be shared between your host and the docker container, you can do text-editing etc. on your host machine, 
and then immediately build JETSCAPE in the docker container. Output files are also immediately accessible on your host machine for analysis. \\

\noindent Some useful commands:
\begin{itemize}
	\item To see the containers you have running, and get their ID: \texttt{docker container ls}  (to also see stopped containers, append \texttt{-a})
	\item To stop the container: \texttt{docker stop <container>} or \texttt{exit}
	\item To re-start the container: \texttt{docker start -ai <container>}
	\item To put a running container into detatched mode: \texttt{Ctrl-p Ctrl-q}, and to re-attach: \texttt{docker attach <container>}
	\item To delete a container: \texttt{docker container rm <container>}
\end{itemize}

\subsection{Installation on Mac}

Getting started on a Mac:

\begin{itemize}
\item Install Xcode and command-line tools
\end{itemize}

For further packages needed (like CMake, Pythia, ROOT, GraphViz) we recommend homebrew for Mac:

\begin{itemize}
\item  Install homebrew (after you install Xcode)

\begin{lstlisting}[language=bash]
/usr/bin/ruby -e "($curl fsSL https://raw.githubusercontent.com/\-Homebrew/install/master/install)"
\end{lstlisting}

\end{itemize}

\begin{itemize}
\item Install CMake via homebrew type:
\begin{lstlisting}[language=bash]
brew install cmake
\end{lstlisting}

\item Install doxygen:
\begin{lstlisting}[language=bash]
brew install doxygen
\end{lstlisting}

\item Tap some repos for further sceintific packages:
\begin{lstlisting}[language=bash]
brew tap davidchall/hep
\end{lstlisting}

\begin{lstlisting}[language=bash]
brew tap homebrew/science
\end{lstlisting}
\item Install Pythia8 for example:

\begin{lstlisting}[language=bash]
brew install pythia8
\end{lstlisting}
\item Install graphViz:

\begin{lstlisting}[language=bash]
brew install graphviz --with-app --with-bindings
\end{lstlisting}
\item Install root6

\begin{lstlisting}[language=bash]
brew install root6
\end{lstlisting}
\item Install graph-tool (python). If done you can create a colored and ``fancy'' graph with the provided python script.

\begin{lstlisting}[language=bash]
brew install graph-tool
\end{lstlisting}
\item Install hdf5

\begin{lstlisting}[language=bash]
brew tap homebrew/science
brew install hdf5 --enable-cxx
\end{lstlisting}
\item Install OpenMPI

\begin{lstlisting}[language=bash]
brew install open-mpi
\end{lstlisting}
\item Install GSL

\begin{lstlisting}[language=bash]
brew install gsl
\end{lstlisting}
\end{itemize}

Remark: So far on brew HepMC is only available in version 2, version 3
is required for the current code, nicer wirter interfaces to root for
example. So one has to install it from: http://hepmc.web.cern.ch/hepmc/

If cmake found other libraries HepMC, ROOT or Pythia8, you might have to
add the library path's in the setup.csh script. For sure works on Mac Os
X 10.11.6.

To make a class documentation using doxygen:

doxygen JetScapeDoxy.conf

\begin{itemize}

\item
  MUSIC support
\end{itemize}

MUSIC is a (3+1)D viscous hydrodynamical code developed at McGill
University. (Official website: http://www.physics.mcgill.ca/MUSIC) MUSIC
can be integrated into the JETSCAPE framework. To download the lastest
version of MUSIC, one can run the shell script under the 3rdparty
folder,

\begin{lstlisting}[language=bash]
    ./get_music.sh
\end{lstlisting}

This shell script will clone the latest version of MUSIC to the 3rdparty
folder. It also setup the environment variables for MUSIC to run.
Specifically, MUSIC needs the folder path for the EOS tables. Please
make sure the environment variable HYDROPROGRAMPATH to be set to the
path for MUSIC code package.

When compiling MUSIC with JETSCAPE, please turn on the MUSIC support
option when generating the cmake configuration file,

\begin{lstlisting}[language=bash]
    mkdir build
    cd build
    cmake -Dmusic=ON ..
    make
\end{lstlisting}

To run JETSCAPE with MUSIC,

\begin{lstlisting}[language=bash]
    ./MUSICTest
\end{lstlisting}

MUSIC uses openMP to perform parallel computations. To use multiple threads to run JETSCAPE with MUSIC, one can set environmental variable 

\begin{lstlisting}[language=bash]
    export OMP_NUM_THREADS=number_of_threads ./MUSICTest
\end{lstlisting}

Here number\_of\_threads stands for the number of threads one wants to use to run the program.

\subsection{Installation on Linux}

In order to install and compile the JETSCAPE framework on a Linux
machine, one need to follow the following steps:

\begin{enumerate}
\def\labelenumi{\arabic{enumi}.}

\item Clone the repository from Github:

  \begin{itemize}
  \item git clone https://github.com/JETSCAPE/JETSCAPE.git
  \end{itemize}

\item Install/update CMake (version  3.0.0 or higher)

  \begin{itemize}
  \item cmake --version
  \item If version is lower than 3.0, then
  \item Download a newer version
\begin{lstlisting}[language=bash]
wget https://cmake.org/files/v3.11/cmake-3.11.0-rc3-Linux-x86_64.sh
\end{lstlisting}
  \item Install CMake
\begin{lstlisting}[language=bash]
sh cmake-3.11.0-rc3-Linux-x86_64.sh
\end{lstlisting}
\item Set path for cmake binary file
\begin{lstlisting}[language=bash]
export PATH=$HOME/cmake-3.11.0-rc3-Linux-x86_64/bin/:$PATH
\end{lstlisting}
  \item Update system setups
\begin{lstlisting}[language=bash]
source ~/.bashrc
\end{lstlisting}

  \end{itemize}

\item Install and configure HepMC (Version 3.0.0 or higher)

  \begin{itemize}
  \item Create a folder for installing HepMC
  \item Go to the created folder
  \item Download HepMC
  \begin{lstlisting}[language=bash]
wget http://hepmc.web.cern.ch/hepmc/releases/hepmc3.0.0.tgz
   \end{lstlisting}
  \item Extract content of the downloaded file
  \begin{lstlisting}[language=bash]
tar -xvzf hepmc3.0.0.tgz
  \end{lstlisting}
  \item Go to the extracted folder
  \begin{lstlisting}[language=bash]
cd hepmc3.0.0/cmake
   \end{lstlisting}
  \item Run cmake command
  \begin{lstlisting}[language=bash]
cmake ..
    \end{lstlisting}

  \item Run make command (the user has to be sudoer)
  \begin{lstlisting}[language=bash]
  make all install
    \end{lstlisting}

  \item
    The default path to install HepMC is /usr/local/lib/
  \end{itemize}

\item
  Change Cmake configuration for HepMC (If Cmake could not find HepMC)

  \begin{itemize}

  \item
    Set ``HEPMC\_DIR'' as the root directory of HepMC
  \item
    Go to ../framework/Modules
  \item
    Open FindHEPMC.cmake

    \begin{itemize}

    \item
      vim FindHEPMC.cmake
    \end{itemize}
  \item
    Make sure that cmake looks for the correct ``include'' and ``lib''
    directories of HepMC

    \begin{itemize}

    \item
      By default is ``\$\{HEPMC\_DIR\}/include/''
    \item
      By default is ``\$\{HEPMC\_DIR\}/lib''
    \item
      Replace ``\$\{HEPMC\_DIR\}'' with ``ENV\{HEPMC\_DIR\}'' if you are
      not root
    \end{itemize}
  \item
    The library file that cmake must find is ``libHepMC.so''
  \end{itemize}
\item
  Install Pythia8

  \begin{itemize}

  \item
    Create a folder for installing Pythia8
  \item
    Got to the created folder
  \item Download Pythia8
  \begin{lstlisting}[language=bash]
http://home.thep.lu.se/~torbjorn/pythia8/pythia8226.tgz
  \end{lstlisting}
  \item Extract content of the downloaded file
  \begin{lstlisting}[language=bash]
tar -xvzf pythia8226.tgz
  \end{lstlisting}
  \item Go to the extracted folder
  \begin{lstlisting}[language=bash]
cd pythia8226
  \end{lstlisting}
  \item Run make command
  \begin{lstlisting}[language=bash]
make
   \end{lstlisting}
  \end{itemize}
  
\item
  Configure the address of Pythia8 in the ../framework/activate\_jetscape.sh (If
  cmake could not find Pythia)

  \begin{itemize}

  \item Open the configuration file
  \begin{lstlisting}[language=bash]
vim activate_jetscape.sh
  \end{lstlisting}
  \item
    Change ``PYTHIAINSTALLDIR'' to the installing folder of Pythia8
  \item
    Update the ``PYTHIA8DIR'' and ``PYTHIA8\_ROOT\_DIR'' address

    \begin{itemize}

    \item
      should be ``\$\{PYTHIAINSTALLDIR\}/pythia8226'' by default
    \end{itemize}
  \item Run the bash file
  \begin{lstlisting}[language=bash]
./activate_jetscape.sh
   \end{lstlisting}
  \item
    Make sure the variable are set into the session
  \item
    Use source if they are not set
  \end{itemize}
\item
  Install Boost libraries (Version 1.5 or higher)

  \begin{itemize}

  \item Download boost
  \begin{lstlisting}[language=bash]
wget https://dl.bintray.com/boostorg/release/1.64.0/source/boost\_1\_64\_0.tar.gz
    \end{lstlisting}
  \item Extract the content of the downloaded file
  \begin{lstlisting}[language=bash]
tar -xvzf boost_1_64_0.tar.gz
    \end{lstlisting}
  \item
    Go to the directory tools/build/.
  \item Run bootstrap.sh
  \begin{lstlisting}[language=bash]
./bootstrap.sh
  \end{lstlisting}
  \item
    Run b2 --prefix=PREFIX where PREFIX is the directory where you want
    Boost to be installed
   \begin{lstlisting}[language=bash]
./b2 install --prefix=PREFIX
    \end{lstlisting}
  \item Set ``BOOST\_ROOT'' variable to the root directory of boost
  \begin{lstlisting}[language=bash]
export BOOST_ROOT= <PREFIX>
  \end{lstlisting}
  \end{itemize}
\item
  Install HDF5 C++ library

  \begin{itemize}

  \item Download the HDF5 file
  \begin{lstlisting}[language=bash]
wget https://support.hdfgroup.org/ftp/HDF5/current18/src/hdf5-1.8.18.tar
   \end{lstlisting}
  \item Extract the content of the downloaded file
  \begin{lstlisting}[language=bash]
tar -xvvf hdf5-1.8.18.tar
  \end{lstlisting}
  \item Run configure
  \begin{lstlisting}[language=bash]
./configure
  \end{lstlisting}
  \item Run make
   \begin{lstlisting}[language=bash]
make
  \end{lstlisting}
  \item
    Set ``HDF5\_ROOT'' variable to the root directory of HDF5
   \begin{lstlisting}[language=bash]
export HDF5\_ROOT= <HDF5_ROOT_DIRECTORY>
    \end{lstlisting}
  \end{itemize}
\item Compiling JETSCAPE code

  \begin{itemize}

  \item Create a build folder in ../src/framework
  \begin{lstlisting}[language=bash]
mkdir build
  \end{lstlisting}
  \item Go to the build folder
  \begin{lstlisting}[language=bash]
cd build
  \end{lstlisting}

  \item Make sure \$LD\_LIBRARY\_PATH is set to the address of dynamic library files

    \begin{lstlisting}[language=bash]
 echo $LD_LIBRARY_PATH
    \end{lstlisting}
    \item If there is nothing to display
    \begin{lstlisting}[language=bash]
export LD_LIBRARY_PATH=<PATH_TO_DYNAMIC_LIBRARY_FILES>
    \end{lstlisting}
    \item If the address is not included in the variable
    \begin{lstlisting}[language=bash]
LD_LIBRARY_PATH=$LD_LIBRARY_PATH:<PATH_TO_DYNAMIC_LIBRARY_FILES>
    \end{lstlisting}
    
  \item Run cmake
   \begin{lstlisting}[language=bash]
cmake ..
   \end{lstlisting}
    \item
      If cmake cannot find HDF5 library, set ``-DCMAKE\_LIBRARY\_PATH''
      and ``-DCMAKE\_INCLUDE\_PATH'' flags when you run cmake

      \begin{lstlisting}[language=bash]
cmake -DCMAKE\_LIBRARY\_PATH= -DCMAKE\_INCLUDE\_PATH= ..
      \end{lstlisting}

  \item run make
    \begin{lstlisting}[language=bash]
make
    \end{lstlisting}
  \end{itemize}
\end{enumerate}

\subsection{Quick Start}

One can compile the JetScape Framework using,
\begin{lstlisting}[language=bash]
    mkdir build
    cd build
    cmake ..
    make
\end{lstlisting}

(Notice: if \textbf{cmake} is not installed, go to the Installation section)

Several test binaries will be generated in the build/ folder. To run the tests, you need to stay in the build/ folder and run:

\begin{lstlisting}[language=bash]
    ./brickTest
\end{lstlisting}

The brickTest simulates the jet energy loss in a medium with constant
temperature by default. The temperature \textbf{T} (in unit GeV) can be
modified in \emph{jetscape\_init.xml} file for the Brick section:

\begin{lstlisting}[language=XML]
    <Brick bjorken_expansion_on="false" start_time="0.6">
        <name>Brick</name>
        <T>0.3</T>
    </Brick>
\end{lstlisting}

If the \emph{bjorken\_expansion\_on} option is switched to
\textbf{true}, the brickTest simulates a more realistic medium expansion
where the temperature is uniform in the transverse plane but decreasing
with time due to the Bjorken expansion along the longitudinal direction.

The jet energyloss module consists of two testing modules, Matter
performing a random in time democratic split and Martini is doing
nothing. Furthermore in ``Matter'' a random ``new graph roots'' is added
for testing (simulating scattering with medium partons). In this test,
the switching criteria is arbitrarily set to 5GeV in pt. An ascii output
file is created which you can read in with the provided reader:

\begin{lstlisting}[language=bash]
    ./readerTest
\end{lstlisting}

which reads in the generated showers does some DFS search and shows the
output. You can generate an output graph format which can be easily
visualized using graphViz or other tools like Gephi (GUI for free for
Mac) or more adanvanced, graph-tools (Python) and many more.
Furthermore, as a ``closure'' test, the FastJet core packackage
(compiled in our JetScape library) is employed to perform a simple
jetfinding on the ``final'' partons (in graph language, incoming partons
in a vertex with no outgoing partons/childs). And since the jet shower
is perfectly collinear the jet pT is identical to the hard process
parton pT (modulo some random new partons/roots in the final state, see
above).

\section{Running JETSCAPE Framework}
To run JETSCAPE framework, we provide some executables as code examples for different types of Physics simulations. We will explain the provided code examples in Section~\ref{sec:examples}. Considering the fact that JETSCAPE is a modular framework, one can build a specific executable program by attaching various Physics modules to the main JETSCAPE task. We explain building a specific program in Section~\ref{sec:addmodule}.

\subsection{Code Examples}
\label{sec:examples}

In this section, we explain the provided code examples that can be run out of the box. Each of these code examples simulate a specific type of Physics that we discuss. For building your own program, look at the Section~\ref{sec:addmodule}.

\begin{lstlisting}[language=bash]
    ./brickTest
\end{lstlisting}

The \textbf{brickTest} simulates the jet energy loss in a static medium with a constant temperature, i.e., a brick. The temperature of the brick medium, as well as the time step and the total time for jet propagation can be set inside ``jetscape\_init.xml". For instance,
\begin{lstlisting}[language=bash]
    <Brick bjorken_expansion_on="false" start_time="0.6">
      <name>Brick</name>
      <T>0.2</T>
    </Brick>
\end{lstlisting}

\begin{lstlisting}[language=bash]
    <Eloss>
      <deltaT>0.1</deltaT>
      <formTime> -0.1</formTime>
      <maxT>20</maxT>
       ...
    <Eloss>
\end{lstlisting}
where $T=0.2$~GeV is set as the brick temperature, $\Delta t = 0.1$~fm is set as the evolution time step, and $t_\mathrm{max}=20$~fm is set as the maximum evolution time.

In ``brickTest", the jet partons are initialized with ``PGun", where their flavors are sampled between $g$, $u$, $d$ and $s$ with equal probability. If a quark flavor is selected, the sign of its flavor (quark or anti-quark) is then determined again with equal probability. The initial transverse momentum of all partons are set in ``jetscape\_init.xml" through
\begin{lstlisting}[language=bash]
    <PGun>
      <name>PGun</name>
      <pT>100</pT>
    </PGun>
\end{lstlisting}
where $p_\mathrm{T}=100$~GeV is set as an example.

One may choose different combinations of energy loss models (MATTER, LBT, MARTINI, AdS) to simulate the parton evolution. Detailed settings will be illustrated in Section~\ref{sec:addmodule}. However, note that when LBT is loaded, one need to download its required data files in "external\_packages/" through
\begin{lstlisting}[language=bash]
    ./get_lbtTab.sh
\end{lstlisting}
before compiling JETSCAPE.

\begin{lstlisting}[language=bash]
    ./PythiaBrickTest
\end{lstlisting}

The \textbf{PythiaBrickTest} simulates the jet energy loss in a brick medium in the same way as ``brickTest" does. The only difference is now partons are initialized with ``PythiaGun" instead of ``PGun". Pythia 8 is utilized to generate the initial hard scatterings in proton-proton collisions, providing a list of partons for their subsequent evolution through JETSCAPE. One may set basic Pythia configurations through ``jetscape\_init.xml". For example,
\begin{lstlisting}[language=bash]
    <PythiaGun>
      <name>PythiaGun</name>
      <FSR_on>0</FSR_on>
      <pTHatMin>110</pTHatMin>
      <pTHatMax>120</pTHatMax>
      <eCM>2760</eCM>
      </PythiaGun>
    </PythiaGun>
\end{lstlisting}
where we set the center-of-mass energy of proton-proton collisions as 2760~GeV, the range of transverse momentum exchange of hard scatterings as 110 to 120~GeV, and final-state shower (FSR) turned off in Pythia.

\begin{lstlisting}[language=bash]
    ./hydroFromFileTest
\end{lstlisting}

The \textbf{hydroFromFileTest} simulates the jet energy loss in a medium simulated by relativistic hydrodynamics. The evolution history of the bulk medium is stored in an external file and is read in at runtime. The temperature and fluid velocity at any given space-time position are available for the jet energy loss modules. The hydro file reader is from a 3rd-party program. Part of the code requires the HDF5 library. Some hydrodynamical profile examples can be downloaded with
\begin{lstlisting}[language=bash]
    ./get_hydroSample_PbPb2760_cen_00-05.sh
    ./get_hydroSample_PbPb2760_cen_20-30.sh
    ./get_hydroSample_PbPb2760_cen_30-40.sh
\end{lstlisting}
They are generated with smooth TRENTO initial conditions, followed by the free-streaming model and the (2+1)-D VISHNEW hydrodynamic model for 0-5\%, 20-30\% and 30-40\% centrality bins of 2.76~TeV Pb-Pb collisions. In this example of ``hydroFromFileTest", jet partons are initialized with ``PythiaGun", as introduced for ``PythiaBrickTest" above, for their momentum space distribution. Their position space distribution are sampled based on the TRENTO initial condition which are also included in the downloaded hydrodynamic profiles. The jet energy loss modules are the same as the ones in ``brickTest" and ``PythiaBrickTest".

\begin{lstlisting}[language=bash]
    ./hydroFromFileTestPGun
\end{lstlisting}

The \textbf{hydroFromFileTestGun} is a simplified version of ``hydroFromFileTest" above. Instead of using ``PythiaGun" to generate the momenta of initial partons, ``PGun" is used as introduced for ``brickTest" above.

After running a given code example, a ``test\_out.dat" file is usually generated, which contains the the evolution history of all particles. For the convenience of data analysis, one may run
\begin{lstlisting}[language=bash]
    ./FinalStatePartons
    ./FinalStateHadrons
\end{lstlisting}
to obtain the data file of final-state partons and hadrons.

\begin{lstlisting}[language=bash]
    mpirun -np 1 ./MUSICTest
\end{lstlisting}

(Notice: several libraries are needed to run MUSICTest, go to the
Installation section for details)

The \textbf{MUSICTest} simulates the jet energy loss in a realistic
medium whose local entropy density, temperature and fluid velocities are
provided by 3+1D viscous hydrodynamic simulations of heavy ion
collisions using MUSIC from McGill group. The fluctuating initial
condition is given by TRENTO Monte Carlo event generator from Duke
group. TRENTO uses parameterization to approximate the MC Glauber, MC
KLN, IP-Glasma and EKRT initial conditions. The default module used in
JETSCAPE is the IP-Glasma-like initial condition, which provides both
the distribution of jet production position, and the entropy density
distribution of the bulk medium. The jet energy loss module is the same
as the one in the brickTest. MUSIC is a 3rd party program outside the
JETSCAPE framework. It requires MPI and GSL libraries. In order to run
MUSICTest, please see the details in the Installation section.

\subsection{Build Your Own Program}
\label{sec:addmodule}

As an event generator, the JETSCAPE framework includes all the necessary Physics modules to simulate heavy ion collisions. However, the main purpose of JETSCAPE framework is to enable physicists to develop their own modules, build their own programs and run them using the JETSCAPE framework. Modules can be developed related to different stages of simulation, from initial state and hydrodynamics to energy loss and hadronization. Programs are executables that include different modules being attached to a main task and produce the results. In this section, we first discuss developing JETSCAPE modules corresponding to different stages of simulation. Next we discuss an example of building a program to incorporate different JETSCAPE modules and get some results.

\subsubsection{Building Programs}
\label{sec:prog}
In order to build a JETSCAPE program, first, one has to define the main JETSCAPE task of type \texttt{JetScape}. Then the other Physics modules that you wish to include, must be declared and attached to the main task. Different Physics modules can be attached to the main task; from initial state and hydrodynamics modules to energy loss and hadronization modules. For energy loss and hadronization module, one has to define a manager, then attaches the corresponding Physics modules to the manager and finally, attaches the manager to the main task. All the necessary header files must be included in the executable programs.

After attaching the modules, one need to call the \texttt{Init()}, \texttt{Exec()} and \texttt{Finish()} methods for the main task. Those functions call the \texttt{Init()}, \texttt{Exec()} and \texttt{Finish()} methods for each module, respectively. The calling continues recursively, for all the attached tasks and sub-tasks. For the output, one can add a writer module to the main task and specifies the name of the output file.

To illustrate this better, let's look at the brickTest.cc code:

\begin{lstlisting}[basicstyle=\footnotesize\ttfamily, language=C++]

#include <iostream>
#include <time.h>

// JetScape Framework includes ...
#include "JetScape.h"
#include "JetEnergyLoss.h"
#include "JetEnergyLossManager.h"
#include "JetScapeWriterAscii.h"
#include "JetScapeWriterAsciiGZ.h"
#include "JetScapeWriterHepMC.h"

// User modules derived from jetscape framework clasess
// to be used to run Jetscape ...
#include "AdSCFT.h"
#include "Matter.h"
#include "LBT.h"
#include "Martini.h"
#include "Brick.h"
#include "GubserHydro.h"
#include "PGun.h"
#include "HadronizationManager.h"
#include "Hadronization.h"
#include "ColoredHadronization.h"
#include "ColorlessHadronization.h"

// Add initial state module for test
#include "TrentoInitial.h"

#include <chrono>
#include <thread>

using namespace Jetscape;

int main(int argc, char** argv)
{
  clock_t t; t = clock();
  time_t start, end; time(&start);
  cout<<endl;
  // DEBUG=true by default and REMARK=false
  // can be also set also via XML file (at least partially)
  JetScapeLogger::Instance()->SetInfo(true);
  JetScapeLogger::Instance()->SetDebug(false);
  JetScapeLogger::Instance()->SetRemark(false);
  //SetVerboseLevel (9 a lot of additional debug output ...)
  //If you want to suppress it: use SetVerboseLevle(0) or max  SetVerboseLevle(9) or 10
  JetScapeLogger::Instance()->SetVerboseLevel(8);
  Show();
  // defining the main task and the number of events
  auto jetscape = make_shared<JetScape>("./jetscape_init.xml",2);
  jetscape->SetId("primary");
  // jetscape->set_reuse_hydro (true);
  // jetscape->set_n_reuse_hydro (10);
  // defining modules as tasks
  auto jlossmanager = make_shared<JetEnergyLossManager> ();
  auto jloss = make_shared<JetEnergyLoss> ();
  auto trento = make_shared<TrentoInitial>();
  auto hydro = make_shared<Brick> ();
  //auto hydro = make_shared<GubserHydro> ();
  auto matter = make_shared<Matter> ();
  auto lbt = make_shared<LBT> ();
  auto martini = make_shared<Martini> ();
  auto adscft = make_shared<AdSCFT> ();
  //matter->SetActive(false);
  //martini->SetActive(false);
  //jloss->SetActive(false);
  auto pGun= make_shared<PGun> ();
  auto hadroMgr = make_shared<HadronizationManager> ();
  auto hadro = make_shared<Hadronization> ();
  auto hadroModule = make_shared<ColoredHadronization> ();
  auto colorless = make_shared<ColorlessHadronization> ();
  // only pure Ascii writer implemented and working with graph output ...
  auto writer= make_shared<JetScapeWriterAscii> ("test_out.dat");
  // autowriter= make_shared<JetScapeWriterAsciiGZ> ("test_out.dat.gz");
  // auto writer= make_shared<JetScapeWriterHepMC> ("test_out.hepmc");
  //writer->SetActive(false);

  // Attaching modules to the main tasks
  jetscape->Add(pGun);
  jetscape->Add(trento);
  jetscape->Add(hydro);
  jloss->Add(matter);
  // go to 3rd party and ./get_lbtTab before adding this module
  //jloss->Add(lbt);
  //jloss->Add(martini);
  //jloss->Add(adscft);
  jlossmanager->Add(jloss);
  jetscape->Add(jlossmanager);
  hadro->Add(hadroModule);
  //hadro->Add(colorless);
  hadroMgr->Add(hadro);
  jetscape->Add(hadroMgr);
  jetscape->Add(writer);

  // Calling recursive functions
  // Intialize all modules tasks
  jetscape->Init();
  // Run JetScape with all task/modules as specified ...
  jetscape->Exec();
  jetscape->Finish();

  INFO_NICE<<"Finished!";
  cout<<endl;

  t = clock() - t;
  time(&end);
  printf ("CPU time: %f seconds.\n",((float)t)/CLOCKS_PER_SEC);
  printf ("Real time: %f seconds.\n",difftime(end,start));
  //printf ("Real time: %f seconds.\n",(start-end));
  return 0;
}

\end{lstlisting}

\section{Benchmarking of JETSCAPE}
In this Appendix, we benchmark the running time and memory consumption for two types of experiments using the JETSCAPE framework. First, we show the running time for simple proton-proton collisions with no medium. Next, we describe the running time numbers for heavy-ion collisions that include hydrodynamics and freestreaming modules.
At the end of this Appendix, we discuss the potential performance gain from optimizing the code to use Graphic Processing Unit (GPU) to run modules inside the JETSCAPE framework.

\subsection{Proton-Proton Collisions}
In Table~\ref{table:ppnumbers}, we benchmark the running time of the JETSCAPE framework for proton-proton collisions. We show the numbers for each attached module to the framework separately. In this benchmark, we simulate 1,000 events for the energy level 2.76 TeV. In total, it takes 150 seconds for all the events, so the runtime for each event is around 0.15 seconds. The memory required for this simulation is around 150MB. The simulation is run on one of the Wayne State University Grid nodes with Intel E5-2643 processor that has four CPUs. Each CPU is a 64-bit AMD quad-core dual-processor with 256GB of RAM. The clock speed of each CPU is 3.3GHz.

\begin{table}[htp]
\footnotesize
\begin{center}
\begin{tabular}{|l|c|c|c|}
\hline
\textbf{Modules} & \textbf{Total Time(Seconds)} & \textbf{Average Time per Event(Seconds)} \\
\hline
TRENTO                      & 3       & 0.003     \\ \hline
PythiaGun                   & 6       & 0.006     \\ \hline
BrickHydro                  & 0.04    & 4x10\^(-5) \\ \hline
MATTER                      & 125     & 0.125     \\ \hline
Colored Hadaronization      & 4       & 0.004     \\ \hline
Colorless Hadaronization    & 2       & 0.002     \\ \hline
ASCII Writer                & 8       & 0.008     \\ \hline
\end{tabular}
\end{center}
\caption{Running time for separate modules included in the JETSCAPE framework for proton-proton collisions.}
\label{table:ppnumbers}
\end{table}%

\subsection{Heavy-Ion Collisions}
In Table~\ref{table:hinumbers}, we benchmark the running time of heavy-ion collisions using the JETSCAPE framework and for a single event. This simulation includes TRENTO plus freestreaming module for initial state. For hydrodynamics module, we attach 2+1D MUSIC. We use Stampede login node as the machine to run this simulation.

\begin{table}[h]
\footnotesize
\begin{center}
\begin{tabular}{|l|c|c|c|}
\hline
\textbf{Modules} & \textbf{Total Memory (MB)} & \textbf{Total Time(Seconds)} \\
\hline
Freestreaming                  & 71     & 1           \\ \hline
MUSIC               & 341    & 112           \\ \hline
\end{tabular}
\end{center}
\caption{Running time for separate modules included in the JETSCAPE framework for heavy-ion collisions (one event).}
\label{table:hinumbers}
\end{table}%

\subsection{GPU Optimization}
There is performance gain to be had once accelerators are incorporated inside the JETSCAPE framework. Parallelization may be introduced to speed-up the most computation-extensive parts of the execution, e.g., hydrodynamics modules. One promising paradigm is Graphic Processing Unit (GPU) for high performance computing. Modern GPUs can have thousands of processing elements than can be utilized to do calculations simultaneously. Using GPUs for a 3+1D hydrodynamics module can lead an order of magnitude speed-up, when compared to standard multi-core CPUs. In a recent work by \cite{Pang:2018zzo}, it was found that AMD GPUs can yield a factor of 6 increase compared to multi-core CPUs (see \cite{Pang:2018zzo} for further details).

\label{benchmark} 

\end{appendices}





\bibliographystyle{elsarticle-num}
\bibliography{references,refs}







\end{document}